\documentclass[12pt]{article}
\usepackage{jheppub}
\pdfoutput=1
\usepackage[utf8]{inputenc}
\usepackage{amssymb}
\usepackage{graphicx}
\usepackage{amsmath}
\usepackage{multirow}
\usepackage{verbatim}
\usepackage{rotating}
\usepackage{pdflscape}

\usepackage{graphics}
\usepackage{pstricks}
\usepackage{bm}
\usepackage{amsfonts}
\usepackage{bbm}
\usepackage[FIGTOPCAP,scriptsize,tight]{subfigure}

\usepackage[textsize=scriptsize,backgroundcolor=red!70,linecolor=red]{todonotes}
\usepackage{paralist}
\usepackage{arydshln}
\usepackage[normalem]{ulem}

\newcommand{\nue}{\ensuremath{\nu_e}}

\newcommand{\anue}{\ensuremath{\bar{\nu}_e}}

\newcommand{\be}{\begin{eqnarray}}
\newcommand{\beq}{\begin{equation}}
\newcommand{\eeq}{\end{equation}}
\newcommand{\ee}{\end{eqnarray}}

\newcommand{\bmp}{\noindent\begin{minipage}{16cm}}
\newcommand{\emp}{\end{minipage}\vskip 7mm} 

\newcommand{\order}[1]{\mathcal{O}(#1)}

\newcommand{\anti}[1]{\ensuremath{\bar{#1}}}

\newcommand{\dm}{\ensuremath{\text{DM}}}

\newcommand{\chisq}{\ensuremath{\chi^{2}}}

\newcommand{\tdm}{\ensuremath{\tau_{\dm}}}
\newcommand{\mdm}{\ensuremath{m_{\dm}}}
\newcommand{\ndm}{\ensuremath{N_{\dm}}}
\newcommand{\nast}{\ensuremath{N_{\rm astro}}}

\newcommand{\phia}{\ensuremath{\phi_{\rm astro}}}

\newcommand{\sv}{\ensuremath{\langle \sigma v \rangle}}

\preprint{{\tt IFIC/19-19}}

\title{Update on decaying and annihilating heavy dark matter with the 6-year IceCube HESE data}
 
\author[a]{Atri Bhattacharya,}
\author[b]{Arman Esmaili,}
\author[c]{Sergio Palomares-Ruiz,} 
\author[d,e]{Ina Sarcevic.}

\affiliation[a]{Space sciences, Technologies and Astrophysics Research (STAR) Institute,
                Universit\'{e} de Li\`{e}ge, B\^{a}t.~B5a, 4000 Li\`{e}ge,
                Belgium}
\affiliation[b]{Departamento de F\'{\i}sica, Pontif\'{\i}cia Universidade Cat\'olica do Rio de Janeiro, C. P. 38071, 22452- 970, Rio de Janeiro, Brazil}
\affiliation[c]{Instituto de F\'{\i}sica Corpuscular (IFIC),
                CSIC-Universitat de Val\`{e}ncia,  
	              Apartado de Correos 22085, E-46071 Valencia, Spain}
\affiliation[d]{Department of Physics, University of Arizona, 1118 E.\ 4th St.\ Tucson, AZ 85704}
\affiliation[e]{Department of Astronomy, University  of Arizona, 933 N.\ Cherry Ave., Tucson, AZ 85721}

\emailAdd{a.bhattacharya@ulg.ac.be}
\emailAdd{arman@puc-rio.br}
\emailAdd{sergio.palomares.ruiz@ific.uv.es}
\emailAdd{ina@physics.arizona.edu}

\date{\today}

\abstract{
In view of the IceCube's 6-year high-energy starting events (HESE) sample, we revisit the possibility that the updated data may be better explained by a combination of neutrino fluxes from dark matter decay and an isotropic astrophysical power-law than purely by the latter. We find that the combined two-component flux qualitatively improves the fit to the observed data over a purely astrophysical one, and discuss how these updated fits compare against a similar analysis done with the 4-year HESE data. We also update fits involving dark matter decay via multiple channels, without any contribution from the astrophysical flux. We find that a DM-only explanation is not excluded by neutrino data alone. Finally, we also consider the possibility of a signal from dark matter annihilations and perform analogous analyses to the case of decays, commenting on its implications.    
}

\begin{document}
\maketitle

\section{Introduction}
\label{sec:intro}

Since the discovery of the first astrophysical high-energy neutrinos at the IceCube neutrino telescope~\cite{Aartsen:2013jdh}, more neutrinos have been steadily detected~\cite{Aartsen:2014gkd, Aartsen:2015rwa, Kopper:2015vzf, Aartsen:2016xlq, Kopper:2017zzm, Haack:2017dxi, Wandkowsky:2018}. This has opened the era of neutrino astronomy and understanding the origin of this high-energy neutrino flux is currently one of the most important problems in astroparticle physics~\cite{Laha:2013lka, Halzen:2013dva, Waxman:2013zda, Anchordoqui:2013dnh, Murase:2014tsa, Meszaros:2017fcs, Ahlers:2018fkn}. 

From the observational point of view, different types of event samples have been considered and a broad picture of the astrophysical neutrino flux has emerged. It can be broadly summarized as follows. The high-energy starting events (HESE), with electromagnetic (EM)-equivalent deposited energies above 10~TeV and produced inside the detector, can be explained as produced by a soft neutrino spectrum with a power-law index ($E_\nu^{-\gamma}$) close to $\gamma \simeq 2.9$~\cite{Kopper:2017zzm, Wandkowsky:2018}. Neutrinos of all flavors contribute to this sample and two topologies have been discriminated so far: muon tracks, mainly produced from muon neutrinos, and showers, produced by all flavors. On the other hand, through-going muon events, produced by up-going muon neutrinos (i.e, from the Northern hemisphere), are sensitive to energies above $\sim$ 200~TeV and can be understood in terms of a harder spectrum, $\gamma \sim 2.2$~\cite{Aartsen:2015rwa, Aartsen:2016xlq, Haack:2017dxi}. This tension has motivated the proposal of two-component fluxes~\cite{Chen:2014gxa, Aartsen:2015knd, Palladino:2016zoe, Vincent:2016nut, Palladino:2016xsy, Palladino:2017qda}, not necessarily isotropic and with a softer spectrum for $E_\nu \lesssim 100$~TeV, or the possible existence of a break or softening in the spectrum~\cite{Winter:2014pya, Anchordoqui:2014hua, Palomares-Ruiz:2015mka, Aartsen:2015zva, Anchordoqui:2016ewn}.

On another hand, based on the near isotropy of the observed events~\cite{Aartsen:2013jdh, Ahlers:2013xia, Aartsen:2014gkd, Anchordoqui:2014rca, Troitsky:2015cnk, Kistler:2015oae, Palladino:2016zoe, Vincent:2016nut, Denton:2017csz, Albert:2017oba, Aartsen:2017ujz, Krings:2017pif}, extragalactic sources are favored. Only a relatively small fraction of the neutrino flux seems to be compatible with a galactic disc origin~\cite{Denton:2017csz, Albert:2017oba, Aartsen:2017ujz, Krings:2017pif}. Note, however, that the limits on the galactic contribution are usually obtained using spatial and energy templates for diffuse neutrino emission, so different conclusions might be reached for other distributions~\cite{Pagliaroli:2017fse}. Thus, currently a galactic contribution to the total flux cannot be excluded, especially if the galactic contribution comes from the halo. Interestingly, the recent observation of the hardening of the $\gamma$-ray flux above $\sim$ 300~GeV in the Fermi-LAT diffuse data at large galactic latitudes seems to point to the existence of a correlated galactic neutrino flux with a hard spectrum~\cite{Neronov:2018ibl} (see also Ref.~\cite{Neronov:2016bnp}). An example of such a spectrum could be the one produced from the decays of heavy dark matter (DM) particles. In this case, a significant fraction of the IceCube neutrinos would come from DM decaying in the halo of our galaxy, while the total neutrino flux from DM decays would result from the sum of the galactic and extragalactic (isotropic) contributions. The consistency, or even mild preference, of the angular distribution of the IceCube events with such a scenario has already been discussed~\cite{Bai:2013nga, Esmaili:2014rma, Troitsky:2015cnk, Chianese:2016opp}, without a conclusive answer yet.

The possibility to explain part or all of the neutrino event spectrum observed by IceCube in terms of heavy DM decays or annihilations has been extensively considered~\cite{Feldstein:2013kka, Esmaili:2013gha, Bai:2013nga, Ema:2013nda, Bhattacharya:2014vwa, Zavala:2014dla, Higaki:2014dwa, Ema:2014ufa, Rott:2014kfa, Esmaili:2014rma, Fong:2014bsa, Daikoku:2015vsa, Murase:2015gea, Esmaili:2015xpa, Aisati:2015vma, Roland:2015yoa, Anchordoqui:2015lqa, Boucenna:2015tra, Ko:2015nma, Troitsky:2015cnk, EsmailiTaklimi:2016bbx, Esmaili:2016swq, Chianese:2016opp, Dev:2016uxj, Fiorentin:2016avj, Dev:2016qbd, DiBari:2016guw, Chianese:2016smc, Chianese:2016kpu, Kuznetsov:2016fjt, Cohen:2016uyg, Borah:2017xgm, Hiroshima:2017hmy, Bhattacharya:2017jaw, Chakravarty:2017hcy, Chianese:2017nwe, Dhuria:2017ihq,  Aartsen:2018mxl, Sui:2018bbh, Chianese:2018ijk, Blanco:2018esa}. Recently, the IceCube collaboration published two analyses using independent datasets~\cite{Aartsen:2018mxl}: 6-year muon data~\cite{Aartsen:2016xlq} and all-sky cascades from the 2-year medium energy starting events (MESE)~\cite{Aartsen:2014muf}. By only considering hard channels ($H + \nu$ and $Z + \nu$), a non-zero DM decay contribution was found for the best fits, including an astrophysical (isotropic) power-law flux. Nevertheless, this was not statistically significant and only limits on the lifetime were derived for these channels as well as for other ones, although in those other cases only using the cascade analysis. On the other hand, the first DM decay analysis using the 6-year HESE data~\cite{Kopper:2017zzm} was performed for fixed values of the spectral index of the astrophysical flux ($\gamma = 2.0, \, 2.2$)~\cite{Chianese:2017nwe}, and a best fit for DM masses of $\sim 400 - 500$~TeV was found. Another recent work~\cite{Sui:2018bbh} made use of the 6-year HESE data~\cite{Kopper:2017zzm} and additionally of the 8-year through-going muon data~\cite{Haack:2017dxi}, which dominates the statistics. This analysis, however, considered only a particular DM decay model, with a final hard neutrino spectrum. It was also found that adding a DM decay contribution, with a DM mass $\sim 300$~TeV, to an astrophysical power-law flux, provides a better fit to data.

In this work we consider just the 6-year HESE data above 60 TeV in EM-equivalent deposited energy,\footnote{Note that preliminary results using 7.5-year HESE data have already been presented~\cite{Wandkowsky:2018}, including limits on the DM contribution~\cite{Yuan:2018}, although details about the charge recalibration and the improved reconstructions are not public yet.} and evaluate whether a DM contribution could provide a good fit to the HESE data and reduce by itself the tension with the through-going muon data. This is an update of our previous detailed analysis, which used the 4-year HESE data~\cite{Bhattacharya:2017jaw}. Thus, one of the main questions is whether our previous conclusions based on 4-year data are substantially altered when additional two years of data are included, which we will discuss thoroughly in this paper. In particular, we also consider the case of DM decaying into two channels (leaving the relative branching ratio as a free parameter),  so that the entire HESE data could be explained without the need for a power-law astrophysical flux. As previously found~\cite{Esmaili:2013gha, Esmaili:2014rma, Bhattacharya:2017jaw}, we confirm that current HESE data could be interpreted in terms of this scenario.\footnote{Nevertheless, it could be in tension with gamma-ray data, as we mention below.} This may seem to be at odds with the claim in Ref.~\cite{Aartsen:2018mxl}, which concluded that a DM-only scenario is disfavored, by neutrino data, with respect to a DM plus a power-law flux. However, apart from the fact that the referred tension is below the 1$\sigma$ confidence level (CL), that conclusion is based on a fixed combination of hard channels for DM decay.

Additionally, we also consider the possibility to explain the HESE data, in part or entirely, in terms of neutrinos from DM annihilations. Due to the dependence on the square of the DM density, this scenario could present a sharper enhancement towards the galactic center. To study this, we leave the normalization of the cosmological DM clumping factor as a free parameter, which modifies the relative importance of the galactic and extragalactic contributions and thus, both the spectral and the spatial distributions of the neutrino flux. Although it is well known that this requires very large annihilation cross sections~\cite{Feldstein:2013kka, Zavala:2014dla, Esmaili:2014rma, Chianese:2016opp, Chianese:2016kpu}, these data could, in principle, probe regions below the unitarity bound~\cite{Griest:1989wd, Hui:2001wy}. We will see, however, that this is not the case with the current HESE data.

The main consideration imposed on our analysis due to the addition of two years worth of additional data, apart from the obvious improvement in statistics, comes from the proliferation of sub-PeV events without a balancing increase in the PeV events. The tension between data and uniform power-law fits, already exaggerated due to this tilting of the spectrum in favor of sub-PeV events, has been quantified as a $2.6\sigma\ (2.1\sigma) $ bump in events between EM-equivalent deposited energies of $60-100$~TeV when compared against a flat power-law spectrum with spectral index $ \alpha = 2.0\ (2.2) $~\cite{Chianese:2017nwe}.

In view of these qualitative changes to the HESE data, we explore the nature of fits from combined DM (decays or annihilations) and power-law astrophysical neutrino fluxes, specifically looking at how these affect our previous 4-year HESE analysis. 

The paper is organized as follows: in section~\ref{sec:analysis} we briefly introduce the HESE sample we use and describe the main ingredients of the analysis we perform. In section~\ref{sec:DMdecays} we present the results for the case of DM decays and in section~\ref{sec:DMann} the corresponding ones for the annihilating scenario. Finally, in section~\ref{sec:conc} we draw our conclusions.

\section{Analysis of the 6-year HESE data}
\label{sec:analysis}

The 6-year HESE data we use in this work correspond to 2078 days of runtime during which, 82 events have been observed (28 new events in the last two years) with EM-equivalent deposited energies above 10~TeV. Two of these events, however, were produced by a coincident pair of background muons from unrelated cosmic-ray air showers and are not included in our analysis. In this work we consider the EM-equivalent deposited energy interval [60~TeV--10~PeV], which contains 50 events (18 from the last two years), classified as either tracks or showers. 

At these energies, cosmic-ray secondaries produced in the atmosphere, muons and neutrinos, are the main sources of background. As done in our previous analysis~\cite{Bhattacharya:2017jaw}, here we do not consider the potential contribution from prompt atmospheric neutrinos from charm decays. For energies above 60~TeV, the number of background atmospheric muon and neutrino events for 2078 days is taken as:\footnote{For the muon background we use the IceCube numbers, whereas for the neutrino background we scale it with respect to the 4-year data~\cite{Kopper:2015vzf}. We do this because in the 6-year analysis slightly larger cross sections have been used~\cite{Kopper:2017zzm} and we have not corrected for this. The difference, though, is only about 0.6 events.} $N_\mu = 1.3$ and $N_\nu = 5.1$.

The two additional contributions we consider in this work for the neutrino flux at Earth are: an isotropic power-law flux and the flux from DM decays or annihilations,
\begin{equation}
\label{eq:flux-osc}
\left. \frac{d \Phi_{\nu_\alpha}}{d E_\nu} \right|_\oplus = \phia \, \left(\frac{E_\nu}{\rm 100~TeV}\right)^{-\gamma} +  \sum_{\beta } \sum_i |U_{\alpha i}|^2 \, |U_{\beta i}|^2 \, \frac{d \Phi_{{\rm DM}, \nu_\beta}}{d E_\nu} ~, 
\end{equation}
where $\gamma$ is the spectral index and $\phia$ is the astrophysical flux normalization (per flavor). Here we assume the canonical flavor composition from hadronic sources after averaged oscillations, $(1 : 1 : 1)_\oplus$, for both neutrinos and antineutrinos, and hence, $\phia$ is flavor-independent. $U$ is the neutrino mixing matrix and the above product represents the oscillation probability of flavor neutrino $\nu_\alpha$ into $\nu_\beta$, after all terms dependent on mass-squared differences are averaged out. The term $\Phi_{{\rm DM}}$ corresponds to the flux of neutrinos from DM decays or annihilations at production. It depends on the DM mass, \mdm, on the decay/annihilation channel(s), and for decays, on the lifetime, \tdm, and for annihilations, on the annihilation cross section, \sv, and on the clumping factor (see below). In both cases, decays and annihilations, the neutrino flux has two contributions: an extragalactic component, which originates from DM particles in halos at all redshifts, and thus it is isotropic; and a galactic contribution, which comes from DM in our galactic halo and it is anisotropic. Whereas the DM decay flux depends linearly on the DM density, the flux from DM annihilations depends on its square. Therefore, the galactic contribution is more sharply peaked towards the galactic center in the case of annihilations. For the details of our computation of the neutrino flux from DM decays we refer the reader to our previous work~\cite{Bhattacharya:2017jaw}. The flux from DM annihilations is described below.

Analogously to Ref.~\cite{Bhattacharya:2017jaw}, to which we refer the reader for details, we perform unbinned extended maximum likelihood analyses using the EM-equivalent deposited energy, the event topology, and the hemisphere of origin of each of the 50 events in the 6-year HESE sample within [60~TeV--10~PeV]. For the description of the computation of the event spectra of both signal and background, we refer the reader to Refs.~\cite{Palomares-Ruiz:2015mka, Vincent:2016nut}.

\section{DM decays}
\label{sec:DMdecays}

The neutrino flux from DM decays depends on the DM lifetime, \tdm, the mass, \mdm, and the decay channels. In this work we only consider two-body final state channels and perform two types of fits:
\begin{inparaenum}[\itshape 1\upshape)]
  \item involving a single channel DM decay plus an isotropic power-law flux and
  \item where the entire spectrum is explained purely by neutrinos from DM decay via two channels.
\end{inparaenum}
In the former case, the total neutrino flux for a given decay channel is described in terms of the set of free parameters $\boldsymbol{\theta} = \{\tdm, \mdm, \phia, \gamma\}$, while in the latter scenario, for each pair of channels, it is described in terms of $\boldsymbol{\theta} = \{\tdm, \mdm, {\rm BR}\}$,
where ${\rm BR}$ is the branching ratio for decay into one of the two channels.

In the case of decays, the relative contribution between the galactic and the extragalactic components is determined by the spatial distribution of the galactic density. The extragalactic component is isotropic and does not depend on the DM clustering properties across redshifts.

\begin{table}[t]
	\begin{center}
		\begin{tabular}{c|cc|cc}
			\hline
			Decay channel & $ \tau_{\rm DM} [10^{28}~{\rm s}] \, \, (\ndm)$  &  $ \mdm $ [TeV] &
			$\phia \, \, (\nast) $  & $\gamma$ \\
			\hline
			$u\bar{u}$                &       0.11 \, \, \, {\it (28.4)} &      1761 &       0.52 \, {\it (13.0)} &      2.34 \\
			$b\bar{b}$                &       0.07 \, \, \, {\it (26.9)} &      1103 &       0.58 \, {\it (14.3)} &      2.35 \\
			$t\bar{t}$                &       0.11 \, \, \, {\it (28.7)} &      598 &       0.45 \, {\it (12.5)} &      2.27 \\
			$\boldsymbol{W^{+}W^{-}}$              &    \textbf{0.37} \, \, \, \textbf{\textit{(28.5)}} &    \textbf{412} &     \textbf{0.47} \, \textbf{\textit{(12.6)}} &    \textbf{2.29} \\
			$Z Z$                     &       0.43 \, \, \, {\it (27.8)} &      407 &       0.52 \, {\it (13.3)} &      2.32 \\
			$h h$                     &       0.12 \, \, \, {\it (28.8)} &      611 &       0.45 \, {\it (12.6)} &      2.27 \\
			$e^{+}e^{-}$              &       2.20 \, \, \, {\it ( 4.0)} &      4160 &       3.53 \, {\it (37.3)} &      3.36 \\
			$\mu^{+}\mu^{-}$          &       9.77 \, \, \, {\it ( 4.9)} &      6583 &       3.51 \, {\it (36.5)} &      3.39 \\
			$\tau^{+}\tau^{-}$        &       0.89 \, \, \, {\it (27.4)} &      472 &       0.59 \, {\it (14.3)} &      2.36 \\
			$\nu_e\bar{\nu}_e$        &       4.12 \, \, \, {\it ( 3.6)} &      4062 &       3.52 \, {\it (37.7)} &      3.33 \\
			$\nu_\mu\bar{\nu}_\mu$    &       4.63 \, \, \, {\it ( 5.0)} &      4196 &       3.52 \, {\it (36.4)} &      3.41 \\
			$\nu_\tau\bar{\nu}_\tau$  &       0.96 \, \, \, {\it (16.6)} &      341 &       1.58 \, {\it (24.9)} &      2.74 \\
			\hline
		\end{tabular}
	\end{center}	 
	\caption{\textbf{\textit{DM decays (single channel) plus astrophysical power-law flux:}} Best-fit values for $\boldsymbol{\theta} = \{\tdm (\ndm), \mdm, \phia (\nast), \gamma\}$, where $\tdm$ is expressed in units of $ 10^{28}$~s, $\mdm$ in TeV and $\phia$ in units of $10^{-18}~{\rm GeV}^{-1}~{\rm cm}^{-2}~{\rm s}^{-1}~{\rm sr}^{-1}$. The corresponding numbers of DM and astrophysical events are also indicated in parenthesis, as $N_{\rm DM}$ and $N_{\rm astro}$. The overall best fit for all those channels is highlighted.}  	  	
	\label{tab:decay-onech}
\end{table}

\subsection{Results: DM decays plus isotropic astrophysical power-law flux}
\label{sec:DMdk-onech}

\begin{figure}[t]
	\centering
	\includegraphics[width=0.75\textwidth]{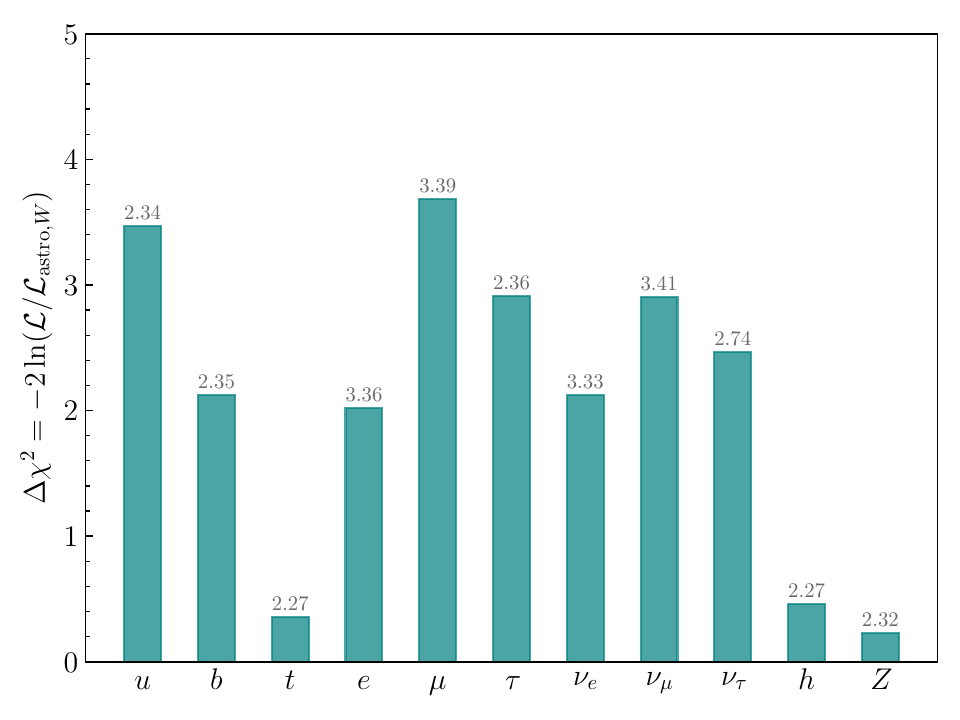}
	\caption{
		\textbf{\textit{DM decays (single channel) plus astrophysical power-law flux: Channel-by-channel comparison of $\boldsymbol{\Delta\chisq}$ at best fit}}, computed against the astrophysical flux plus $\dm \to W^+ \, W^-$ channel, which gives the overall best fit. Best-fit values of the spectral index $\gamma$ for each channel are displayed above the corresponding bar to indicate that the best fits prefer flat astrophysical spectra, and consequently low-\mdm\ values.}
	\label{fig:all-ch-like}
\end{figure}

\begin{figure}[t]
	\begin{center}
		\subfigcapskip=-4pt
		\subfigure[{\scriptsize DM $\to W^{+} \, W^{-}$}]{\includegraphics[width=0.49\linewidth]{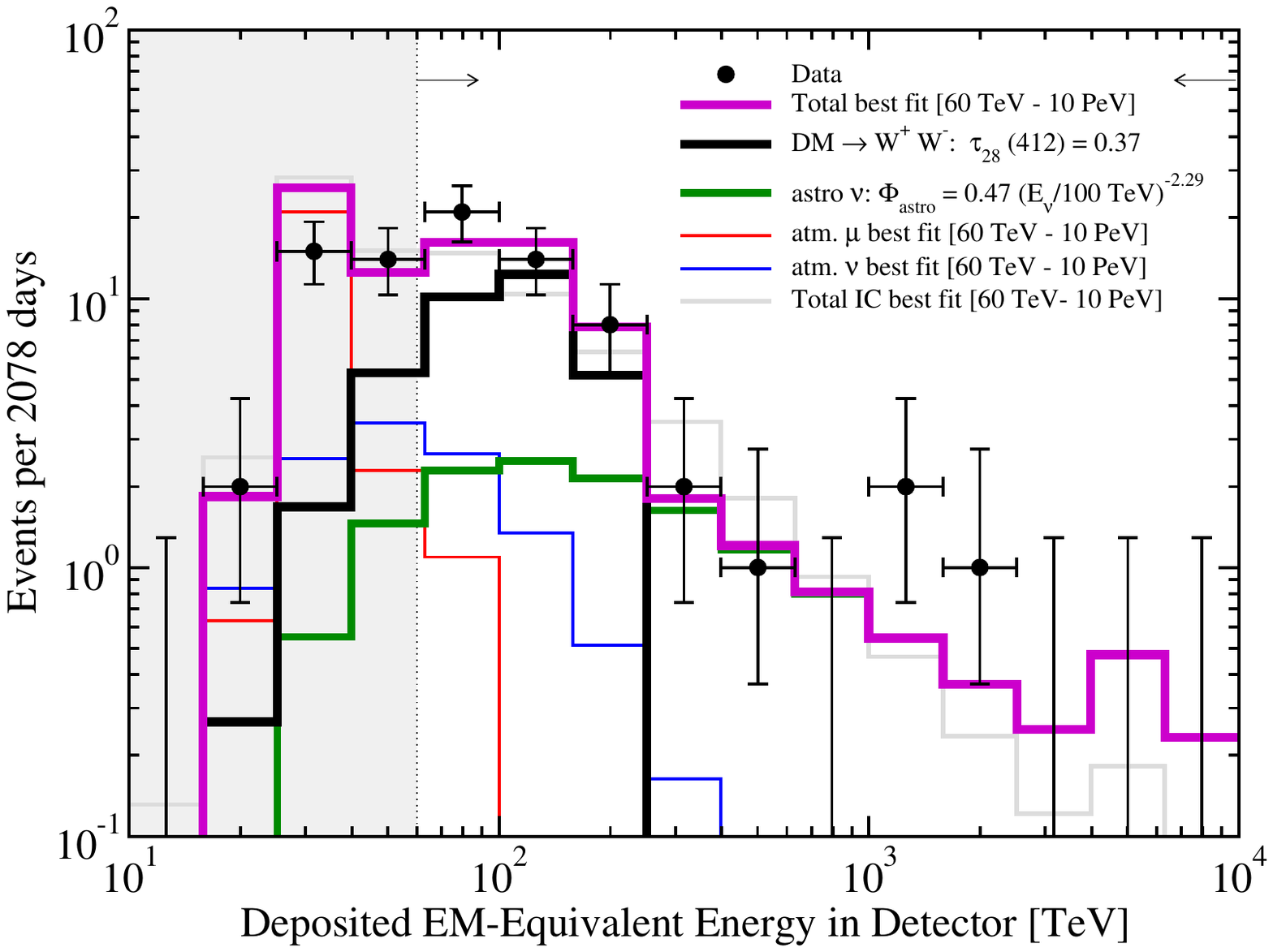}}
		\subfigure[{\scriptsize DM $\to \nu_e \, \bar{\nu}_e$}]{\includegraphics[width=0.49\linewidth]{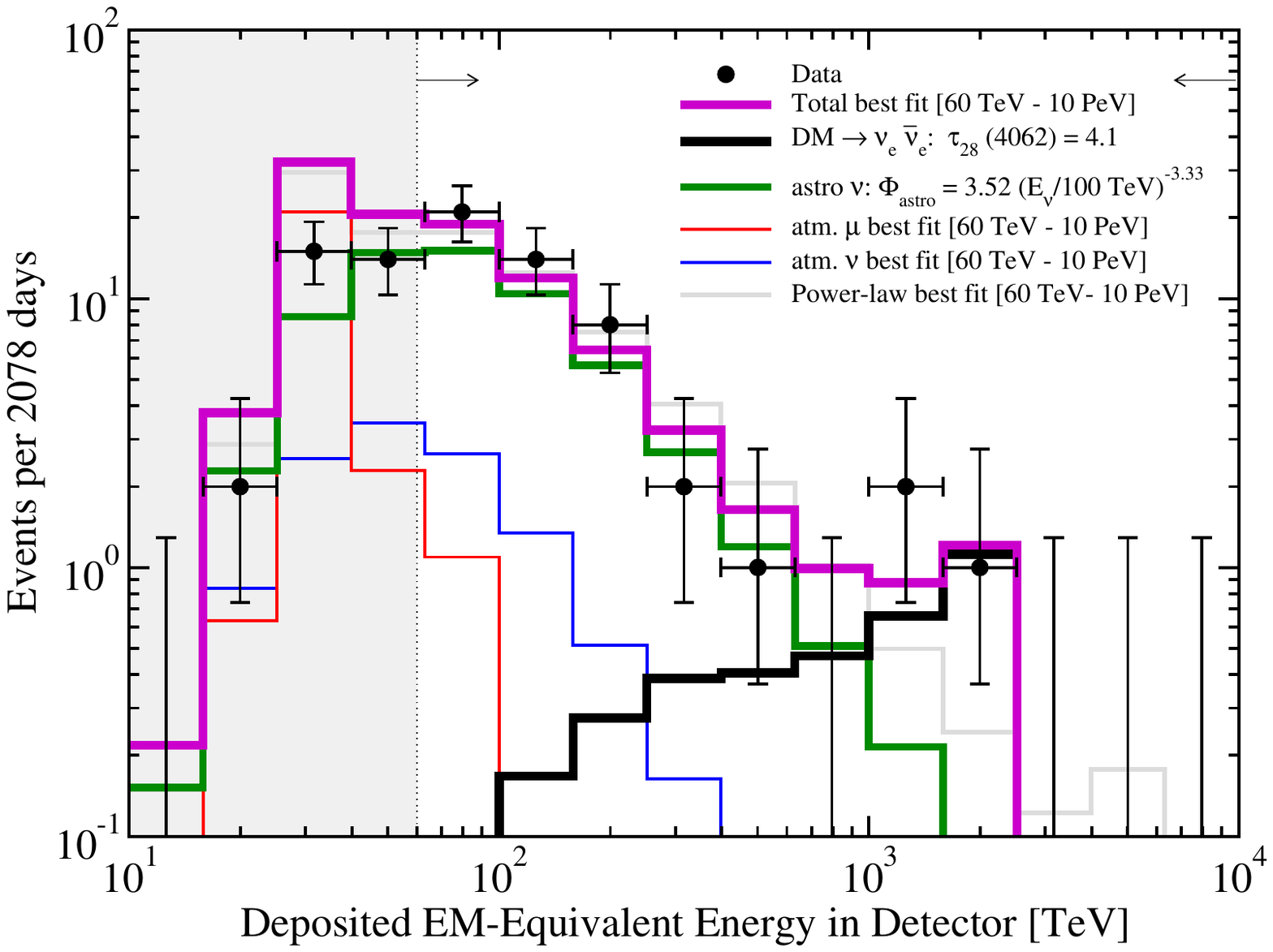}}
	\end{center}
	\caption{
		\textbf{\textit{DM decays (single channel) plus astrophysical power-law flux: Event spectra in the IceCube detector after 2078 days}} for the best fits for two DM decay channels: $\dm \to W^{+} \, W^{-}$ (left) and $\dm \to  \nue \anue$ (right). In all panels: atmospheric muon events (red histogram), conventional atmospheric neutrino events (blue histogram), astrophysical neutrino events (green histogram), neutrino events from DM decays (black histogram), and total event spectrum (purple histogram). We indicate the best fit values of the DM lifetime and mass [$\tau_{28} (\mdm)$] in units of $10^{28}$~s and TeV, and the per-flavor normalization of the power-law flux ($\phia$) in units of  $10^{-18} \, {\rm GeV}^{-1} \, {\rm cm}^{-2} \, {\rm s}^{-1} \, {\rm sr}^{-1}$. We also show the spectrum obtained using the 6-year IceCube best fit for a single power-law flux (gray histogram), $E_\nu^2 \, d\Phi/dE_\nu = 2.46 \times 10^{-8} \, (E_\nu/100 \, {\rm TeV})^{-0.92}  \, {\rm GeV} \, {\rm cm}^{-2} \, {\rm s}^{-1} \, {\rm sr}^{-1}$ (per flavor) and the binned high-energy neutrino event data (black dots)~\cite{Kopper:2017zzm} with Feldman-Cousins errors~\cite{Feldman:1997qc}.}
	\label{fig:events}
\end{figure}

The best-fit parameters corresponding to a total signal flux comprising neutrinos from a power-law astrophysical spectrum determined by \phia\ and $\gamma$, and from DM decays, as a function of $\tdm$ and \mdm, for different two-body decay channels, are indicated in Table~\ref{tab:decay-onech}. The comparison of the best-fit likelihoods for all channels is shown in Figure~\ref{fig:all-ch-like}. On top of each bar, we indicate the best fit obtained for the astrophysical index $\gamma$. In Figure~\ref{fig:events} we show the event spectra for two channels: in one, DM decay explains the low-energy data ($\dm \to W^+ W^-$) while in the other it explains the PeV events ($\dm \to \nue \anue$).

The overall best fit comes from decays to bosons $ W,\ Z,\text{ and } h $ (also top quarks), which prefer $\mdm \sim (400-650)$~TeV, thereby contributing only to the sub-PeV spectrum. In these cases, a relatively flat astrophysical flux, with $\gamma \sim 2.3$, accounts for the high-energy events (left panel of Figure~\ref{fig:events}) and supplements the contribution from DM decays to sub-PeV energies therefore, making the total sub-PeV spectrum steeply falling.

Fits to hard channels with DM mass in the high PeV (e.g., decays to charged lepton pairs and neutrinos) do comparatively worse, yet the statistical significance is weak. In these cases, for the best fits (except $\dm \to \tau^+ \tau^-$), events from DM decays explain the higher end of the spectrum, while a soft astrophysical power-law flux fills in the sub-PeV part of the event spectrum, falling quickly to have no impact on the high energy events (right panel of Figure~\ref{fig:events}).\footnote{From the comparison of the two panels of Figure~\ref{fig:events}, it might seem that the fit for $\dm \to \nue \anue$ is slightly better. However, a note of caution is in order. The event spectra is shown in bins, so a particular choice of binning can have a misleading impact. We avoid this problem we perform an unbinned extended maximum likelihood analysis to determine the best fits. Moreover, important parameters that influence the fits, such as the type of event topology or the neutrino direction, cannot be illustrated in this event-rate plot. As it turns out, the full likelihood analysis reveals the overall best fit comes from decays to $ W^{+}W^{-} $ and not from those to $ \nue\bar{\nu}_e $ (Figure~\ref{fig:all-ch-like}).} As in the 4-year HESE analysis~\cite{Bhattacharya:2017jaw}, the lack of high-energy tracks in the data influences the fit, and the flux from $\dm \to \mu^+ \mu^-$ turns out to be slightly disfavored compared to cases with other hard channels.

Given that the best-fit spectral index for several channels --- indeed for those giving better fits --- naturally turns out to be within $1\sigma$ of the best fit obtained by the latest IceCube through-going muon analysis, $\gamma_{\rm TG} = 2.19$~\cite{Haack:2017dxi}, we forego the prior-added analysis we carried out in our previous work. This is one of the major takeaways from the current analysis: some specific combinations of DM decay and astrophysical flux naturally lead to best-fit spectral indices consistent with that from the IceCube through-going muon analysis, sensitive only to the highest energy neutrinos. This is in contrast to what occurred with the analysis of the 4-year HESE data~\cite{Bhattacharya:2017jaw}, where we obtained a best fit for a hard channel ($\dm \to e^+ e^-$) with PeV mass, explaining the high-energy part of the event spectrum and thus, with a very soft astrophysical spectrum to describe the low-energy side.

\begin{figure}[t]
	\begin{center}
		\includegraphics[width=0.99\textwidth]{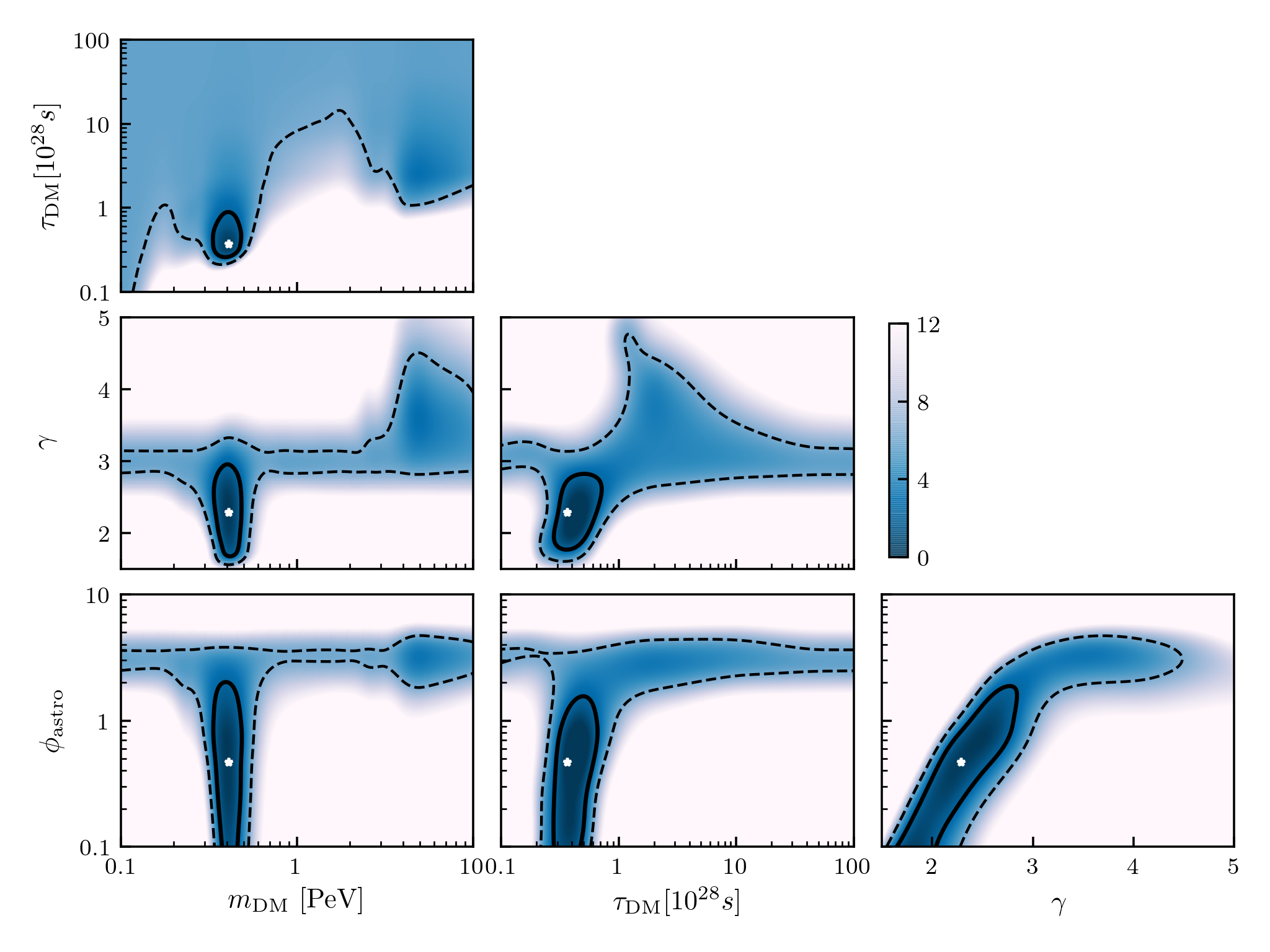}
	\end{center}
	\caption{
		\textbf{\textit{DM decays (single channel) plus astrophysical power-law flux: Correlations between all fit parameters for the overall best-fit channel}} $\boldsymbol{{\rm DM} \to W^{+}W^{-}}$. The contours indicated by the solid black curves represent the $1\sigma$~CL preferred regions around the best fit (indicated by a white `$\star$' sign), while the corresponding $2\sigma$~CL regions are indicated by the dashed black curves. We express $\tdm$ in units of $ 10^{28}$~s, $\mdm$ in PeV and $\phia$ in units of $10^{-18}~{\rm GeV}^{-1}~{\rm cm}^{-2}~{\rm s}^{-1}~{\rm sr}^{-1}$.}
	\label{fig:dk-corr-WW}
\end{figure}

\subsubsection{Parameter correlations and preferred regions}
\label{sec:DMdk-param-corr}

\begin{figure}[t]
	\begin{center}
		\includegraphics[width=0.99\textwidth]{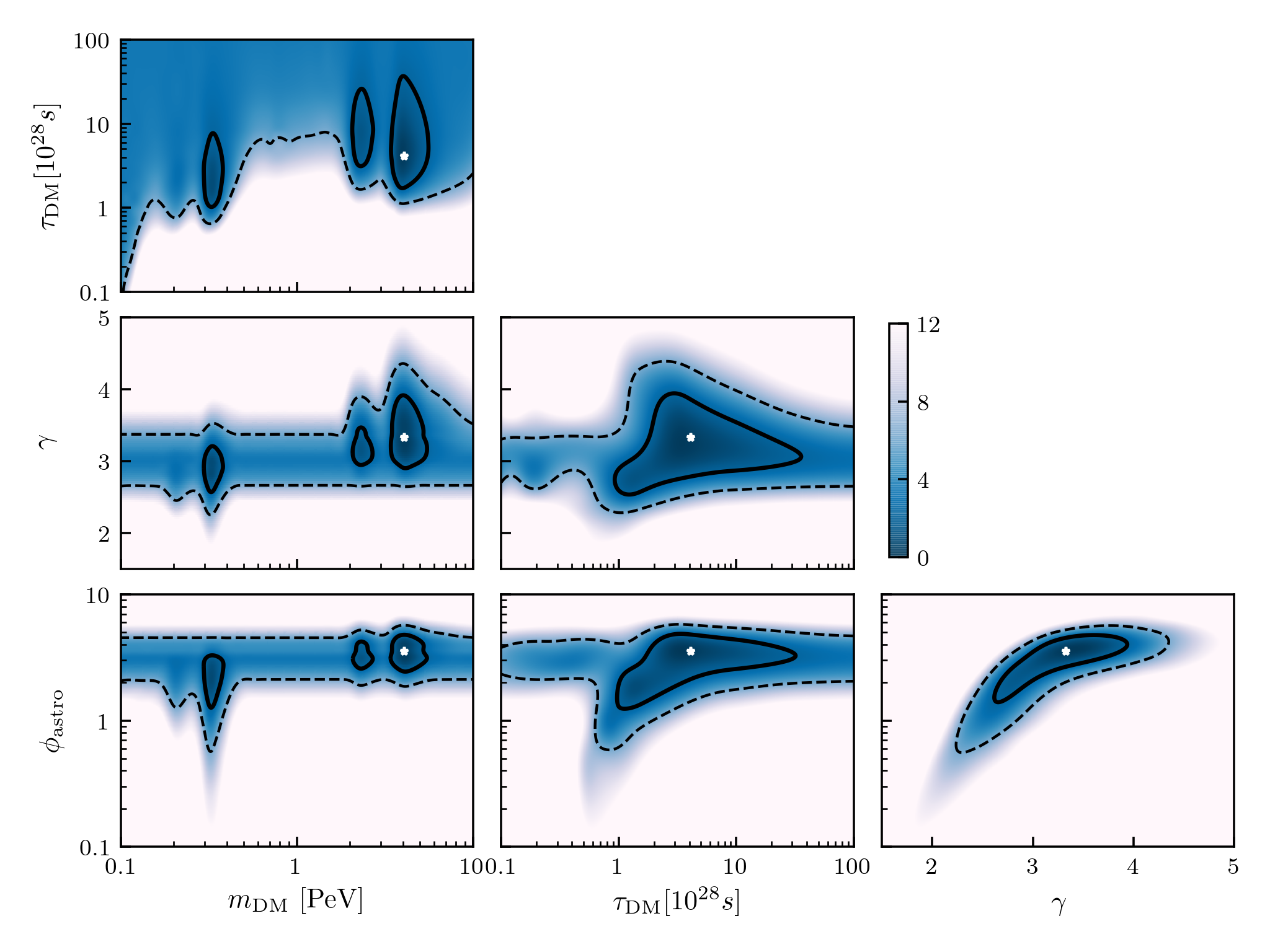}
	\end{center}
	\caption{
		\textbf{\textit{DM decays (single channel) plus astrophysical power-law flux: Correlations between all fit parameters for the hard channel}} $\boldsymbol{{\rm DM} \to \nue\anue}$. Same as Figure~\ref{fig:dk-corr-WW}.}
	\label{fig:dk-corr-nuenue}
\end{figure}

Here we discuss the correlations between the parameters and compute the preferred region of parameter space to the statistical 1$\sigma$ and 2$\sigma$ CL against the corresponding best-fit point (Figures~\ref{fig:dk-corr-WW} and~\ref{fig:dk-corr-nuenue}). Correlations between two parameters corresponding to different flux components (e.g., between \tdm\ and $\gamma$) demonstrate the careful balancing act performed by these two complementary fluxes while fitting to the data. On the other hand, correlations between parameters representing the same flux (e.g., \tdm\ and \mdm) reflect the sensitivity of the corresponding flux component to the best fit.

As noted above, for a decay channel like $\dm \to W^+ W^-$, the majority of the total signal events comes from DM decays, with all of these events populating the sub-PeV energies. The flatter astrophysical spectrum then fills in the missing events, with most of them lying at high energies. Correlation plots in Figure~\ref{fig:dk-corr-WW} are indicative of this nature. In particular, the correlation between \mdm\ and \tdm\ on one hand and the astrophysical spectral index $\gamma$ on the other shows the preference for a relatively flat spectrum ($1\sigma$ CL region lies below $\gamma = 3$) and a large DM contribution (close to the lower edge of the illustrated lifetime range) at low energies (narrow $1\sigma$ CL region around low \mdm). It is interesting to compare these results to those obtained with the 4-year HESE data~\cite{Bhattacharya:2017jaw}. Then, there were two $1\sigma$ CL regions in the $(\mdm, \tdm)$ parameter space, but the best fit was in the high-mass one. Now, with the additional events detected around $\sim 100$~TeV, only the region around $\mdm \approx 400$~TeV remains, which substantially changes the event spectrum corresponding to the best fit for this channel. 

For channels with harder spectra (e.g., $\dm \to \nue \anue $), multiple near-degenerate $1\sigma$ regions in terms of \mdm\ open up (Figure~\ref{fig:dk-corr-nuenue}). While the best fit for this channel lies in the high-mass region, comparative $1\sigma$ CL regions are also allowed in the low-\mdm\ region. This is in contrast to what happened with the 4-year HESE data~\cite{Bhattacharya:2017jaw}, where only the high-mass region was preferred at $ 1\sigma $ CL. Again, this is a consequence of the increase in low-energy events in the 6-year HESE data vis-à-vis the 4-year sample, while leaving the PeV spectrum unchanged. Clearly, the $1\sigma$ CL contour corresponding to low-\mdm\ allows for a flatter astrophysical flux, while the allowed $1\sigma$ CL contour corresponding to high-\mdm\ extends to the high-$\gamma$ region, indicating a steeply falling astrophysical flux. It should be noted that for these hard channels, the constraints on the astrophysical flux parameters are more restrictive than in the case of $\dm \to W^+ W^-$.

\subsubsection{Limits on the DM lifetime}
\label{sec:DMdk-limits}

\begin{figure}[t]
	\subfigure[][]{\includegraphics[width=0.32\textwidth]{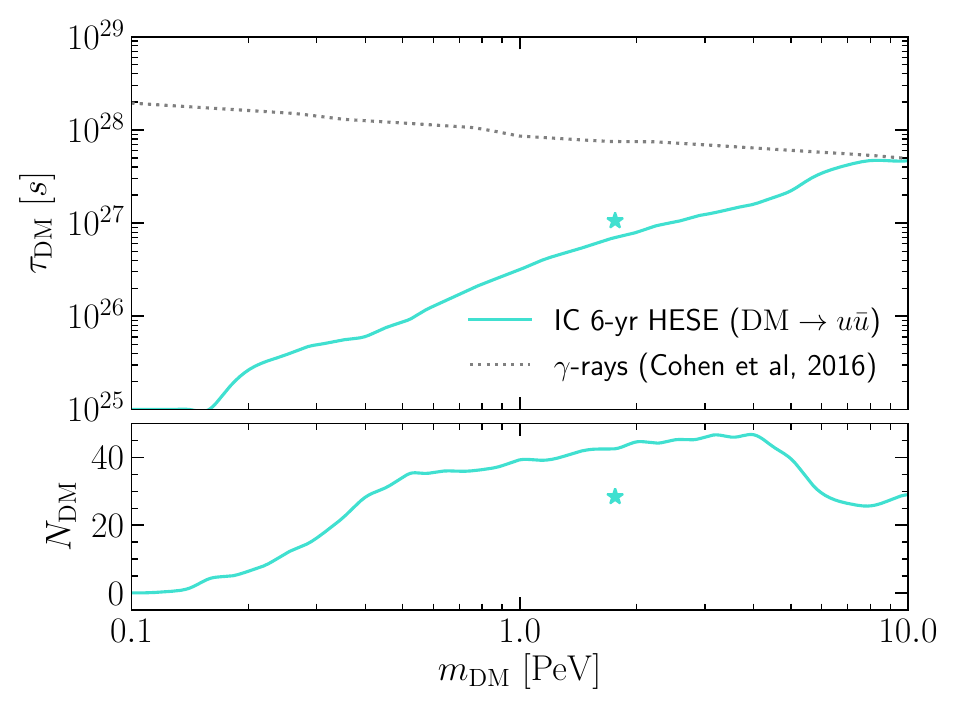}}
	\subfigure[][]{\includegraphics[width=0.32\textwidth]{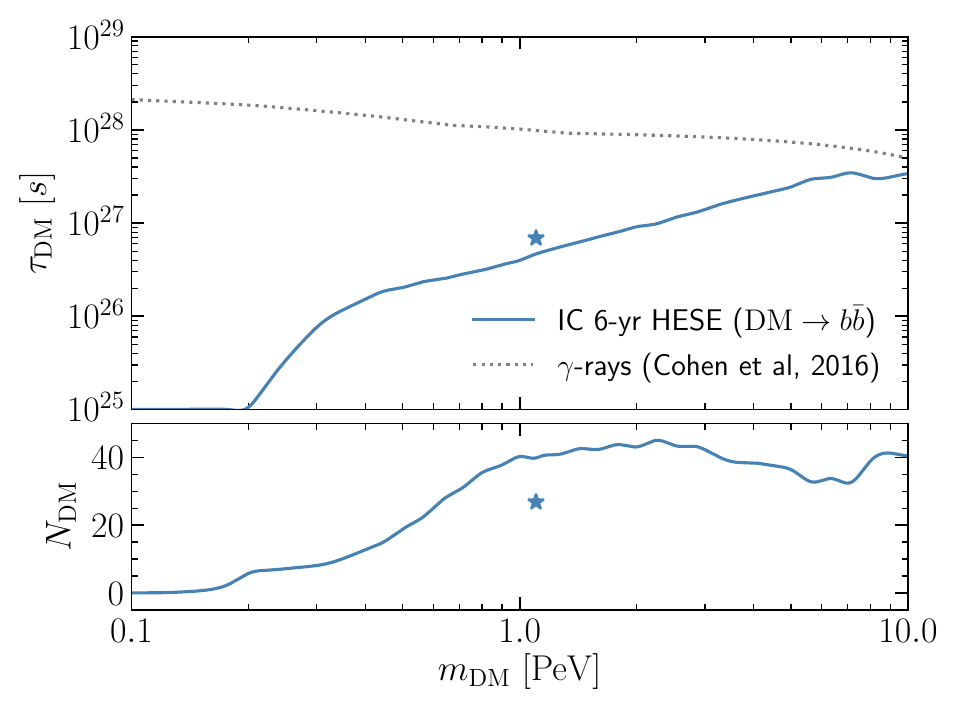}}
	\subfigure[][]{\includegraphics[width=0.32\textwidth]{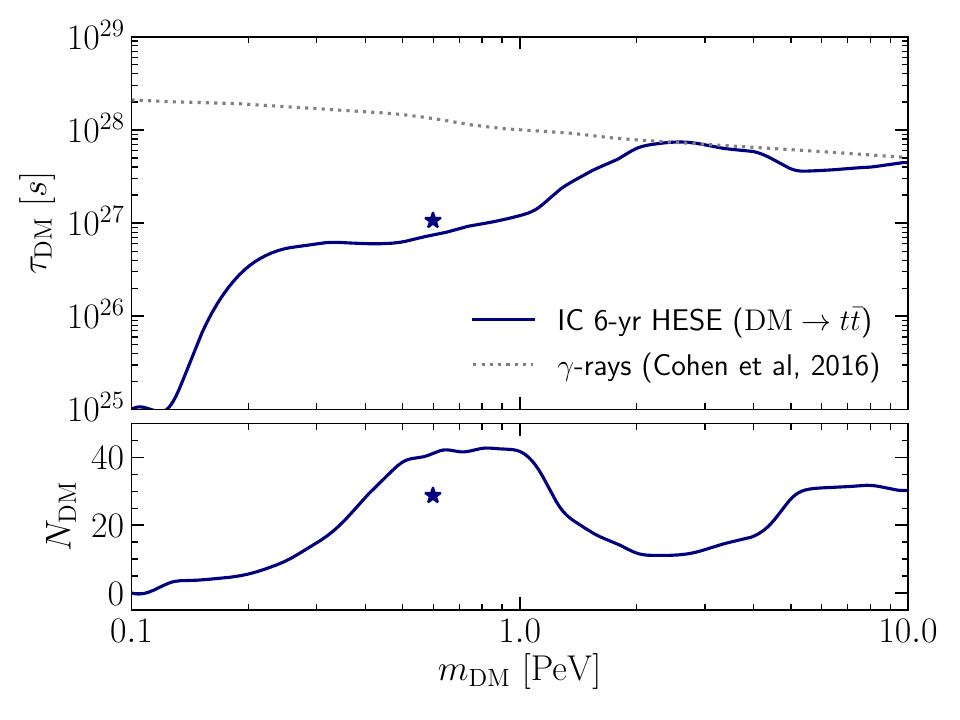}}\\
	\subfigure[][]{\includegraphics[width=0.32\textwidth]{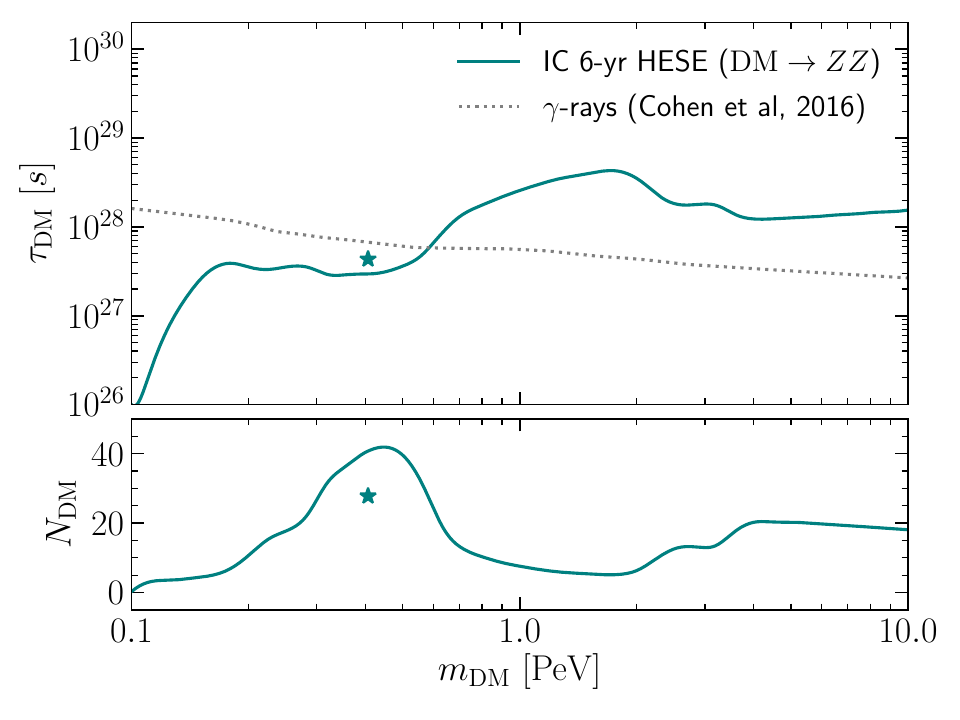}}
	\subfigure[][]{\includegraphics[width=0.32\textwidth]{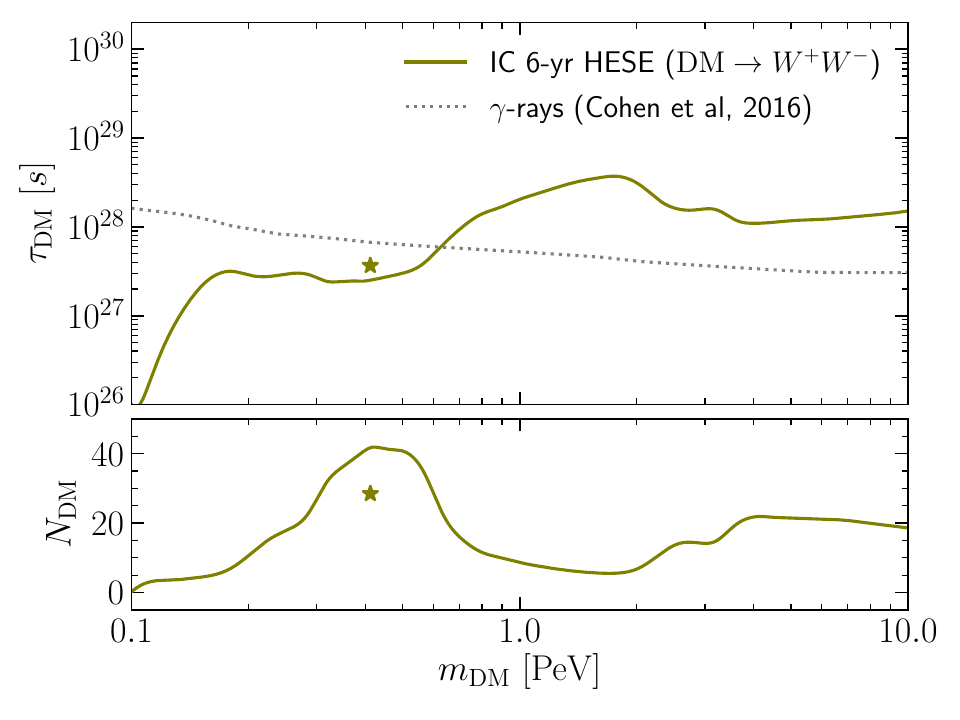}}
	\subfigure[][]{\includegraphics[width=0.32\textwidth]{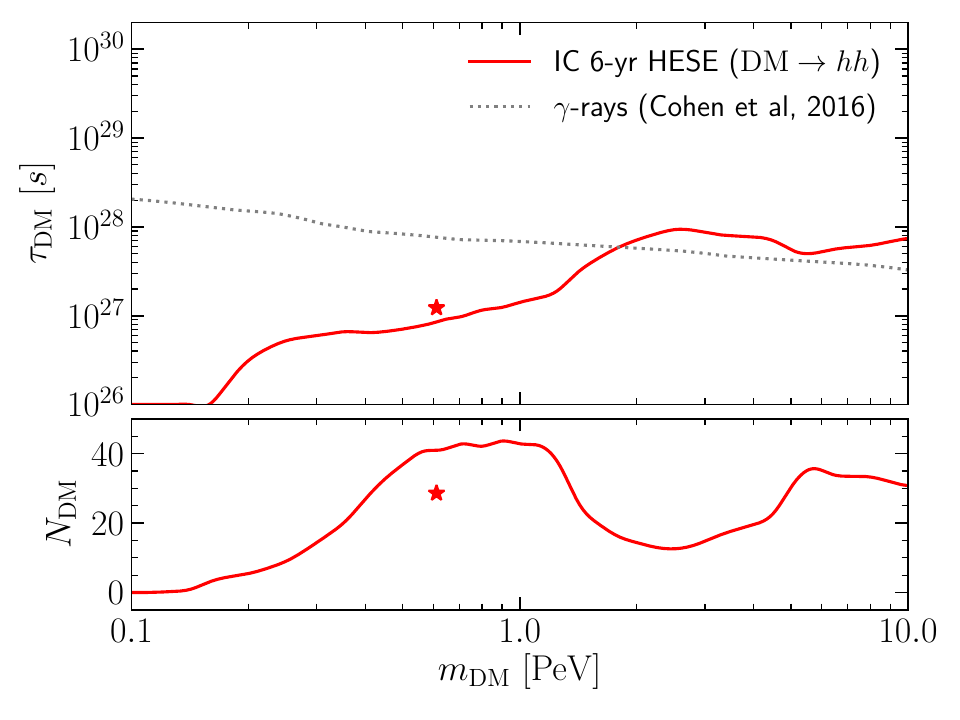}}\\
	\subfigure[][]{\includegraphics[width=0.32\textwidth]{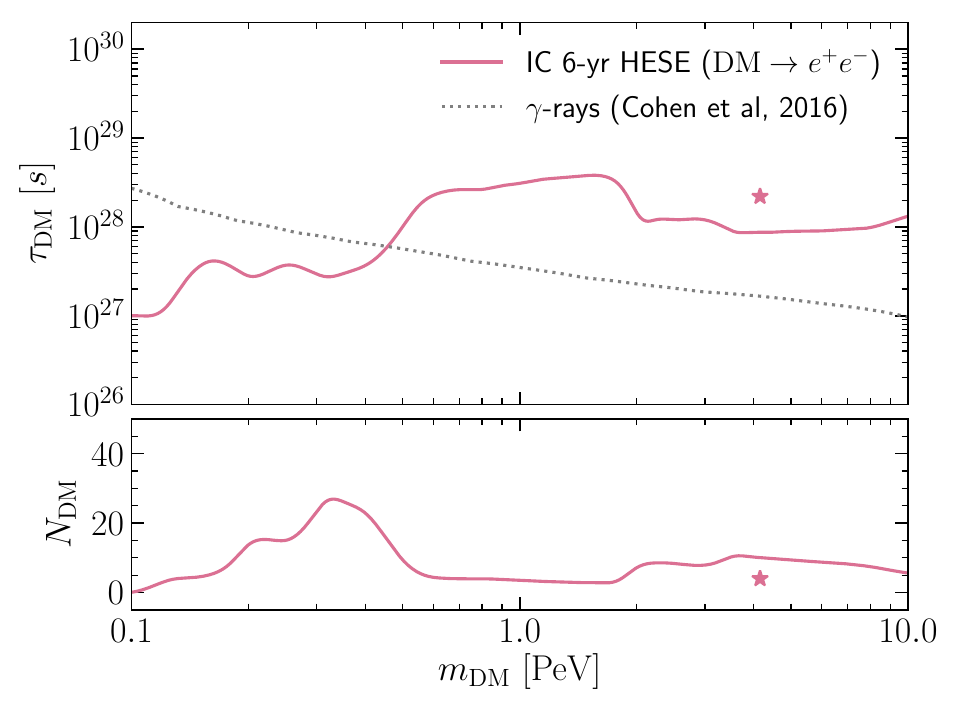}}
	\subfigure[][]{\includegraphics[width=0.32\textwidth]{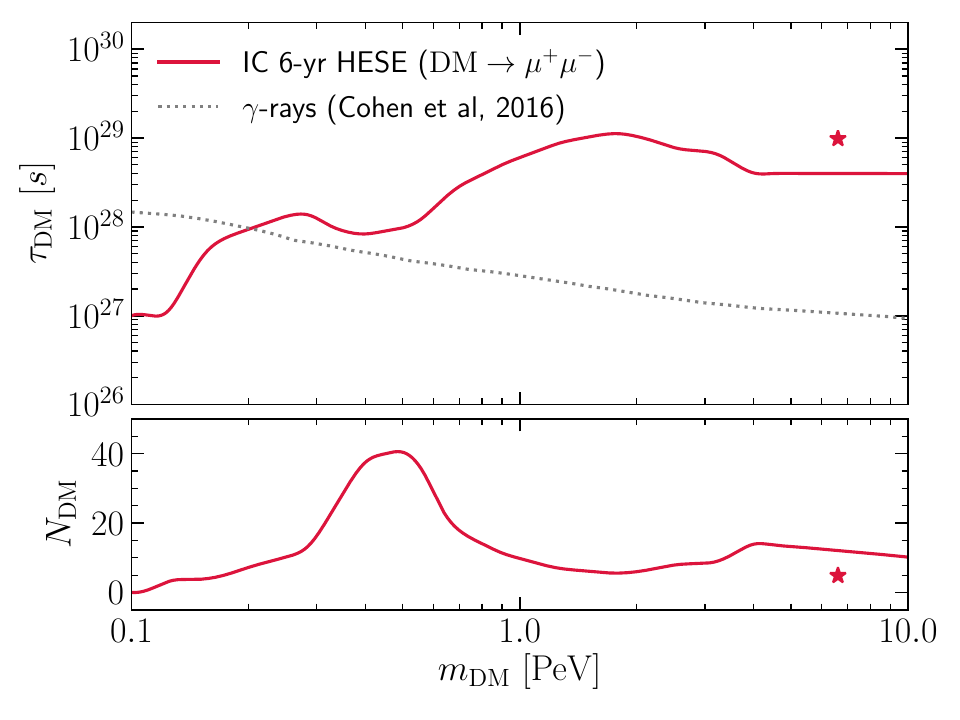}}
	\subfigure[][]{\includegraphics[width=0.32\textwidth]{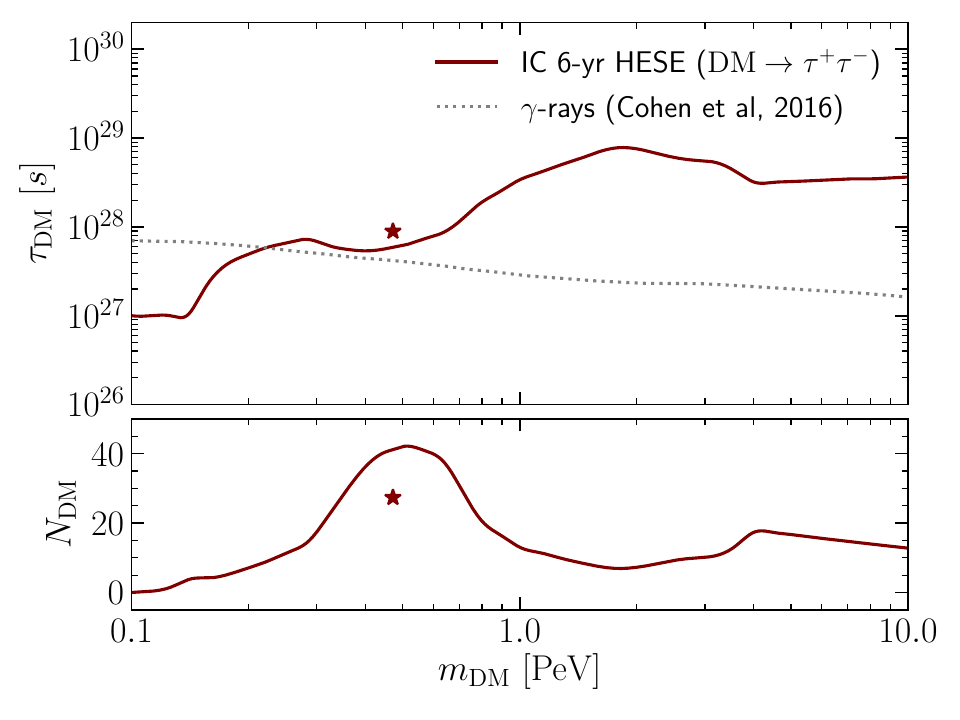}}\\
	\subfigure[][]{\includegraphics[width=0.32\textwidth]{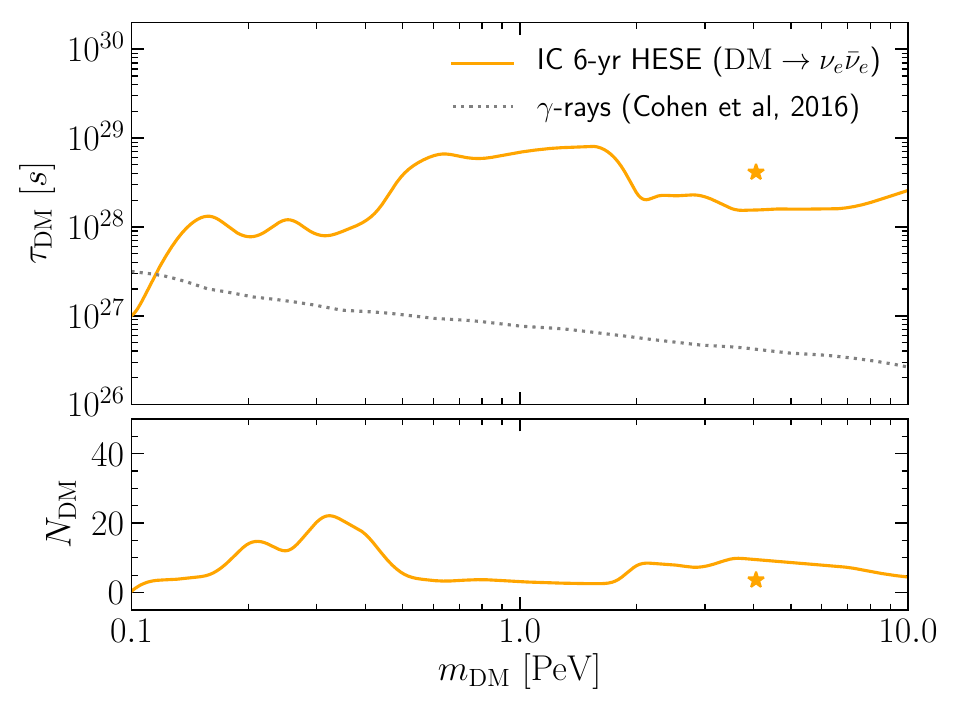}}
	\subfigure[][]{\includegraphics[width=0.32\textwidth]{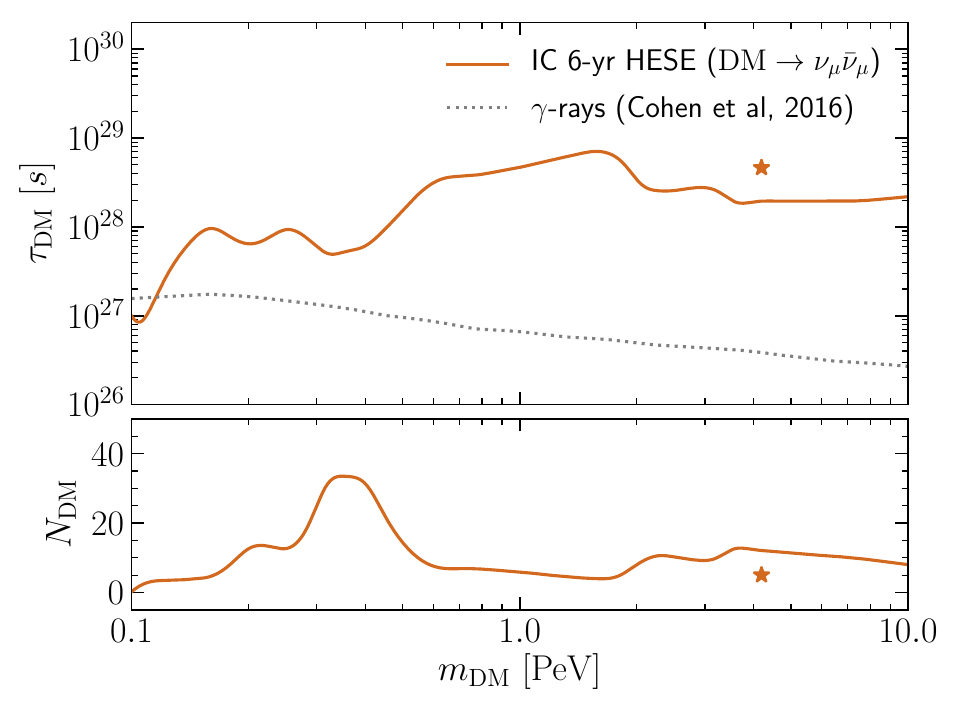}}
	\subfigure[][]{\includegraphics[width=0.32\textwidth]{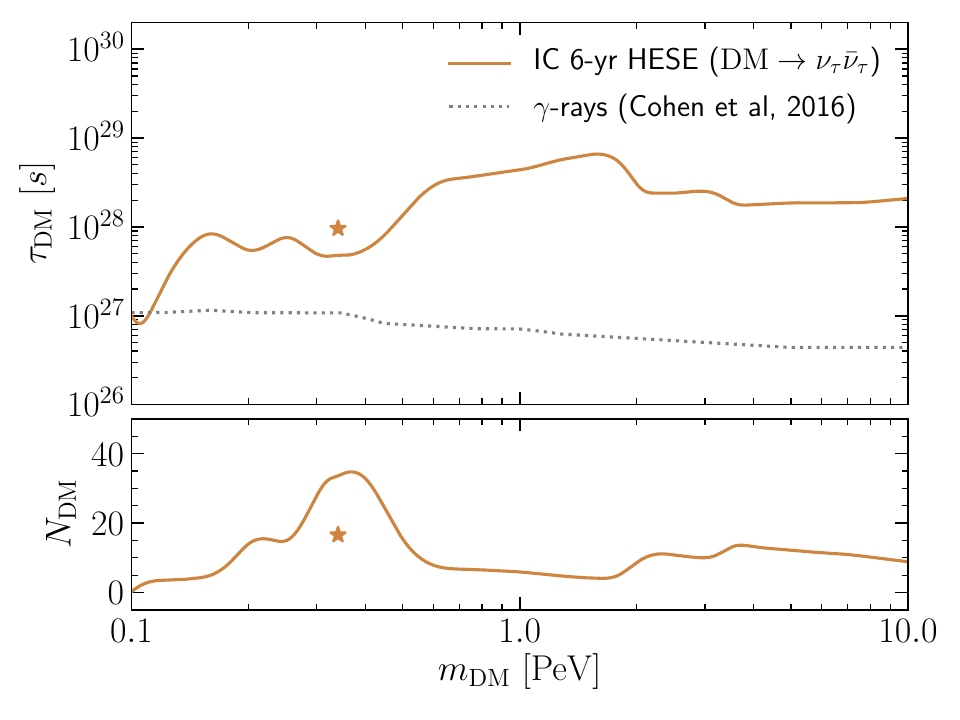}}  
	\caption{
		\textbf{\textit{DM decays (single channel) plus astrophysical power-law flux: Limits on the DM lifetime}} and \ndm\ at 95\%~CL as a function of the DM mass, for all single-channel DM decays in Table~\ref{tab:decay-onech}, plus an astrophysical flux. The best-fit values for $(\mdm, \tdm)$ and $(\mdm, \ndm)$ are indicated in each case by the `$\star$' sign. The dotted curve shows gamma-ray bounds on DM decay from Ref.~\cite{Cohen:2016uyg}.}
	\label{fig:ltlims-allch-app}
\end{figure}

For each decay channel, we also estimate the maximum allowed contribution from DM decays and thus, obtain limits on the corresponding values of \tdm\ as a function to \mdm. The 95\%~CL limits on the DM lifetime and on the number of DM events (\ndm), for all single-channel DM decays in Table~\ref{tab:decay-onech}, are shown in Figure~\ref{fig:ltlims-allch-app} as a function of the \mdm. These results are depicted along the bounds obtained from $\gamma$-ray observations in Ref.~\cite{Cohen:2016uyg}.\footnote{Note that a factor of a few stronger bounds have been recently obtained~\cite{Blanco:2018esa}.}

In comparison with analogous limits obtained in our previous analysis using 4-year HESE data~\cite{Bhattacharya:2017jaw}, they are not significantly altered by the two years of additional data except at the low-\mdm\ region for neutrino and charged lepton channels, where the bounds strengthen by about an order of magnitude at \mdm\ values between 100--200~TeV.

Since the fits to the gauge boson channels shift toward lower values of \mdm, the corresponding best fit is somewhat in tension with the gamma-ray bounds on DM lifetime obtained in Ref.~\cite{Cohen:2016uyg}. On the other hand, both because these bounds are weaker for leptons and because they weaken with increasing \mdm, best fits for decays to charged leptons and neutrinos with \mdm\ in the PeV region can evade these bounds. As already happened with the 4-year HESE data~\cite{Bhattacharya:2017jaw}, decays to light quark pairs are in strong tension with gamma-ray data: neutrino limits are weaker than the corresponding gamma-ray bounds, and the best fits run afoul of them by an order of magnitude. In all other cases, neutrino limits are stronger than the corresponding gamma-ray bounds in the high-\mdm\ regions.

When understanding the tension between the best-fit DM parameters and gamma-ray constraints, a note of caution is in order. The gamma-ray bounds on DM decays obtained in Ref.~\cite{Cohen:2016uyg}, and recently reevaluated and slightly strengthened in Ref.~\cite{Blanco:2018esa}, are obtained from the DM contribution to the diffuse gamma-ray background\footnote{Also known as isotropic gamma-ray background (IGRB).} (DGRB) observed by Fermi-LAT~\cite{Ackermann:2014usa}. A heavy decaying DM produces high-energy gamma-rays that initiate an electromagnetic cascade en route to the Earth and contributes to the DGRB in the energy range $\lesssim$ a few 100 GeV. On the other hand, low-energy astrophysical contributions to the DGRB could in fact entirely explain the observed DGRB~\cite{Fornasa:2015qua}.\footnote{The theoretical expectation for the contribution from these sources at some energies can completely explain the DGRB data, or even overshoot it.} Obviously, by accounting for these contributions from astrophysical sources to the DGRB, as is done in Refs.~\cite{Cohen:2016uyg, Blanco:2018esa}, there remains very little room for any contribution from the high-energy cascaded flux. Nevertheless, the same approach would strongly constrain any type of source of high-energy neutrinos, including the conventional $pp$ and $p\gamma$ astrophysical sources, since in these scenarios, an accompanying high-energy gamma-ray flux would also be produced at the source and initiate an electromagnetic cascade (except if the source is opaque to high-energy gamma-rays). In particular, the interpretation of the HESE data in terms of a single power-law flux with spectral index $\gamma \simeq 2.9$ is in strong tension with the DGRB (see Ref.~\cite{Murase:2013rfa}, where an upper limit of $\gamma\lesssim2.2$ was derived for $pp$ sources by considering the whole DGRB data; using the DGRB data after subtracting low-energy astrophysical source contributions would lead to even stronger limits). When taking the more conservative gamma-ray limits derived from the whole DGRB data~\cite{Murase:2015gea}, and not the low-energy source-subtracted one, all the best fit points in Figure~\ref{fig:ltlims-allch-app} are actually allowed.

\subsection{Results: DM decays via multiple channels}
\label{sec:DMdk-multich}

\begin{table}[t]
	\begin{center}
		\begin{tabular}{cl|ccc}
			\hline
			\multicolumn{2}{c|}{Decay channels}         & $ \tau_{\rm DM} [10^{27}~{\rm s}] $   &  $ \mdm $ [PeV] &	 BR \\
			\hline
			$b\bar{b}$                & $e^{+}e^{-}$              &      1.82 &      4.001 &       0.91 \\
			$b\bar{b}$                & $\mu^{+}\mu^{-}$          &      2.00 &      4.517 &       0.97 \\
			$b\bar{b}$                & $\nu_e\bar{\nu}_e$        &      1.79 &      3.942 &       0.97 \\
			$b\bar{b}$                & $\nu_\mu\bar{\nu}_\mu$    &      1.82 &      4.015 &       0.97 \\
			$u\bar{u}$                & $e^{+}e^{-}$              &      1.76 &      3.898 &       0.92 \\
			$u\bar{u}$                & $\mu^{+}\mu^{-}$          &      1.85 &      4.148 &       0.98 \\
			$\boldsymbol{u\bar{u}}$     & $\boldsymbol{\nu_e\bar{\nu}_e}$    &    \textbf{1.73} &   \textbf{3.845} &    \textbf{0.97} \\
			$u\bar{u}$                & $\nu_\mu\bar{\nu}_\mu$    &      1.76 &      3.906 &       0.97 \\
			$h h$                     & $e^{+}e^{-}$              &      3.78 &      4.184 &       0.92 \\
			$h h$                     & $\mu^{+}\mu^{-}$          &      4.54 &      6.132 &       1.00 \\
			$h h$                     & $\nu_e\bar{\nu}_e$        &      3.72 &      4.076 &       0.97 \\
			$h h$                     & $\nu_\mu\bar{\nu}_\mu$    &      4.54 &      6.132 &       1.00 \\
			$W^{+}W^{-}$              & $e^{+}e^{-}$              &      6.32 &      4.498 &       1.00 \\
			$W^{+}W^{-}$              & $\mu^{+}\mu^{-}$          &      6.32 &      4.498 &       1.00 \\
			$W^{+}W^{-}$              & $\nu_e\bar{\nu}_e$        &      6.32 &      4.498 &       1.00 \\
			$W^{+}W^{-}$              & $\nu_\mu\bar{\nu}_\mu$    &      6.32 &      4.498 &       1.00 \\
			\hline
		\end{tabular}
	\end{center}	 
	\caption{
		\textbf{\textit{DM-only two-channel decays: Best-fit values}} for $\dm \to p_{1} \, \bar{p}_{1},\, p_{2} \, \bar{p}_{2}$ defined by $\boldsymbol{\theta} = \{\tdm, \mdm, \text{BR}\}$, where $\tdm$ is expressed in units of $10^{27}$~s, $\mdm$ in TeV, and ${\rm BR} = \Gamma_{\dm \to p_{1} \, \bar{p}_{1}} / \left(\Gamma_{\dm \to p_{1} \, \bar{p}_{1}} + \Gamma_{\dm \to p_{2} \, \bar{p}_{2}}\right)$. The overall best fit for all those combinations of channels is highlighted.}  	  	
	\label{tab:fits-mulch-60}
\end{table}

We also analyze the consequences of turning off the astrophysical flux entirely and instead allowing DM to decay via multiple channels. This analysis proceeds in the same vein as the previous analysis with the 4-year HESE data (see Sec.~6 in Ref.~\cite{Bhattacharya:2017jaw}). Allowing the DM to decay via two distinct channels allows us to parameterize the resulting flux in terms of the following physical quantities:
\begin{inparaenum}[\itshape a\upshape)]
  \item the DM mass, \mdm,
  \item the  DM lifetime \tdm, and 
  \item the branching ratio in favor of the (arbitrarily ordered) first
channel: ${\rm BR} = \Gamma_{\dm \to p_{1} \, \bar{p}_{1}} / \left(\Gamma_{\dm
\to p_{1} \, \bar{p}_{1}} + \Gamma_{\dm \to p_{2} \, \bar{p}_{2}}\right)$.
\end{inparaenum}

\begin{figure}[t]
	\centering
	\includegraphics[width=0.75\textwidth]{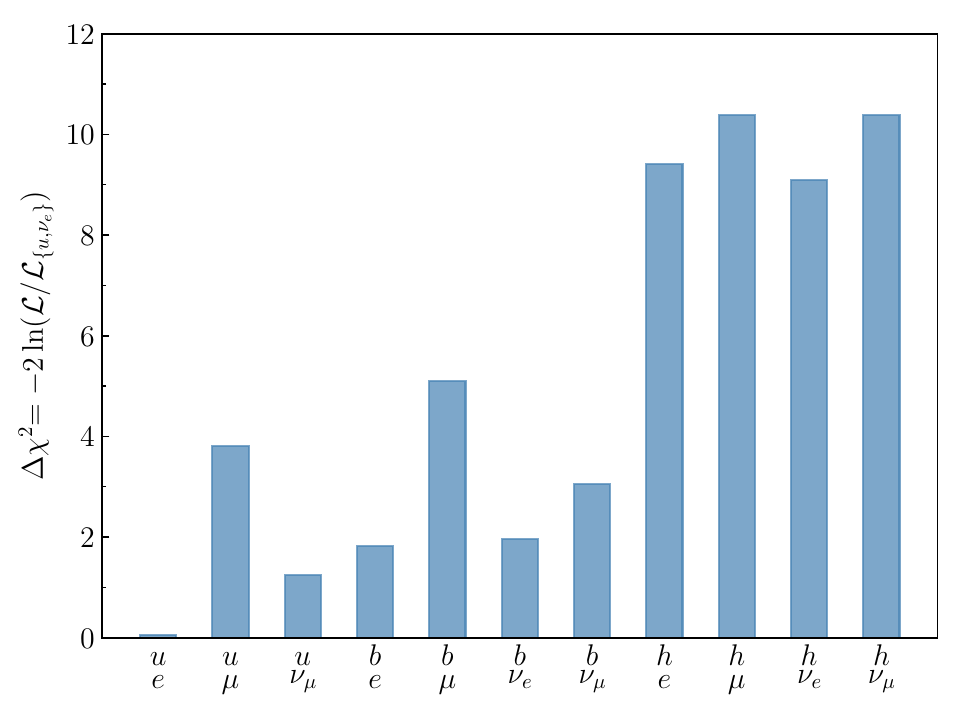}
	\caption{
		\textbf{\textit{DM-only two-channel decays: Channel-by-channel comparison of $\boldsymbol{\Delta\chisq}$ at best fit}}, computed against the $\dm \to \{u \bar{u}, \, \nu_e \bar{\nu}_e\}$ combination, which gives the overall best fit. Channel combinations from Table \ref{tab:fits-mulch-60} not shown in this figure represent extremely poor fits to the data with $\Delta \chisq \gtrsim 20$.} 
	\label{fig:all-mulch-chi}
\end{figure}

In this scenario, the DM mass is necessarily pushed to PeV values to accommodate both PeV and sub-PeV events. Results for a few selected combinations of decay channels are shown in Table~\ref{tab:fits-mulch-60} and the $\Delta \chi^2$ for each combination with respect to the best fit is shown in Figure~\ref{fig:all-mulch-chi}. The overall best fits come from the combination of $u\bar{u}$ and $ e^{+}e^{-} $ or $\nu_e\bar{\nu}_e$ channels with the DM mass around 4~PeV and a 92--97\% branching ratio in favor of decays to the quark (soft) channel. The hard-channel decay, either to $\nue\bar{\nu}_e$ or to $ e^{+}e^{-} $, explains the PeV events while the softer secondary neutrino flux from decays to $u\bar{u}$ explains the sub-PeV events. Other combinations of soft and hard channels provide similar results. Note, however, that if both channels are hard (e.g., $W^+ W^-$ and $\nu_e \bar{\nu}_e$), the fit worsens and prefers the spectrum to be dominated by the softer of the two spectra.

\begin{figure}[t]
	\begin{center}
		\includegraphics[width=0.75\linewidth]{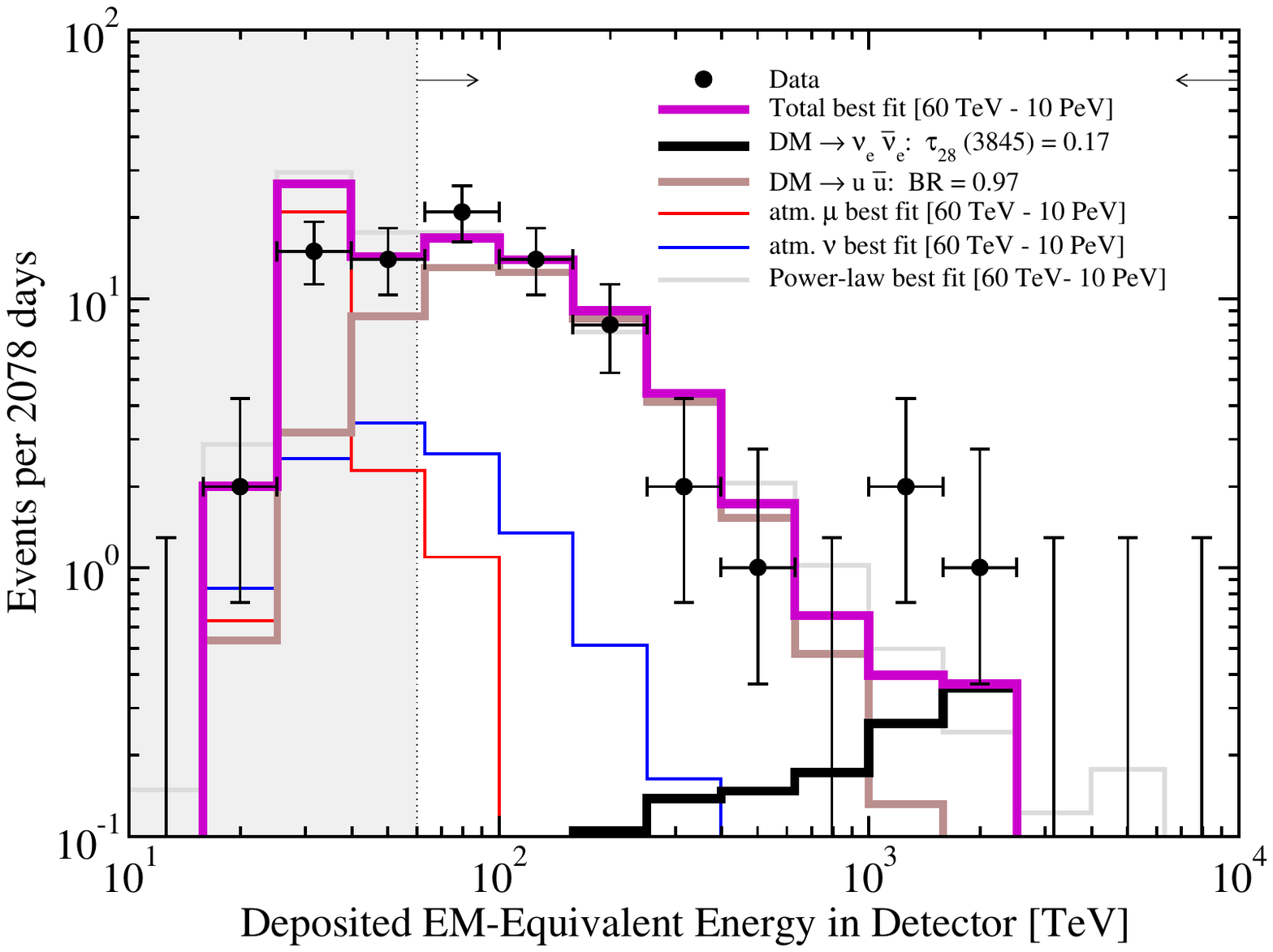}
	\end{center}
	\caption{\label{fig:events-mulch}
		\textbf{\textit{DM-only two-channel decays: Event spectra in the IceCube detector after 2078 days}} for DM decays into the best-fit two-channel combination, DM $\to \lbrace u \bar{u}, \, \nu_e \bar{\nu}_e \rbrace$, with their corresponding branching fraction into the quark channel also indicated. The histograms represent: atmospheric muon events (red histogram), conventional atmospheric neutrino events (blue histogram), neutrino events from DM decays into the quark channel (brown histogram) and into the lepton channel (black histogram), and total event spectrum (purple histogram). We indicate the best fit values of the DM lifetime and mass [$\tau_{28} (\mdm)$] in units of $10^{28}$~s and TeV. We also show the spectrum obtained using the 6-year IceCube best fit for a single power-law flux (gray histogram), $E_\nu^2 \, d\Phi/dE_\nu = 2.46 \times 10^{-8} \, (E_\nu/100 \, {\rm TeV})^{-0.92}  \, {\rm GeV} \, {\rm cm}^{-2} \, {\rm s}^{-1} \, {\rm sr}^{-1}$ (per flavor) and the binned high-energy neutrino event data (black dots)~\cite{Kopper:2017zzm} with Feldman-Cousins errors~\cite{Feldman:1997qc}.}
\end{figure}

Qualitatively, these patterns are similar to the results obtained with the 4-year HESE data~\cite{Bhattacharya:2017jaw}, albeit with slightly larger branching ratios into soft channels. In relative terms, when compared to the DM single channel decay plus astrophysical flux, we find that the overall best-fit two-channel combination does somewhat worse with the updated 6-year HESE data. Yet, this is not statistically significant.\footnote{The DM-only scenario here considered fits the data using three degrees of freedom, as opposed to four in the scenario with DM plus astrophysical fluxes, considered in the previous section. Consequently, a straight comparison to determine which of these two scenarios provides a comparatively better fit is not straightforward and is out of the scope of this work.} The main reason for this slight swing is due to the fact that the 6-year HESE data strongly prefers DM decay fluxes with low-\mdm, and the requirement in this scenario of having to explain PeV events also from DM decay necessitates the addition of at least one hard component. This ends up worsening the fit slightly in comparison to the two-channel best fit obtained in the previous work. Nevertheless, unlike Ref.~\cite{Aartsen:2018mxl}, we cannot conclude that the DM-only scenario is disfavored by neutrino data. This is in agreement with previous findings using earlier HESE data sets~\cite{Esmaili:2013gha, Esmaili:2014rma, Bhattacharya:2017jaw}. Moreover, as already happened for the 4-year HESE data, for all these soft-hard combinations with a large branching ratio into the soft channel, the best fit for the DM lifetime is in tension with the gamma-ray limits from Refs.~\cite{Cohen:2016uyg, Blanco:2018esa}, while compatible with the limits from Ref.~\cite{Murase:2015gea} (see the comment at the end of section~\ref{sec:DMdk-limits}).

As an illustration, in Figure~\ref{fig:events-mulch} we show the best-fit event spectrum (among the combinations in Table~\ref{tab:fits-mulch-60}), where we clearly see the interplay between the spectra of the two decay channels. In comparison with results obtained with the 4-year HESE data, the increase in events at EM-equivalent deposited energies of $\sim 100$~TeV makes it more difficult for the DM-only scenario to explain the PeV events simultaneously, making the overall fit slightly worse.

\section{DM annihilations}
\label{sec:DMann}

The majority of the studies on the possible DM contribution to the IceCube high-energy neutrino flux have focused on the decay scenario, mainly because the required annihilation cross section, \sv, for heavy DM ($\mdm \gtrsim 100$~TeV)~\cite{Feldstein:2013kka, Zavala:2014dla, Esmaili:2014rma, Chianese:2016opp, Chianese:2016kpu} violates the unitary bound $\sv \leq 4\pi/(\mdm^2 \, v)$ for thermal production~\cite{Griest:1989wd, Hui:2001wy}. Yet, even if not produced thermally, the IceCube data could, in principle, test values below the local unitarity bound~\cite{Esmaili:2014rma, Chianese:2016kpu}.\footnote{Note that the unitarity bound is less stringent locally, as the relative velocity of DM particles is smaller than in the early Universe.} Moreover, the effect of DM annihilation in substructures, which would boost the signal with respect to the smooth contribution, along with a potential dependence of the relative velocity on negative powers, as in Sommerfeld-enhanced models~\cite{Hisano:2004ds, Profumo:2005xd, Lattanzi:2008qa, Feng:2010zp}, could give rise to DM fluxes that can account for the observed number of high-energy neutrino-induced events~\cite{Zavala:2014dla}. In this section, however, we only consider a constant DM annihilation cross section and study the values that would give a potential contribution to the observed IceCube neutrino flux, in analogy to the case of decay presented above.

While in the case of the decaying DM scenario the resulting neutrino flux is linearly proportional to the (integral of the) DM number density, the neutrino flux from DM annihilations is proportional to the square of the DM density. Since the number density decreases with increasing \mdm, obtaining a neutrino flux comparable to that required for the IceCube event rate, requires a rather large annihilation cross section. As in the decaying DM case, the neutrino flux from DM annihilations at Earth receives both galactic and extra-galactic contributions. Nevertheless, there are important differences between these two cases.

The (smooth) galactic contribution of the neutrino flux from DM annihilation is given by
\begin{equation}
\frac{{\rm d}\Phi_{\rm DM, G}^{\rm ann}}{{\rm d}E_\nu} (E_\nu,b,l) = \frac{\sv}{2} \frac{1}{4\pi m_{\rm DM}^2} \frac{{\rm d}N}{{\rm d}E_\nu} \int_{0}^{\infty} \rho^2 [r(s,b,l)]~{\rm d}s ~,
\end{equation}  
where $\rho$ is the galactic DM density, $r(s,b,l) = \sqrt{s^2 + R_\odot^2 - 2 s R_\odot\cos b\cos l}$, is the distance to the Galactic center with $(b,l)$ being the Galactic coordinates and $R_\odot=8.5$~kpc. For the DM density we use an NFW profile \cite{Navarro:1995iw,Navarro:1996gj}, with the same parameters as in the decay case~\cite{Bhattacharya:2017jaw}. Given that the closer to the galactic center, the larger the DM density, this flux is more sharply peaked in that direction than in the case of DM decays. 

In addition to the smooth galactic contribution, hierarchical structure formation in the standard cold DM scenario implies that DM halos must contain a large amount of smaller subhalos, which would further boost the annihilation flux. In this work, though, we do not account for this potential overall enhancement of the galactic DM annihilation signal. This boost factor depends on the direction of observation, as subhalos are not distributed homogeneously~\cite{Springel:2008cc, Hellwing:2015upa, Rodriguez-Puebla:2016ofw} and their properties also depend on the position within the main halo~\cite{Springel:2008cc, Diemand:2008in, Pieri:2009je, Moline:2016pbm}. Nevertheless, we do not expect this to have an important impact on our results, as this enhancement is partly degenerated with the annihilation cross section and the ratio of the extragalactic to galactic contributions, which we keep as free parameters (see below).

The isotropic extragalactic contribution from DM annihilations is given by
\begin{equation}
\frac{{\rm d}\Phi^{\rm ann}_{\rm DM, EG}}{{\rm d}E_\nu} (E_\nu) = \frac{\sv}{2} \frac{\Omega_{\rm DM}^2 \, \rho_c^2}{4\pi \, \mdm^2} \int_{0}^\infty \frac{dz}{H(z)} \, (1+z)^3 \, \zeta(z) \, \frac{{\rm d}N}{{\rm d}E_\nu} \left[\left(1+z\right)E_\nu\right] ~,
\end{equation}        
where $H(z) = H_0 \sqrt{\Omega_m (1+z)^3 + \Omega_\Lambda}$, and the cosmological parameters $H_0$, $\rho_c$, $\Omega_m$ and $\Omega_\Lambda$ are the same used for the decay scenario~\cite{Bhattacharya:2017jaw}. The dimensionless quantity $\zeta(z)$ accounts for the non-homogeneous DM clustering. It represents the variance of the DM density fluctuations at redshift $z$, so can be expressed in terms of the non-linear matter power spectrum~\cite{Serpico:2011in, Sefusatti:2014vha}. Equivalently, within the halo model approach~\cite{Cooray:2002dia}, it is proportional to the cumulative sum of the enhancement in each individual halo at redshift $z$~\cite{Bergstrom:2001jj, Ullio:2002pj, Taylor:2002zd}, and thus it depends on the abundance and internal properties of DM halos. Both approaches require extrapolations to small scales and high redshifts of the findings of N-body numerical simulations. This results in relatively large uncertainties in the computation of the clumping factor $\zeta(z)$, which can amount up to two or three orders of magnitude~\cite{Sefusatti:2014vha, Moline:2014xua}. In this work we consider as our default $\zeta(z)$ the revised \texttt{halofit} plus stable clustering (RHF-SC) prescription of Ref.~\cite{Sefusatti:2014vha}, such that $(1 + z)^3 \, \zeta(z) \, H_0/H(z) \sim 10^5$. To take into account uncertainties, we introduce a constant normalization, $\xi$, where $\xi = 1$ represents the default case and we vary it within $[10^{-3}, 10^3]$. Note, however, that the estimated uncertainty is likely smaller~\cite{Sefusatti:2014vha, Moline:2014xua}. The total flux is given by
\begin{equation}
\frac{{\rm d}\Phi^{\rm ann}_{\rm DM}}{{\rm d}E_\nu} (E_\nu)
  = \frac{{\rm d}\Phi^{\rm ann}_{\rm DM, G}}{{\rm d}E_\nu} (E_\nu) + \xi \,
    \frac{{\rm d}\Phi^{\rm ann}_{\rm DM, EG}}{{\rm d}E_\nu} (E_\nu) ~.
\end{equation}
Unlike what happens for the case of DM decays, for which there is no enhancement of the flux due to the linear dependence on the density, in the DM annihilating scenario, the enhancement could be significant and determines the importance of the extragalactic contribution: the larger $\xi \, (1 + z)^3 \, \zeta(z) \, H_0/H(z)$, the more isotropic the flux. Therefore, the value of $\xi$ serves as a measure for the anisotropy level in the total flux. As a reference, for $\xi = 1$, the number of galactic events is about a factor of ten larger than the number of extragalactic ones, although the actual relative factor depends on the DM mass and annihilation channel. 

As with DM decays, we consider two distinct scenarios here too:
\begin{inparaenum}[\itshape 1\upshape)]
  \item neutrinos from DM annihilations into a single channel plus an isotropic astrophysical power-law flux, and
  \item neutrinos solely from DM annihilations into two channels.
\end{inparaenum}
The flux in the former case is defined in terms of the set of free parameters $\boldsymbol{\theta} = \{\sv, \mdm, \xi, \phia, \gamma\}$, while in the latter scenario, for each pair of channels, it is defined in terms of $\boldsymbol{\theta} = \{\sv, \mdm, \xi, {\rm BR}\}$, where ${\rm BR}$ is the branching ratio for annihilations into the first of the two channels.

\subsection{Results: DM annihilations plus isotropic astrophysical power-law flux}
\label{sec:DNann-onech}

\begin{table}[t]
	\thispagestyle{empty}
	\centering
	\begin{tabular}{c|rcccc|ccc}
		\hline
		Ann.\ channel & $\sv_{22}$  &  $\mdm$ [TeV] & $\xi$ & $\phia $  & $\gamma$ & $N_{\rm DM, G}^{\rm ann}$  & $N_{\rm DM, EG}^{\rm ann}$ & $\nast$  \\
		\hline
		$u\bar{u}$                &    52.24 &     260 &      0.001 &      1.02 &      2.52 &    {\it 20.6} &     {\it 0.0} &    {\it 20.2} \\
		
		$b\bar{b}$                &    24.10 &     491 &      0.001 &      0.81 &      2.45 &    {\it 23.2} &    {\it 0.0} &    {\it 17.3} \\
		
		$t\bar{t}$                &     8.20 &     270 &      0.001 &      0.69 &      2.40 &    {\it 24.8} &     {\it 0.0} &    {\it 15.8} \\
		
		$\boldsymbol{W^{+}W^{-}}$              &     \textbf{1.51} &    \textbf{178} &    \textbf{0.001} &   \textbf{0.87} &   \textbf{2.48} &    \textbf{\textit{22.5}} &   \textbf{\textit{0.0}} &   \textbf{\textit{18.1}}  \\
		
		$Z Z$                     &     1.27 &     177 &      0.001 &      0.91 &      2.50 &   {\it 22.2} &    {\it 0.0} &     {\it 18.4} \\
		
		$h h$                     &     7.46 &     278 &      0.001 &      0.69 &      2.40 &    {\it 24.9} &     {\it 0.0} &     {\it 15.8} \\
		
		$e^{+}e^{-}$              &     1.03 &     159 &      0.635 &      1.65 &      2.75 &    {\it 13.5} &    {\it 1.3} &     {\it 25.8} \\
		
		$\mu^{+}\mu^{-}$          &     0.63 &     205 &      0.001 &      0.71 &      2.41 &    {\it 24.6} &  {\it 0.0} & {\it 15.9} \\
		
		$\tau^{+}\tau^{-}$        &     0.96 &     218 &      0.001 &      0.66 &      2.39 &    {\it 25.5} &   {\it 0.0} &   {\it 15.4} \\
		
		$\nu_e\bar{\nu}_e$        &     0.33 &     158 &      3.388 &      1.67 &      2.76 &    {\it 10.8} &   {\it 3.8} &   {\it 26.0} \\
		
		$\nu_\mu\bar{\nu}_\mu$    &     0.70 &     159 &      1.791 &      0.96 &    2.52 &   {\it 19.0} &   {\it 3.1} &   {\it 18.9} \\
		
		$\nu_\tau\bar{\nu}_\tau$  &     0.70 &     159 &      1.945 &      0.96 &      2.52 &   {\it 18.8} &   {\it 3.4} &  {\it 18.9} \\
		
		\hline
	\end{tabular}
	\caption{\textbf{\textit{DM annihilations (single channel) plus astrophysical power-law flux: Best-fit values}} for $\boldsymbol{\theta} = \{\sv, \mdm, \xi, \phia, \gamma\}$, where \sv\ is expressed in units of $10^{-22}~{\rm cm}^3~{\rm s}^{-1}$, $\mdm$ in TeV and $\phia$ in units of $10^{-18}~{\rm GeV}^{-1}~{\rm cm}^{-2}~{\rm s}^{-1}~{\rm sr}^{-1}$. The corresponding numbers of galactic and extragalactic DM and astrophysical events are also indicated as $N_{\rm DM, G}^{\rm ann}$, $N_{\rm DM, EG}^{\rm ann}$ and $N_{\rm astro}$. The overall best fit for all those channels is highlighted.}
	\label{tab:ann-onech}
\end{table}

\begin{figure}[t]
	\centering
	\includegraphics[width=0.75\textwidth]{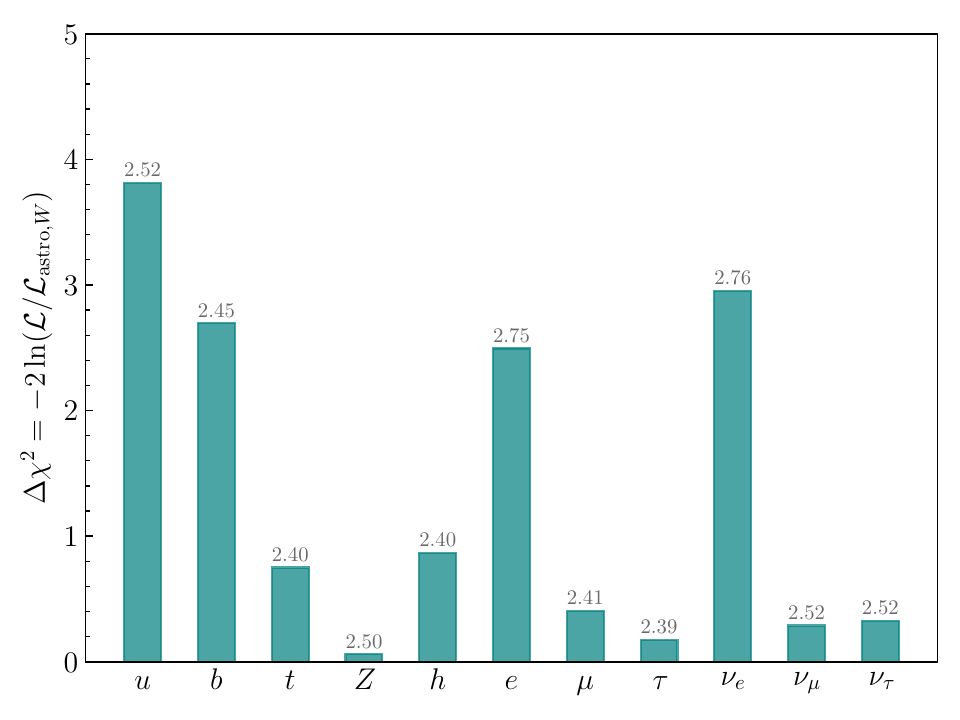}
	\caption{
		\textbf{\textit{DM annihilations (single channel) plus astrophysical power-law flux: Channel-by-channel comparison of $\boldsymbol{\Delta\chisq}$ at best fit}}, computed against the astrophysical flux plus $\dm \, \dm \to W^+ W^-$ channel, which gives the overall best fit. Best-fit values of the spectral index $\gamma$ for each channel are displayed above the corresponding bar to indicate that the best fits prefer flat astrophysical spectra, and consequently low-\mdm\ values.}
	\label{fig:all-ch-like_ann}
\end{figure}

\begin{figure}[t]
	\begin{center}
		\includegraphics[width=0.75\linewidth]{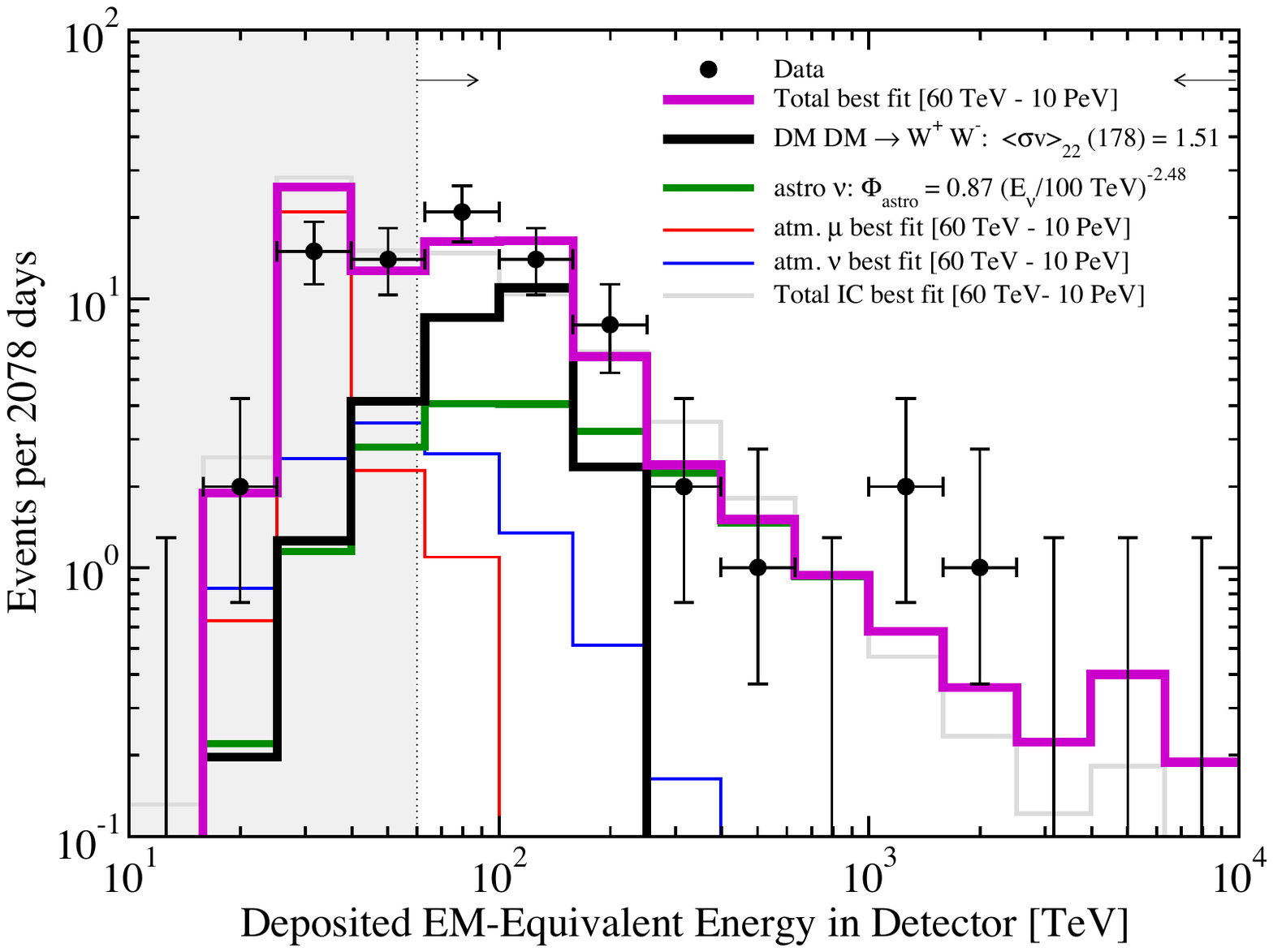}
	\end{center}
	\caption{
		\textbf{\textit{DM annihilations (single channel) plus astrophysical power-law flux: Event spectra in the IceCube detector after 2078 days}} for the best fit channel: $\dm \, \dm \to W^{+} W^{-}$, with $\xi = 10^{-3}$. The histograms represent: atmospheric muon events (red histogram), conventional atmospheric neutrino events (blue histogram), astrophysical neutrino events (green histogram), neutrino events from DM decays (black histogram), and total event spectrum (purple histogram). We indicate the best fit values of the DM annihilation cross section (\sv) and mass ($\mdm$) in units of $10^{-22}~{\rm cm}^{3}~{\rm s}^{-1}$ and TeV respectively, and the per-flavor normalization of the power-law flux ($\phia$) in units of  $10^{-18} \, {\rm GeV}^{-1} \, {\rm cm}^{-2} \, {\rm s}^{-1} \, {\rm sr}^{-1}$. We also show the spectrum obtained using the 6-year IceCube best fit for a single power-law flux (gray histogram), $E_\nu^2 \, d\Phi/dE_\nu = 2.46 \times 10^{-8} \, (E_\nu/100 \, {\rm TeV})^{-0.92}  \, {\rm GeV} \, {\rm cm}^{-2} \, {\rm s}^{-1} \, {\rm sr}^{-1}$ (per flavor) and the binned high-energy neutrino event data (black dots)~\cite{Kopper:2017zzm} with Feldman-Cousins errors~\cite{Feldman:1997qc}.}
	\label{fig:ann-events}
\end{figure}

First we consider the annihilations of DM particles into a single channel and their possible contribution to the observed neutrino flux by IceCube, in addition to an astrophysical power-law flux. The set of free parameters in the fit is $\boldsymbol{\theta} =  \{\sv , \mdm, \xi, \phia, \gamma\}$. The best-fit parameters corresponding to a total signal flux comprising neutrinos from a power-law astrophysical spectrum determined by \phia\ and $\gamma$, and from DM annihilations, as a function of \sv, \mdm\ and $\xi$, for different two-body annihilation channels, are indicated in Table~\ref{tab:ann-onech}. The comparison of the best-fit likelihoods for all channels is shown in Figure~\ref{fig:all-ch-like_ann}. As in the case of DM decays, on top of each bar, we indicate the best fit obtained for the astrophysical index $\gamma$. In Figure~\ref{fig:ann-events} we show the event spectra for the best-fit channel, $\dm \, \dm \to W^+ W^-$.

Qualitatively similar results are obtained for all channels. The preferred value for the DM mass lies in the range $\sim$ (160--500)~TeV, and with an annihilation cross section of the order of $\sim (3 \times 10^{-23} - 5 \times 10^{-21})~{\rm cm}^3~{\rm s}^{-1}$. Thus, the DM annihilation signal tends to explain the low-energy part of the event spectrum, whereas a relatively hard astrophysical flux (although softer than the through-going muon best fit~\cite{Haack:2017dxi}) explains the highest energy events (Figure~\ref{fig:ann-events}). In Table~\ref{tab:ann-onech} we also show the number of DM events from the galactic and extragalactic contributions separately, whose relative importance is governed by the parameter $\xi$. For most channels, especially for the soft channels, the preferred value of $\xi$ reaches our lower boundary (i.e, $10^{-3}$), which results in a negligible amount of events produced from the DM extragalactic neutrino flux. In these cases, the astrophysical power-law flux is the main contribution to events from the Northern hemisphere. Note, however, that in the analysis we perform here the angular information is only taken into account at the hemisphere level, which could reduce the sensitivity to the $\xi$ parameter. In any case, the upper limit on the anisotropy parameter $\xi$ can be explained by data preferring the isotropic component to be decoupled from the DM contribution, even if this is not statistically significant yet. This is related to the combination of a preference for a mild anisotropy, as shown for the 3-year HESE data~\cite{Esmaili:2014rma}, and the typically hard spectrum of the DM signal. Therefore, below some value of $\xi$, the number of events from extragalactic DM annihilations would be very small and data is not sensitive anymore to this parameter.

\subsubsection{Parameter correlations and preferred regions}
\label{sec:DMann-param-corr}

\begin{figure}[t]
	\includegraphics[width=1.0\textwidth]{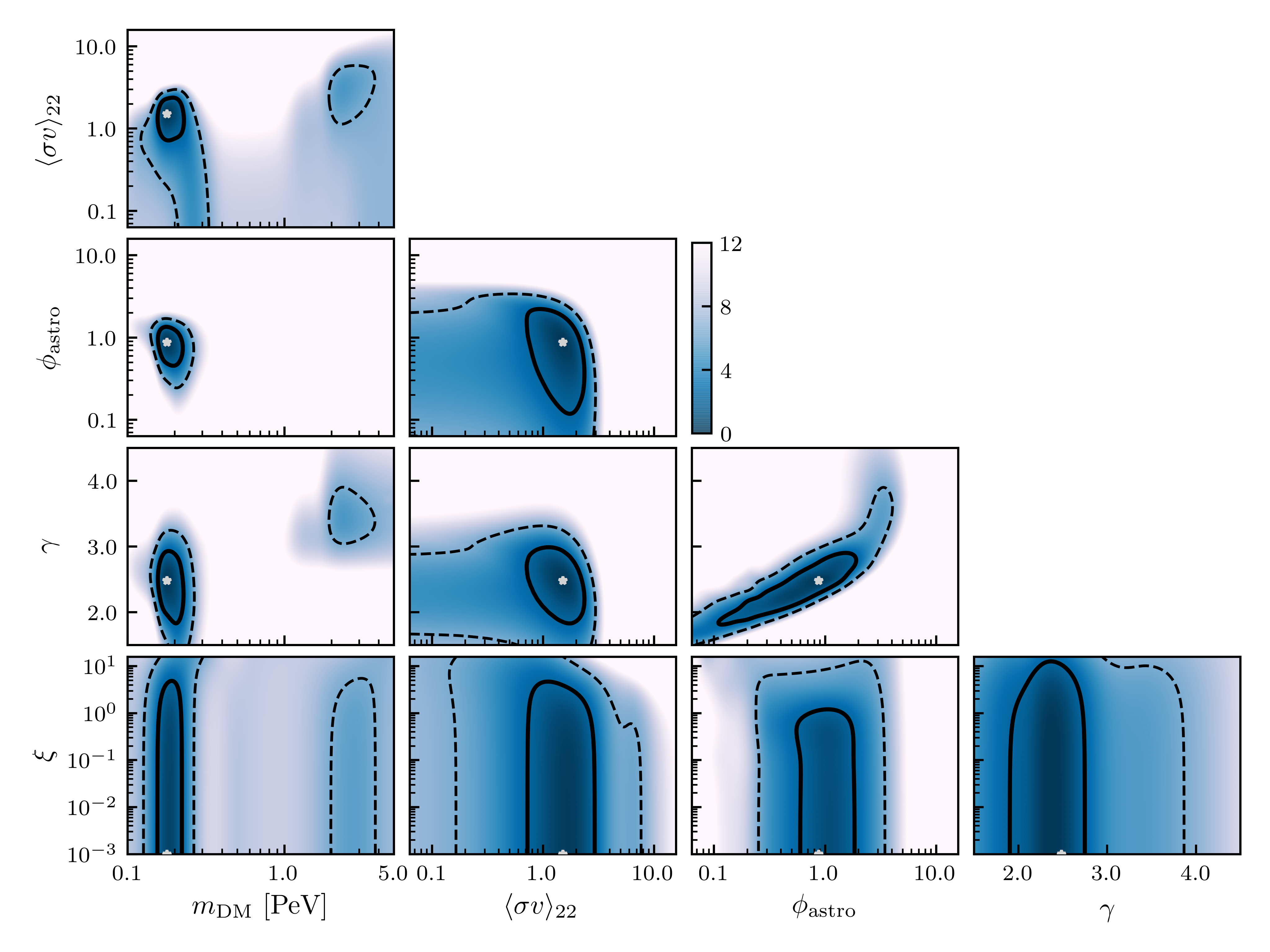}
	\caption{
		\textbf{\textit{DM annihilations (single channel) plus astrophysical power-law flux: Correlations between all fit parameters for the hard channel}} $\boldsymbol{{\rm DM \, DM} \to W^+ W^-}$. Analogous to Figures~\ref{fig:dk-corr-WW} and~\ref{fig:dk-corr-nuenue}, with \sv\ in units of $10^{-22}~{\rm cm}^{3}~{\rm s}^{-1}$.}
	\label{fig:ann-corr-WW}
\end{figure}

The preference for small values of the parameter $\xi$ can be seen from the plots in the bottom row of Figure~\ref{fig:ann-corr-WW}, which show different correlations for the $W^+ W^-$ annihilation channel plus and astrophysical isotropic power-law flux. The vertical area in those plots visualizes the lack of sensitivity for small values of $\xi$.\footnote{Note that for DM decays, the relative contribution from the galactic and extragalactic components only depends on the galactic DM profile, so there is no freedom analogous to the $\xi$ parameter.} We can also clearly see the preference --- although at less than 2$\sigma$~CL --- for $\mdm \sim 150$~TeV and annihilations cross section orders of magnitude larger than the common thermal freeze-out value, and for a relatively hard astrophysical power-law flux, in acceptable agreement with the through-going muon spectrum~\cite{Haack:2017dxi}.

\begin{figure}[t]
	\subfigure[][]{\includegraphics[width=0.32\textwidth]{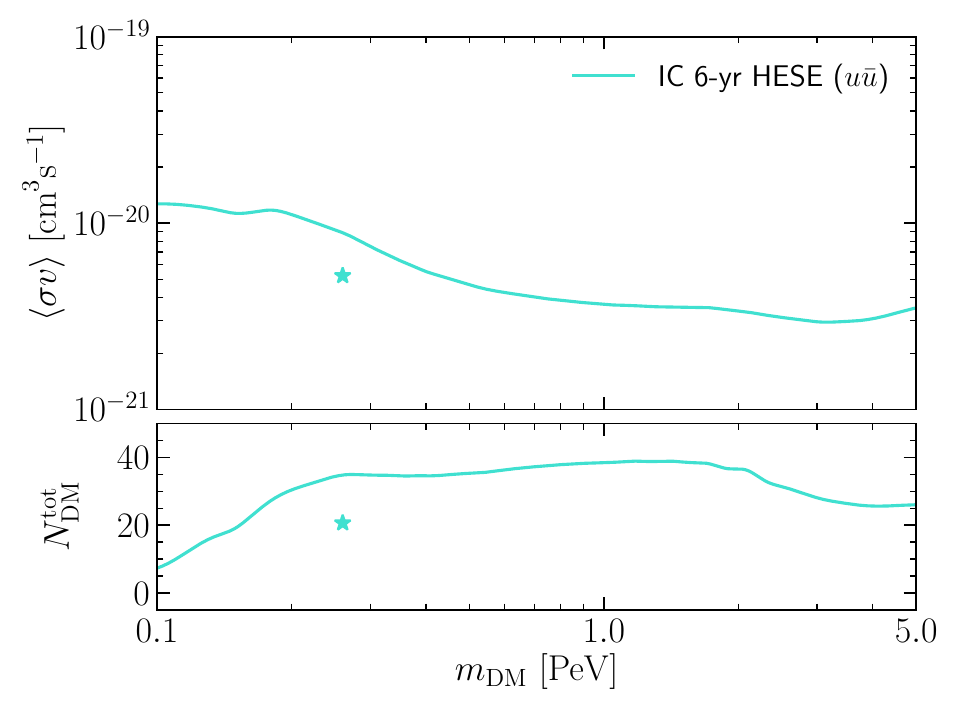}}
	\subfigure[][]{\includegraphics[width=0.32\textwidth]{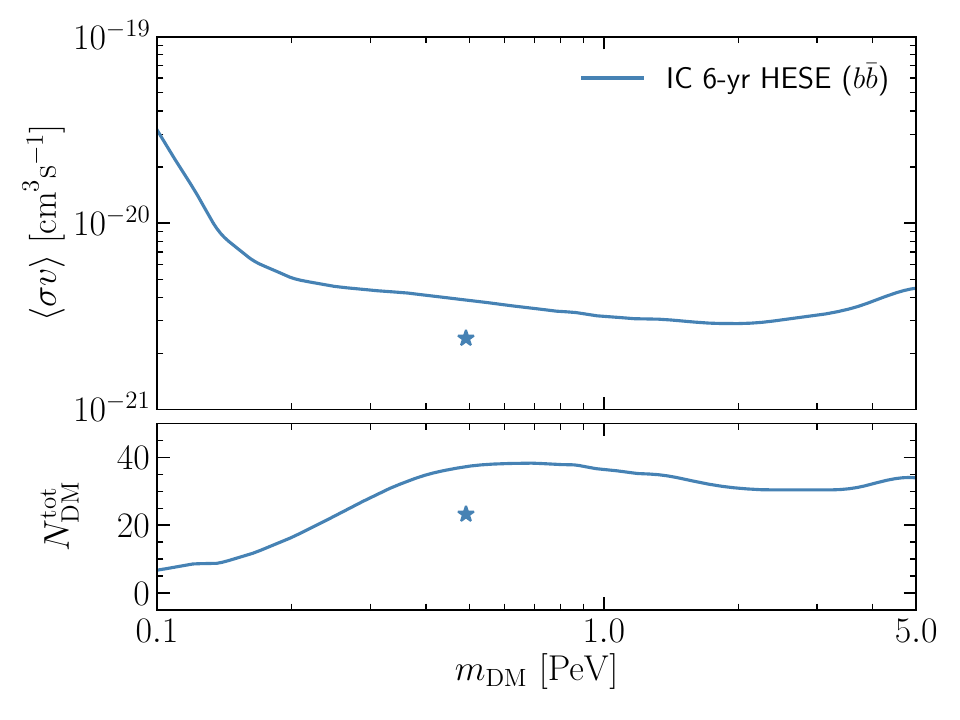}}
	\subfigure[][]{\includegraphics[width=0.32\textwidth]{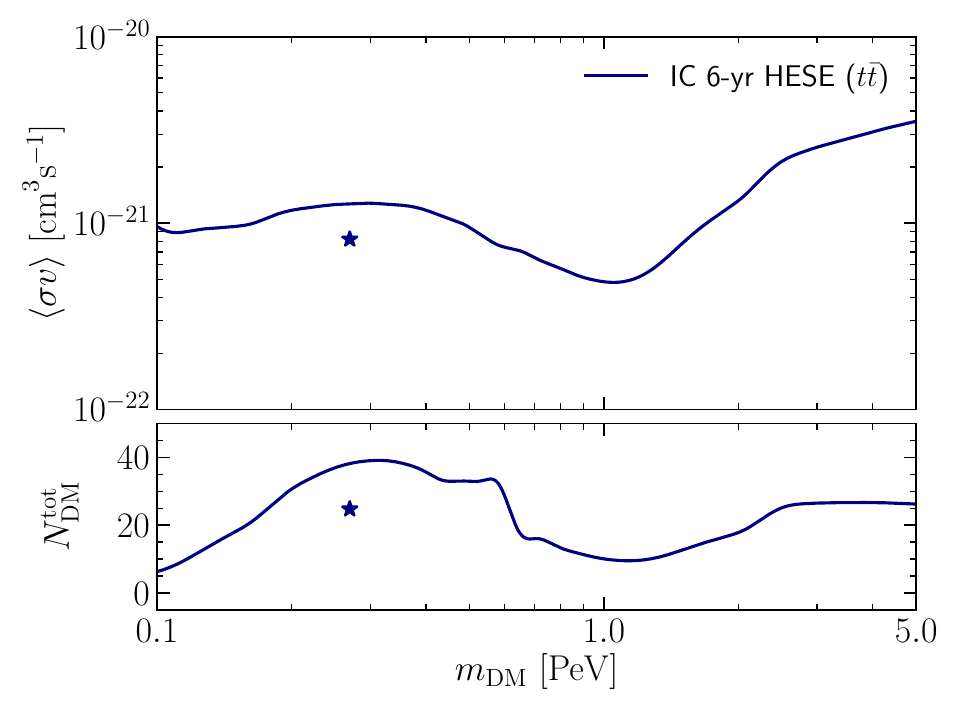}}\\
	\subfigure[][]{\includegraphics[width=0.32\textwidth]{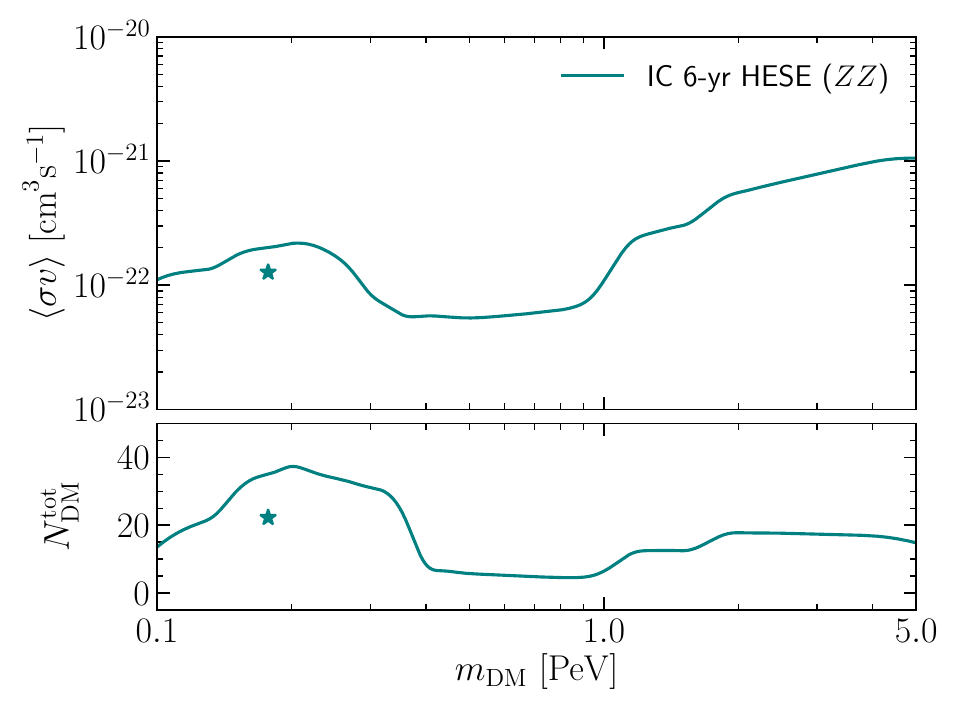}}
	\subfigure[][]{\includegraphics[width=0.32\textwidth]{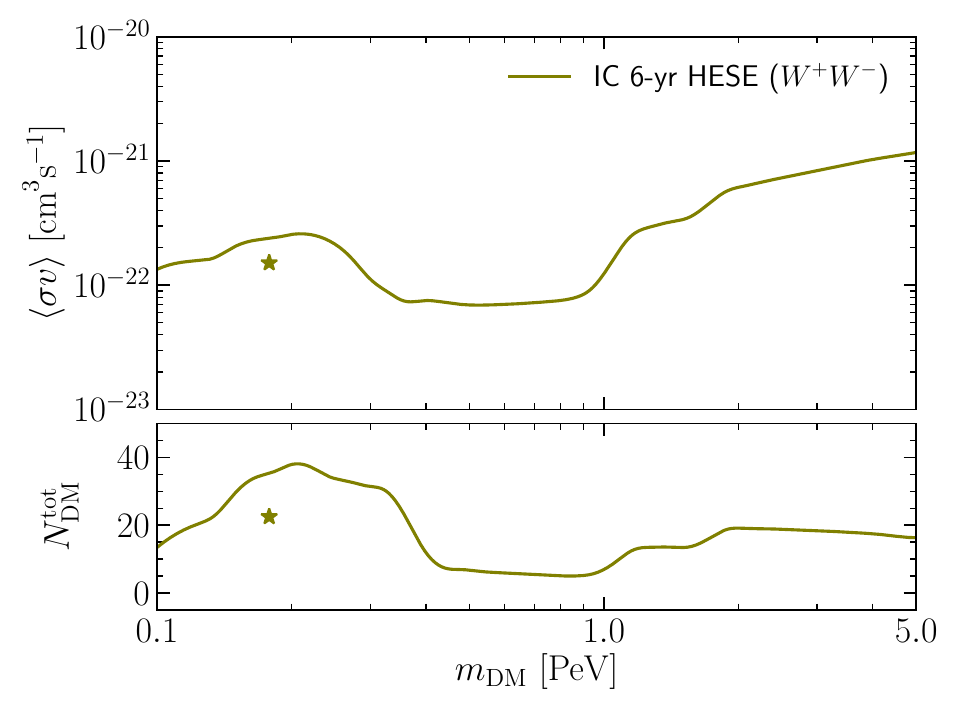}}
	\subfigure[][]{\includegraphics[width=0.32\textwidth]{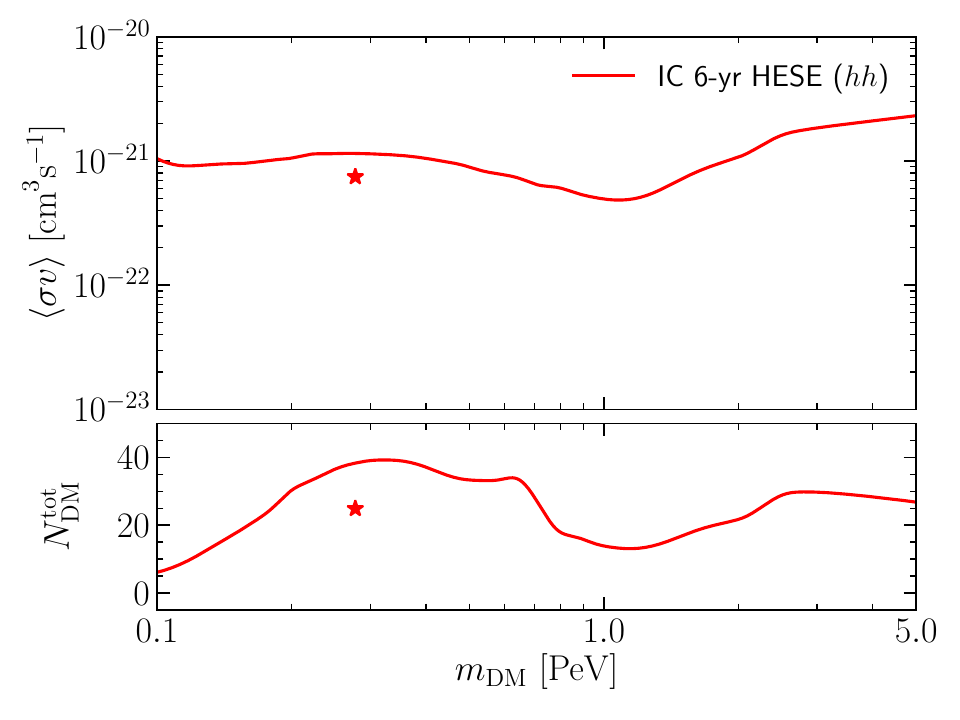}}\\
	\subfigure[][]{\includegraphics[width=0.32\textwidth]{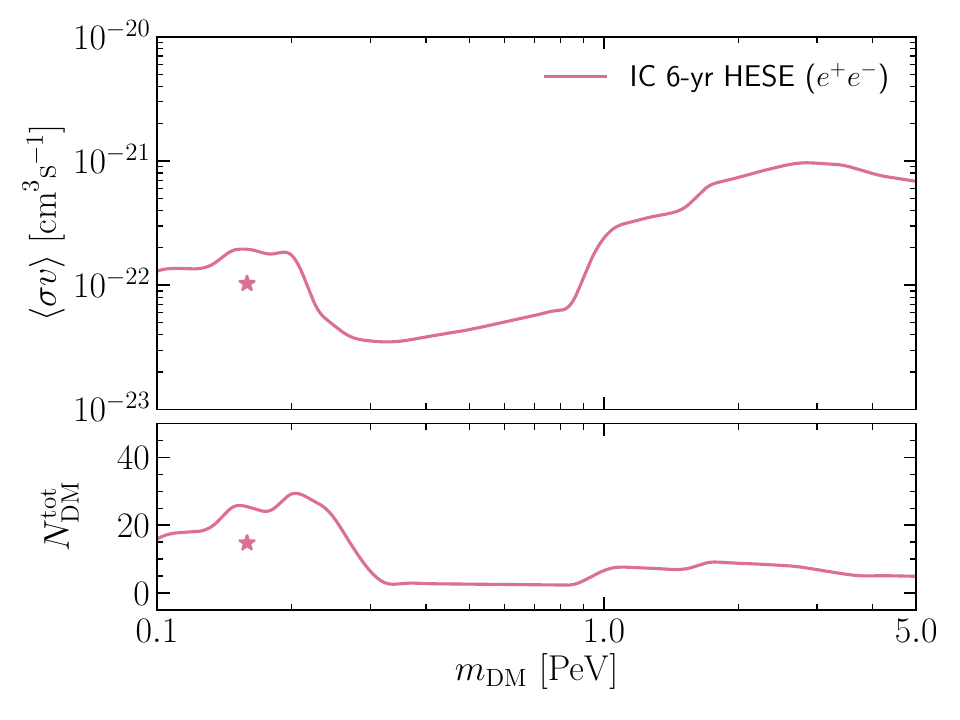}}
	\subfigure[][]{\includegraphics[width=0.32\textwidth]{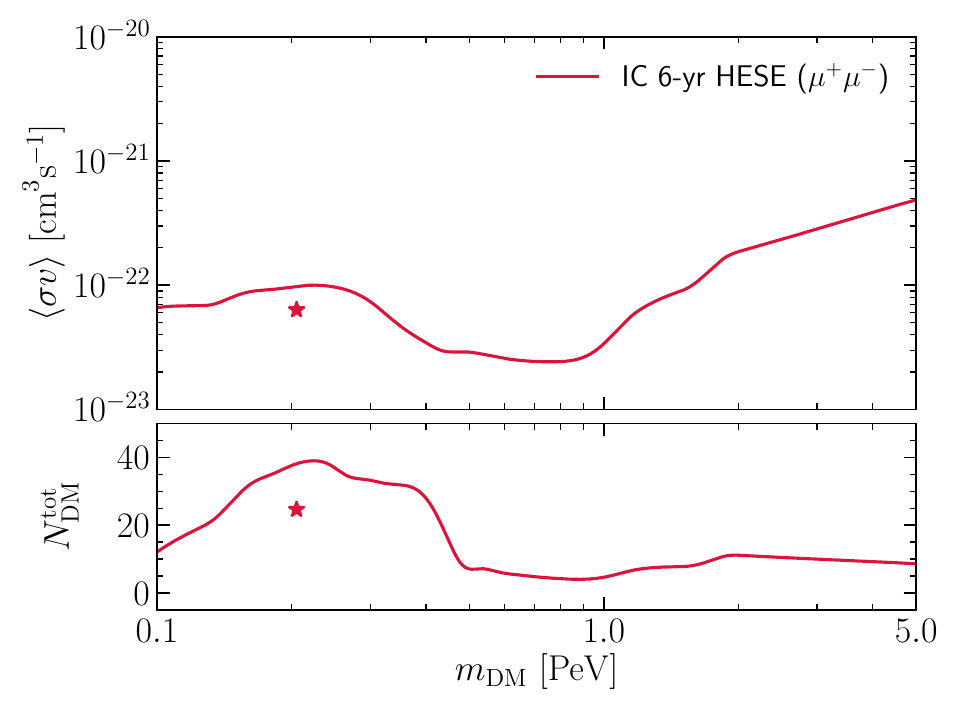}}
	\subfigure[][]{\includegraphics[width=0.32\textwidth]{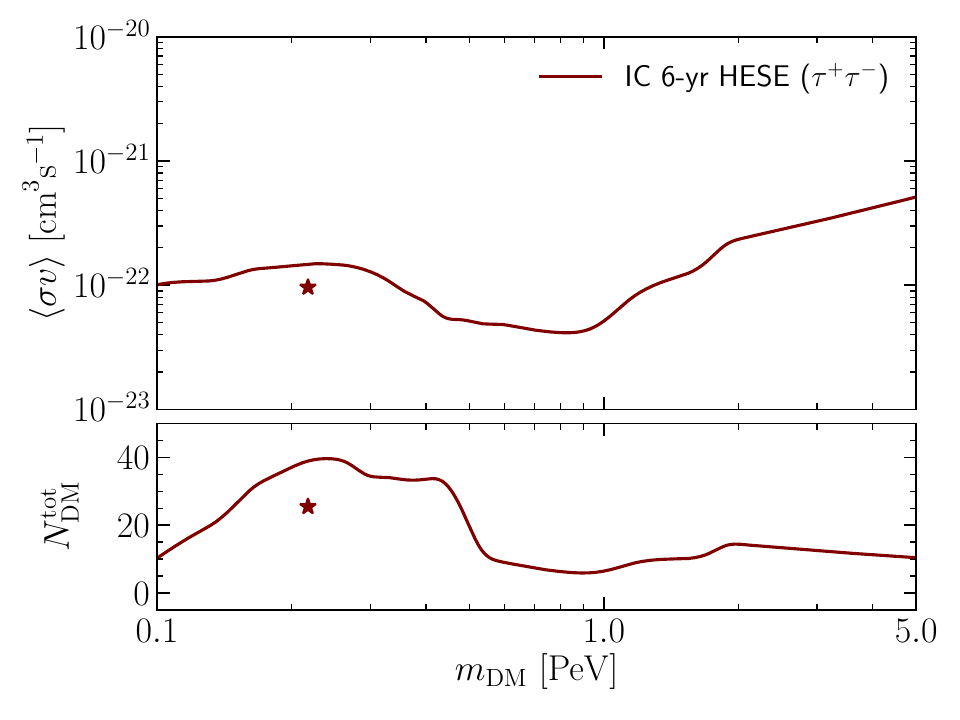}}\\
	\subfigure[][]{\includegraphics[width=0.32\textwidth]{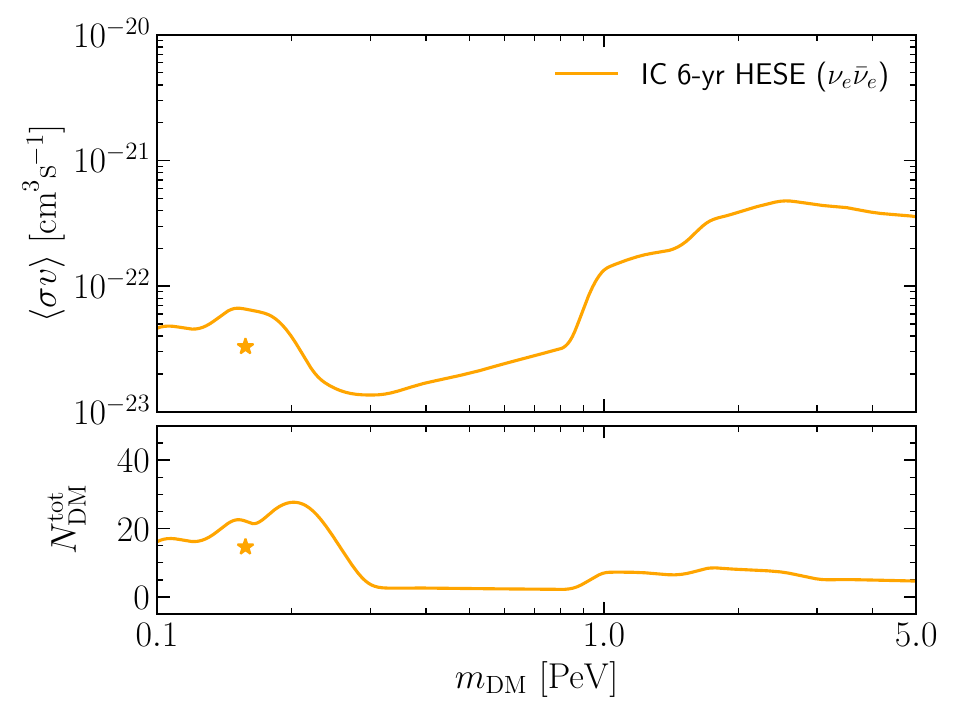}}
	\subfigure[][]{\includegraphics[width=0.32\textwidth]{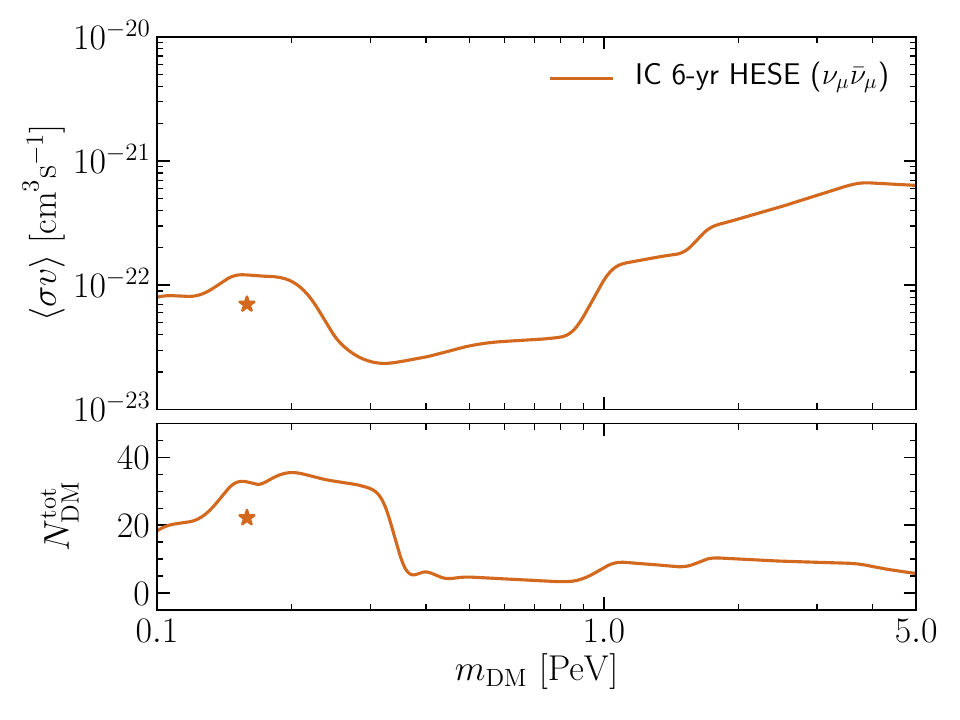}}
	\subfigure[][]{\includegraphics[width=0.32\textwidth]{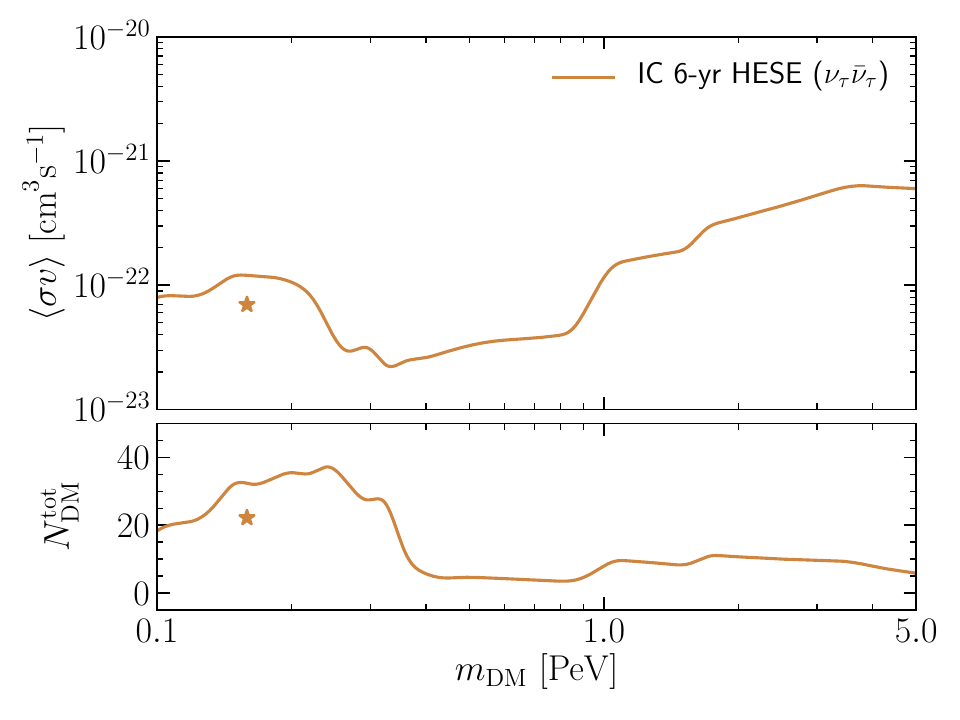}}  
	\caption{
		\textbf{\textit{DM annihilations (single channel) plus astrophysical power-law flux: Limits on the DM cross-section}} \sv\ and $N^\text{ann}_\text{DM} = N^\text{ann}_\text{DM,G} + N^\text{ann}_\text{DM,EG}$ at 95\%~CL, as a function of the DM mass. The best-fit values for $(\mdm, \sv)$ are indicated in each case by the `$\star$' sign. The unitarity bound in the halo, with a typical local relative velocity of DM particles of $v_{\rm local} = 10^{-3} \, c$, lies below the best-fit point in all cases and is not shown.}
	\label{fig:siglims-allch-app}
\end{figure}

In summary, a common feature in all cases is the preference for a negligible or subdominant flux of extragalactic neutrinos from DM annihilations (i.e., small $\xi$) and with galactic DM annihilations only contributing to the low-energy part of the spectrum.

\subsubsection{Limits on the DM annihilation cross section}
\label{sec:DMann-limits}

As done for the case of DM decays, we can also use the 6-year HESE data to evaluate the maximum contribution to the event spectrum that could come from DM annihilations. Thus, we also compute the limits on the annihilation cross section as a function of the DM mass, for the two-body annihilation channels indicated in Table~\ref{tab:ann-onech}. All these results are shown in Figure~\ref{fig:siglims-allch-app} and can be easily understood by comparison with Figure~\ref{fig:ltlims-allch-app}. On the other hand, the unitarity bound in the local halo~\cite{Griest:1989wd, Hui:2001wy} is approximately given by
\begin{equation}
\label{eq:unitarity}
\sv \lesssim \frac{4 \pi}{\mdm^2 \, v_{\rm local}} \simeq 1.5 \times 10^{-23}~{\rm cm}^3/{\rm s} \, \left(\frac{100~{\rm TeV}}{\mdm}\right)^2 \, \left(\frac{10^{-3} \, c}{v_{\rm local}}\right) ~,
\end{equation}
where $v_{\rm local}$ is the typical local relative velocity of DM particles.

For all the channels we consider, this bound (not shown in the figure) is below the best fit point. In particular, soft channels (upper panels in Figure~\ref{fig:siglims-allch-app}) are in very strong tension with unitarity constraints, with differences of about two orders of magnitude. And even in the case of the hardest channels, data prefers values of the annihilation cross section a factor of a few above the unitarity limit. Note, however, that this bound should be taken as an order of magnitude estimate. There are different effects that could modify it, as the average over the galactic distribution or the redshift dependence when applied to the cosmological contribution.

\subsection{Results: DM annihilations via multiple channels}
\label{sec:DMann-multich}

\begin{table}[t]
	\begin{center}
		\begin{tabular}{cc|rcrc|rccc}
			\hline
			\multicolumn{2}{c|}{Ann.\ channels}
			& $\sv_{22}$  &  $ \mdm $ [PeV] & $\xi$ &  BR & $N^a_{\rm G}$  & $N^a_{\rm EG}$ & $N^b_{\rm G}$ & $N^b_{\rm EG}$   \\
			\hline
			$\boldsymbol{u\bar{u} }$   & \boldsymbol{$e^{+}e^{-}}$ & \textbf{18.32} & \textbf{1.99} & \textbf{4.52} & \textbf{0.92 } & {\it\textbf{17.9 }} & {\it\textbf{18.3 }} & {\it\textbf{1.7 }} & {\it\textbf{2.8 }} \\
			$u\bar{u}$                 & $\mu^{+}\mu^{-}$          &     19.41 &       2.13 &       3.24 &       0.98 & {\it 19.8 } & {\it 14.9 } & {\it  2.6 } & {\it  3.5 } \\
			$u\bar{u}$                 & $\nu_\mu\bar{\nu}_\mu$    &     17.95 &       1.99 &       4.15 &       0.95 & {\it 18.2 } & {\it 17.0 } & {\it  2.5 } & {\it  3.2 } \\
			$u\bar{u}$                 & $\nu_\tau\bar{\nu}_\tau$  &     17.42 &       1.99 &       4.33 &       0.95 & {\it 17.6 } & {\it 17.2 } & {\it  2.6 } & {\it  3.5 } \\
			$u\bar{u}$                 & $\tau^{+}\tau^{-}$        &     17.86 &       2.16 &       3.64 &       0.97 & {\it 18.1 } & {\it 15.2 } & {\it  3.1 } & {\it  4.3 } \\
			$u\bar{u}$                 & $Z Z$                     &     18.07 &       2.15 &       3.64 &       0.93 & {\it 17.7 } & {\it 14.9 } & {\it  3.6 } & {\it  4.6 } \\
			$b\bar{b}$                 & $e^{+}e^{-}$              &     17.14 &       2.07 &       5.02 &       0.92 & {\it 16.6 } & {\it 19.4 } & {\it  1.7 } & {\it  3.0 } \\
			$b\bar{b}$                 & $\mu^{+}\mu^{-}$          &     19.23 &       2.40 &       3.47 &       0.97 & {\it 19.1 } & {\it 15.8 } & {\it  2.3 } & {\it  3.3 } \\
			$b\bar{b}$                 & $\nu_\mu\bar{\nu}_\mu$    &     17.10 &       2.06 &       4.44 &       0.95 & {\it 17.2 } & {\it 17.7 } & {\it  2.4 } & {\it  3.4 } \\
			$b\bar{b}$                 & $\nu_\tau\bar{\nu}_\tau$  &     16.48 &       2.05 &       4.66 &       0.95 & {\it 16.6 } & {\it 17.9 } & {\it  2.6 } & {\it  3.7 } \\
			$b\bar{b}$                 & $\tau^{+}\tau^{-}$        &     17.91 &       2.60 &       4.04 &       0.97 & {\it 17.3 } & {\it 16.9 } & {\it  2.5 } & {\it  4.0 } \\
			$b\bar{b}$                 & $Z Z$                     &     17.74 &       2.29 &       3.77 &       0.93 & {\it 17.0 } & {\it 15.2 } & {\it  3.6 } & {\it  4.9 } \\
			$\mu^{+}\mu^{-}$           & $\nu_\mu\bar{\nu}_\mu$    &      0.36 &       1.06 &      26.91 &       0.65 & {\it  3.0 } & {\it 28.6 } & {\it  0.8 } & {\it  5.7 } \\
			$\mu^{+}\mu^{-}$           & $\nu_\tau\bar{\nu}_\tau$  &      0.55 &       1.78 &      32.81 &       0.49 & {\it  1.9 } & {\it 24.4 } & {\it  1.0 } & {\it  9.7 } \\
			$\tau^{+}\tau^{-}$         & $\nu_\mu\bar{\nu}_\mu$    &      0.70 &       1.90 &      20.28 &       1.00 & {\it  4.5 } & {\it 34.3 } & {\it  0.0 } & {\it  0.0 } \\
			$\tau^{+}\tau^{-}$         & $\nu_\tau\bar{\nu}_\tau$  &      0.71 &       1.89 &      20.61 &       0.95 & {\it  4.4 } & {\it 33.6 } & {\it  0.1 } & {\it  0.7 } \\
			$W^{+}W^{-}$               & $e^{+}e^{-}$              &      4.85 &       2.20 &       5.09 &       1.00 & {\it 13.9 } & {\it 24.8 } & {\it  0.0 } & {\it  0.0 } \\
			$Z Z$                      & $e^{+}e^{-}$              &      4.92 &       2.20 &       4.64 &       1.00 & {\it 14.4 } & {\it 23.9 } & {\it  0.0 } & {\it  0.0 } \\
			\hline
		\end{tabular}
	\end{center}	 
	\caption{
		\textbf{\textit{DM-only two-channel annihilations: Best-fit values}} for $\dm \, \dm \to a \anti{a}, \, b\anti{b}$ defined by $\boldsymbol{\theta} = \{\sv, \mdm, \xi, \text{BR}\}$, where \sv\ is expressed in units of $10^{-22}~{\rm cm}^3~{\rm s}^{-1}$ and $\mdm$ in PeV. Galactic and extragalactic DM event numbers corresponding to channel \textit{a} are indicated as $N^a_{\rm G}$, $N^a_{\rm EG}$, respectively. The overall best fit for all those channels is highlighted.}  	  	
	\label{tab:twoch-ann-fits-mulch-60}
\end{table}

In a scenario where all the HESE (non-background) events come from DM annihilations, necessarily the constraints on the $\xi$ parameter get modified. In this case, the purely isotropic component must come from the extragalactic DM neutrino flux and hence, the number of extragalactic events from DM annihilation must be at the level of or dominant in comparison to those from the galaxy, underpinned by larger values of $ \xi $. To verify this, we consider a two-channel annihilation scenario without an astrophysical flux, similar to the case for DM decays in section~\ref{sec:DMdk-multich}. The set of free parameters for each pair of channels is $\boldsymbol{\theta} = \{\sv, \mdm, \xi, {\rm BR}\}$.

\begin{figure}[t]
  \centering
  \includegraphics[width=0.7\textwidth]{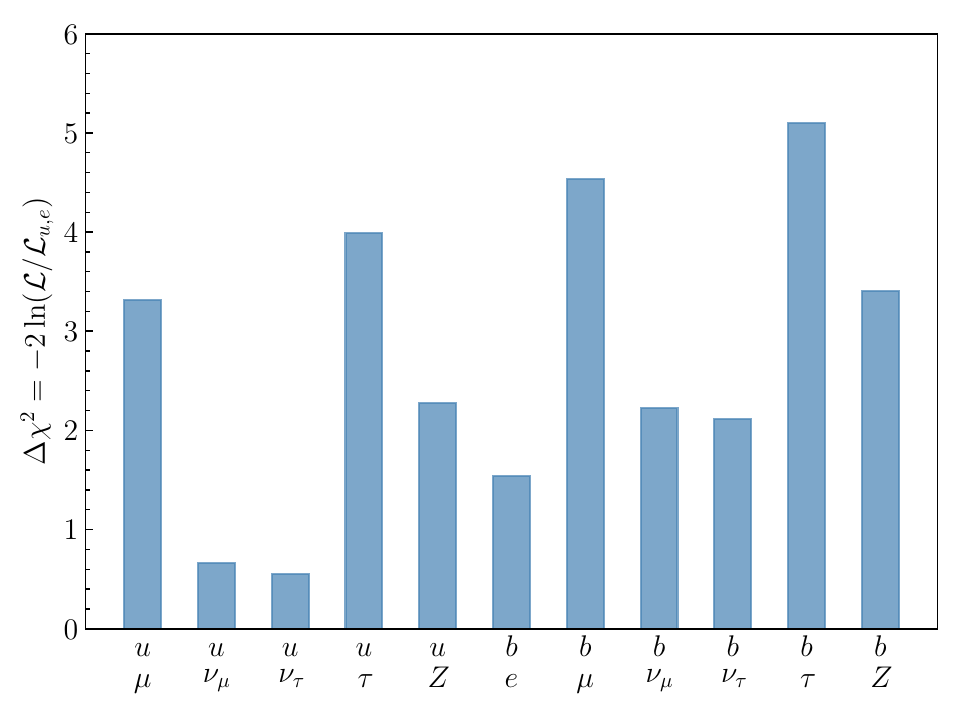}
	\caption{
		\textbf{\textit{DM-only two-channel annihilation: Channel-by-channel comparison of $\boldsymbol{\Delta\chisq}$ at best fit}}, computed against the overall best-fit channel: $\dm\ \dm \to  \lbrace u\bar{u}, e^{+}e^{-}\rbrace$. Channel combinations from Table~\ref{tab:twoch-ann-fits-mulch-60} not shown in this plot represent extremely poor fits with $\Delta\chi^{2} \gtrsim 15$.}
	\label{fig:twoch-all-ch-like_ann}
\end{figure}

\begin{figure}[t]
  \centering
  \subfigure[{\scriptsize $\dm \, \dm \to \lbrace u\bar{u},e^+e^- \rbrace$}]{\includegraphics[width=0.48\textwidth]{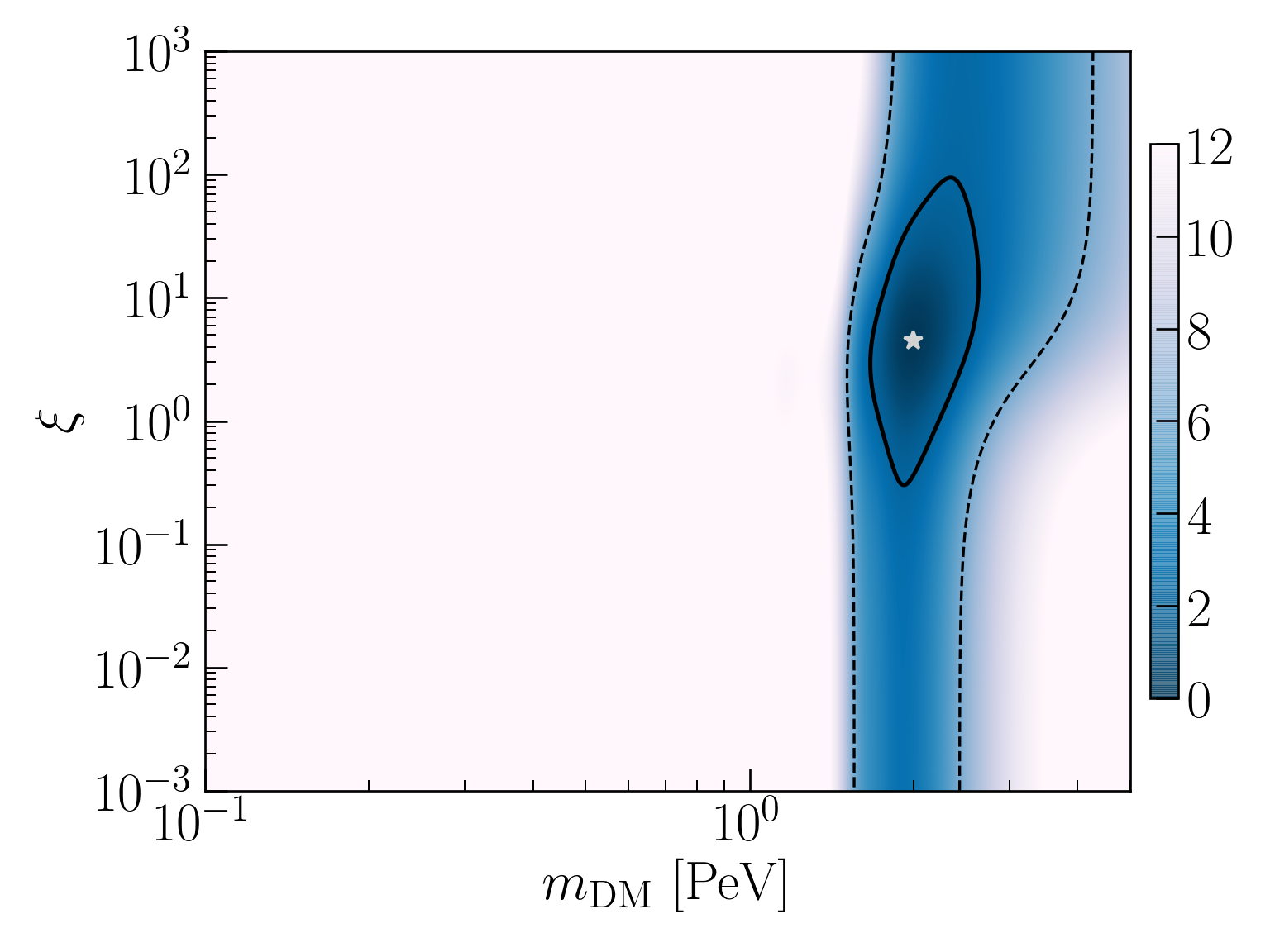}}
  \subfigure[{\scriptsize $\dm \, \dm \to \lbrace u\bar{u},\nu_\tau\bar{\nu}_\tau \rbrace$}]{\includegraphics[width=0.48\textwidth]{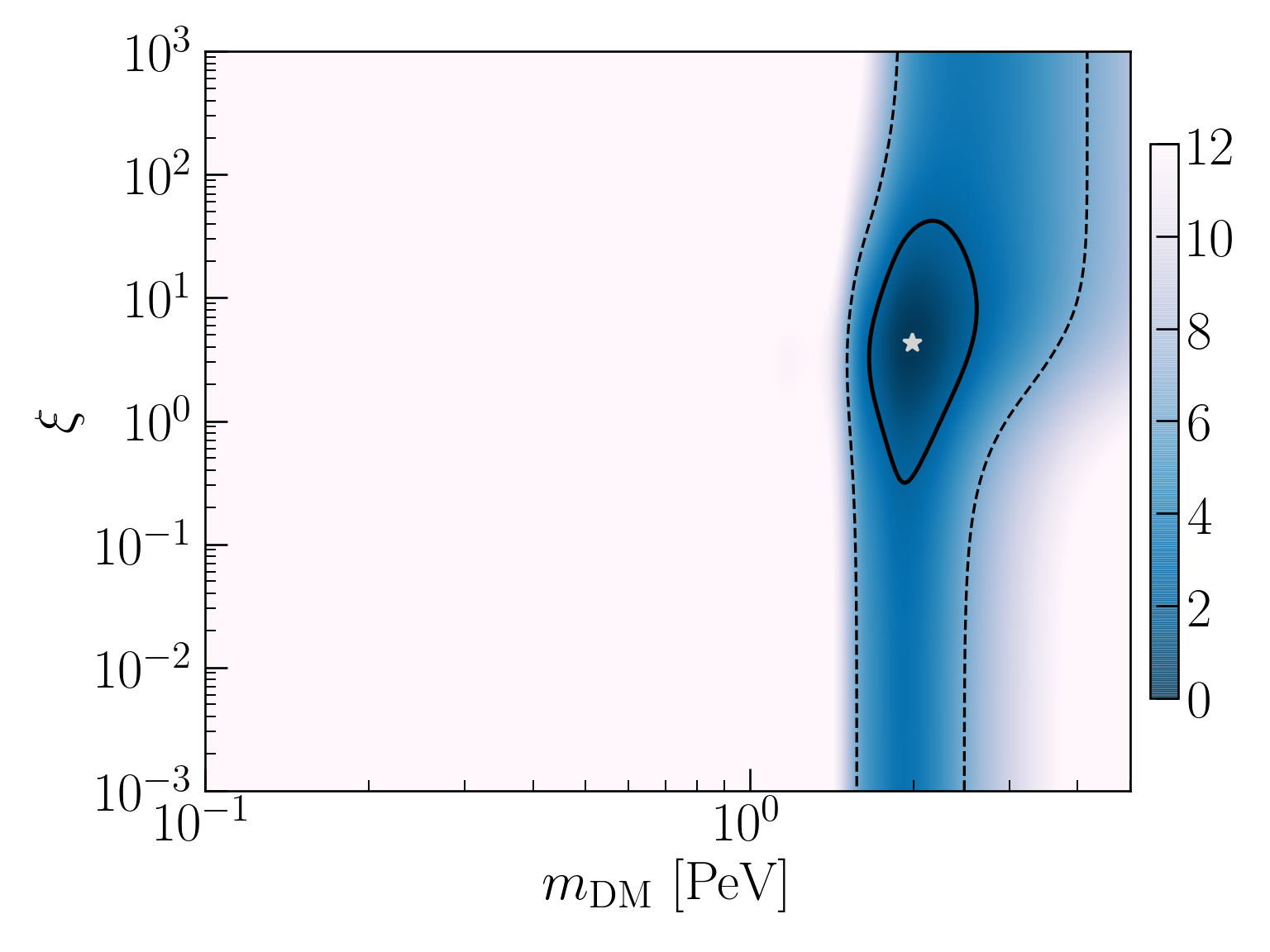}}
  \caption{
  	\textbf{\textit{DM-only two-channel annihilation: Correlations}} between the parameters \mdm\ and $\xi$ for DM annihilating via two different channels without astrophysical neutrinos: $\dm \, \dm \to \lbrace u\bar{u},e^+e^- \rbrace$ (lefpt panel) and $\dm \, \dm \to \lbrace u\bar{u},\nu_\tau\bar{\nu}_\tau \rbrace$ (right panel).}
  \label{fig:ann-twoch}
\end{figure}

Results for a few selected combinations of annihilation channels are shown in Table~\ref{tab:twoch-ann-fits-mulch-60} and the $\Delta \chi^2$ for each combination with respect to the best fit is shown in Figure~\ref{fig:twoch-all-ch-like_ann}. The overall best fit in this case is obtained for the combination of channels $\dm \, \dm \to u\bar{u}$ and $ \dm \, \dm \to e^{+}e^{-}$, with a branching ratio of about 92\% in favor of the former. As expected, in the absence of the astrophysical flux, the \dm\ mass shifts to around 2~PeV (half that for the case of DM decays) in order to accommodate the entire event spectrum, with events from annihilation into $e^{+}e^{-}$ explaining the PeV-energy events, and those into $u\bar{u}$ producing the soft spectrum that fits the sub-PeV energy signal. Furthermore, as anticipated, there is a clear preference for larger values of $\xi$ compared to those in the previous section. Figure~\ref{fig:ann-twoch}, which shows the correlation of $\xi$ and the DM mass for two different annihilation channels:
\begin{inparaenum}[\itshape 1\upshape)]
	\item $u\bar{u}$ + $e^{+}e^{-}$ (left panel) and
	\item $u\bar{u}$ + $\nu_\tau \bar{\nu}_\tau$ (right panel),
\end{inparaenum}
reveals $1\sigma$~CL preferential regions starting just below $ \xi = 1 $ and extending beyond $ \xi = 10 $.

As the comparison amongst different combinations of channels shows (Figure~\ref{fig:twoch-all-ch-like_ann}), qualitatively good fits are obtained when pairing \dm\ annihilation to a hard channel, such as a lepton pair, with a very soft one ($ u\bar{u} $ or $ b\bar{b} $). In other cases (e.g., with two hard channels), the fits worsen rather strongly, sometimes ending up with the best-fit branching fraction wholly in favor of a single channel, and invariably leading to a large $\Delta\chisq$ when computed against the overall best fit. On the other hand, as it happens for the case of decays, the large contribution from soft channels is likely to be in tension with (the most restrictive) gamma-ray limits. Nevertheless, a dedicated analysis, taking into account the full angular dependence of the signal, has not been performed and lies beyond the scope of this paper. In turn, the requirement for PeV masses also implies that for all cases the required annihilation cross section is many orders of magnitude above the unitarity bound (see eq.\ \eqref{eq:unitarity}).

\section{Discussion and conclusions}
\label{sec:conc}

As IceCube progressively records new data and improves its statistics, it is imperative to keep track of consistency between data on one hand and the many theories proposed to explain the origins of these high-energy particles on the other. As the standard explanation involving a uniform power-law fit becomes increasingly fraught with problems, including tensions with other observations such as IceCube's 8-year through-going muon set and increasing differences between sub-PeV and PeV spectral features, alternative explanations for the origins of the flux must be explored. Multicomponent explanations, involving neutrinos with two or more dynamically different origins have been invoked to yield better fits~\cite{Chen:2014gxa, Aartsen:2015knd, Palladino:2016zoe, Vincent:2016nut, Palladino:2016xsy, Palladino:2017qda, Kopper:2017zzm, Wandkowsky:2018}. In this vein, in Ref.~\cite{Bhattacharya:2017jaw} we explored two alternative scenarios that were found to qualitatively improve the fit to the 4-year HESE data. The first scenario involved a neutrino flux from the decay of PeV-scale DM complemented by a power-law flux of astrophysical origin. The second scenario involved DM decays via two different channels, thus producing complementary spectra to explain the sub-PeV and PeV events without an astrophysical flux. 

In the current work, we have reanalyzed both these scenarios against the 6-year HESE data. The two-year worth of additional data is entirely clustered in the sub-PeV region, with no change to the PeV events. Additionally, it strengthens the apparent bump in the $\sim 100$~TeV region, making the overall spectrum difficult to fit with a uniform power-law spectrum. Notwithstanding these additional events, we find that a combined flux from DM decays and an astrophysical power-law still improves the fit. The best overall fits are obtained in the cases where a relatively flat power-law spectrum explains a fraction of the sub-PeV events and the entire PeV data, while neutrinos from an $\mathcal{O}(400)$~TeV DM particle decaying to gauge bosons complement sub-PeV events from the power-law flux, thereby enhancing its softness and reproducing the $\order{100}$~TeV bump. In turn, by using only HESE data, the resulting flat astrophysical power-law flux is in good agreement with the 8-year through-going muon analysis~\cite{Haack:2017dxi}, sensitive to high energies. 

In the two-channel DM decay scenario, we also find improvements to the fit in comparison to single power-law scenarios for decays to a combination of soft and hard channels. In this case, the PeV events typically come from the hard spectrum, such as neutrinos, while the lower energy events come from decays to quarks. The overall best fit in this scenario is obtained for the decay $\dm \to \lbrace u\bar{u},\nue\anti{\nu}_e \rbrace$ with a branching ratio of 97\% in favor of the quark channel.

\begin{figure}[t]
	\centering
	\includegraphics[width=0.9\textwidth]{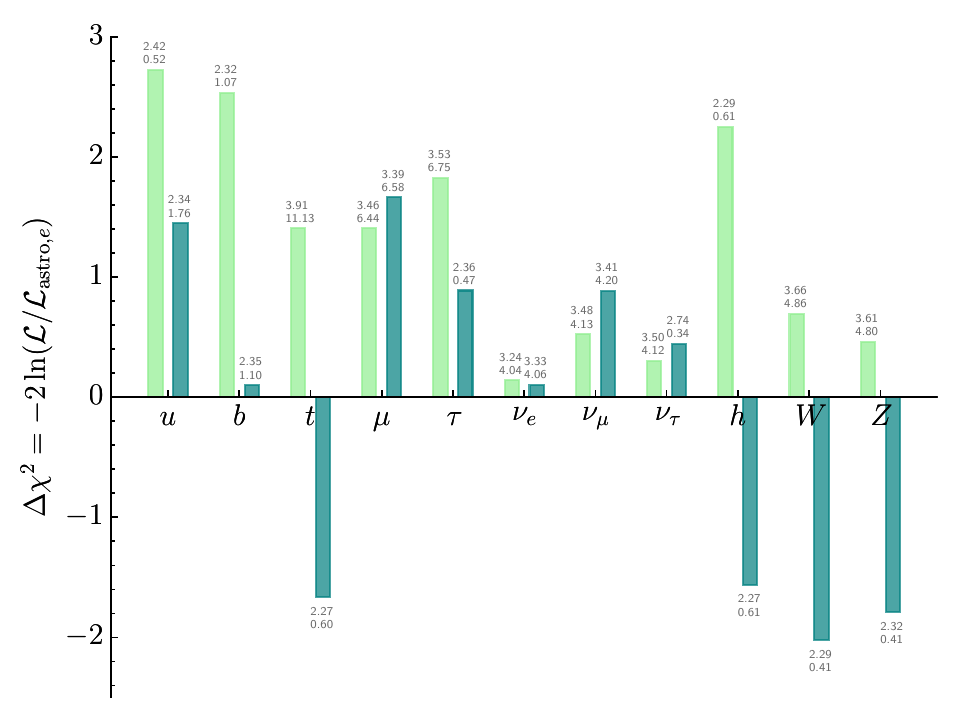}
	\caption{
		\textbf{\textit{DM decays (single channel) plus astrophysical power-law flux: Comparison of the 4-year and 6-year results.}} Channel-by-channel comparisons of $\Delta\chisq$ at best fit for the current 6-year (dark green bars) and for the 4-year (light green bars)~\cite{Bhattacharya:2017jaw} HESE data analyses, computed against the best-fit astrophysical flux plus flux from $\dm \to e^+ \, e^-$ corresponding to each dataset. $\dm \to e^+ \, e^-$ has been chosen as the reference channel for this figure as it was the overall best-fit in the 4-year HESE data analysis. Negative values of $\Delta\chisq$ in the current analysis for $h,\ t,\ W,\text{ and }Z$ channels indicate that these are better fits to the data than decays to $e^+ e^-$. Best-fit values of the DM mass (lower number, in PeV) and astrophysical spectral index (upper number) for each channel are displayed above the corresponding bars.}
	\label{fig:all-ch-like-vs-4yr}
\end{figure}

It is apparent that the 6-year HESE data prefers fits involving a DM particle of a few hundred TeV that decays via bosonic channels and populates the sub-PeV region of the spectrum. This might seem to contrast with our conclusions of Ref.~\cite{Bhattacharya:2017jaw}, where we found a general trend of hard-spectrum, high-mass DM channels involving decays to neutrinos or $e^{\pm}$ to fit the 4-year HESE data better. In that case, the neutrino flux from DM decays was found to explain the PeV events while a steeply dropping astrophysical power-law flux filled in the sub-PeV energies to complement the spectrum. On the other hand, as Fig.~\ref{fig:all-ch-like-vs-4yr} attests, with the 6-year HESE data we find that a flatter astrophysical spectrum accounts for the PeV events, while it is the component from DM decay that explains the low-energy excess and bump which are enhanced by the additional data. We attribute these qualitative changes among the nature of the fits obtained using the 4-year data and 6-year data to the strengthening of the $\order{100}$~TeV bump in the event spectrum. Moreover, note that, for the best-fit channel, the 6-year $1\sigma$ CL region was already part of the 4-year $1\sigma$ CL region (compare Figure~\ref{fig:ann-corr-WW} above with Figure~3 in Ref.~\cite{Bhattacharya:2017jaw}). 

Additionally, by looking at the maximum number of events from DM decay to within 95\% CL of the best-fit, we have drawn improved constraints on the DM lifetime. While these limits are largely unchanged from our 4-year HESE data analysis, they become stronger for the neutrino and charged leptonic channels, which, as noted before, are not as good fits to the 6-year HESE data than they were to the 4-year data.

Furthermore, in this work we have similarly analyzed the potential contribution from DM annihilations to the 6-year HESE event spectrum. It is well known that to explain a few events per year, the required annihilation cross section is very large~\cite{Feldstein:2013kka, Zavala:2014dla, Esmaili:2014rma, Chianese:2016opp, Chianese:2016kpu} and even violates the unitarity bound~\cite{Griest:1989wd, Hui:2001wy} in most cases. Nevertheless, for low DM masses and for final state leptons, previous works found results more restrictive than the unitarity bound~\cite{Esmaili:2014rma, Chianese:2016kpu}. Thus, we have revisited this scenario, assuming a constant annihilation cross section, and have performed analogous analyses to the ones for DM decay.
In this case, we have also accounted for uncertainties in the non-homogeneous DM clustering properties by adding an extra parameter that determines the relative contribution of the galactic and extragalactic DM components. Due to the more abundant low-energy events in the extra two years of data, now the best-fit values for the annihilation cross section are above the unitarity bound even for leptonic final states (Figure~\ref{fig:siglims-allch-app}), and if the data is to be explained only in terms of DM annihilations, the required cross section is many orders of magnitude larger than the unitarity limit. Therefore, we conclude that the results about the potential contribution from heavy DM annihilations obtained from the 6-year HESE data are of limited interest.

\section*{Acknowledgments}
AB expresses gratitude to Jean-Ren\'{e} Cudell and Maxim Laletin for helpful discussions and to support from the Fonds de la Recherche Scientifique-FNRS, Belgium, under grant No.~4.4503.19. AB is also thankful to the computational resource provided by Consortium des Équipements de Calcul Intensif (CÉCI), funded by the Fonds de la Recherche Scientifique de Belgique (F.R.S.-FNRS) under Grant No. 2.5020.11 where a part of the computation was carried out. AE thanks the computing resource provided by CCJDR, of IFGW-UNICAMP with resources from FAPESP Multi-user Project 09/54213-0. AE thanks the partial support by the CNPq fellowship No.~310052/2016-5. SPR is supported by a Ram\'on y Cajal contract, by the Spanish MINECO under grants FPA2017-84543-P and SEV-2014-0398, and by the European Union's Horizon 2020 research and innovation program under the Marie Sk\l odowska-Curie grant agreements No. 690575 and 674896. SPR is also partially supported by the Portuguese FCT through the CFTP-FCT Unit 777 (UID/FIS/00777/2019). IS was supported in part by the Department of Energy under Grant DE-FG02-13ER41976 (DE-SC0009913).

\bibliographystyle{JHEP}
\bibliography{dmastrefs}

\providecommand{\href}[2]{#2}\begingroup\raggedright\begin{thebibliography}{100}

\bibitem{Aartsen:2013jdh}
{\scshape IceCube} collaboration, M.~G. Aartsen et~al., \emph{{Evidence for
  high-energy extraterrestrial neutrinos at the IceCube detector}},
  \href{http://dx.doi.org/10.1126/science.1242856}{\emph{Science} {\bf 342}
  (2013) 1242856}, [\href{http://arxiv.org/abs/1311.5238}{{\tt 1311.5238}}].

\bibitem{Aartsen:2014gkd}
{\scshape IceCube} collaboration, M.~G. Aartsen et~al., \emph{{Observation of
  high-energy astrophysical neutrinos in three years of IceCube data}},
  \href{http://dx.doi.org/10.1103/PhysRevLett.113.101101}{\emph{Phys. Rev.
  Lett.} {\bf 113} (2014) 101101}, [\href{http://arxiv.org/abs/1405.5303}{{\tt
  1405.5303}}].

\bibitem{Aartsen:2015rwa}
{\scshape IceCube} collaboration, M.~G. Aartsen et~al., \emph{{Evidence for
  astrophysical muon neutrinos from the Northern sky with IceCube}},
  \href{http://dx.doi.org/10.1103/PhysRevLett.115.081102}{\emph{Phys. Rev.
  Lett.} {\bf 115} (2015) 081102}, [\href{http://arxiv.org/abs/1507.04005}{{\tt
  1507.04005}}].

\bibitem{Kopper:2015vzf}
{\scshape IceCube} collaboration, C.~Kopper, W.~Giang and N.~Kurahashi,
  \emph{{Observation of astrophysical neutrinos in four years of IceCube
  data}}, {\emph{PoS} {\bf ICRC2015} (2016) 1081}.

\bibitem{Aartsen:2016xlq}
{\scshape IceCube} collaboration, M.~G. Aartsen et~al., \emph{{Observation and
  characterization of a cosmic muon neutrino flux from the Northern hemisphere
  using six years of IceCube data}},
  \href{http://dx.doi.org/10.3847/0004-637X/833/1/3}{\emph{Astrophys. J.} {\bf
  833} (2016) 3}, [\href{http://arxiv.org/abs/1607.08006}{{\tt 1607.08006}}].

\bibitem{Kopper:2017zzm}
{\scshape IceCube} collaboration, C.~Kopper, \emph{{Observation of
  astrophysical neutrinos in six years of IceCube data}},
  \href{http://dx.doi.org/10.22323/1.301.0981}{\emph{PoS} {\bf ICRC2017} (2018)
  981}.

\bibitem{Haack:2017dxi}
{\scshape IceCube} collaboration, C.~Haack and C.~Wiebusch, \emph{{A
  measurement of the diffuse astrophysical muon neutrino flux using eight years
  of IceCube data.}}, \href{http://dx.doi.org/10.22323/1.301.1005}{\emph{PoS}
  {\bf ICRC2017} (2018) 1005}.

\bibitem{Wandkowsky:2018}
{\scshape IceCube} collaboration, N.~Wandkowsky, ``{\it Results on
  astrophysical neutrinos using 7.5 years of high-energy events with contained
  vertices}.'' Poster at Neutrino 2018, Heidelberg, Germany, June 4-9, 2018.

\bibitem{Laha:2013lka}
R.~Laha, J.~F. Beacom, B.~Dasgupta, S.~Horiuchi and K.~Murase,
  \emph{{Demystifying the PeV cascades in IceCube: Less (energy) is more
  (events)}}, \href{http://dx.doi.org/10.1103/PhysRevD.88.043009}{\emph{Phys.
  Rev.} {\bf D88} (2013) 043009}, [\href{http://arxiv.org/abs/1306.2309}{{\tt
  1306.2309}}].

\bibitem{Halzen:2013dva}
F.~Halzen, \emph{{The highest energy neutrinos: first evidence for cosmic
  origin}}, \href{http://dx.doi.org/10.1393/ncc/i2014-11772-8,
  10.1002/asna.201412058}{\emph{Nuovo Cim.} {\bf C037} (2014) 117--132},
  [\href{http://arxiv.org/abs/1311.6350}{{\tt 1311.6350}}].

\bibitem{Waxman:2013zda}
E.~Waxman, \emph{{IceCube's Neutrinos: The beginning of extra-galactic neutrino
  astrophysics?}},
\newblock \href{http://arxiv.org/abs/1312.0558}{{\tt 1312.0558}}.

\bibitem{Anchordoqui:2013dnh}
L.~A. Anchordoqui et~al., \emph{{Cosmic neutrino Pevatrons: A brand new pathway
  to astronomy, astrophysics, and particle physics}},
  \href{http://dx.doi.org/10.1016/j.jheap.2014.01.001}{\emph{JHEAp} {\bf 1-2}
  (2014) 1--30}, [\href{http://arxiv.org/abs/1312.6587}{{\tt 1312.6587}}].

\bibitem{Murase:2014tsa}
K.~Murase, \emph{{On the origin of high-energy cosmic neutrinos}},
  \href{http://dx.doi.org/10.1063/1.4915555}{\emph{AIP Conf. Proc.} {\bf 1666}
  (2015) 040006}, [\href{http://arxiv.org/abs/1410.3680}{{\tt 1410.3680}}].

\bibitem{Meszaros:2017fcs}
P.~M{\'e}sz{\'a}ros, \emph{{Astrophysical sources of high energy neutrinos in
  the IceCube era}},
  \href{http://dx.doi.org/10.1146/annurev-nucl-101916-123304}{\emph{Ann. Rev.
  Nucl. Part. Sci.} {\bf 67} (2017) 45--67},
  [\href{http://arxiv.org/abs/1708.03577}{{\tt 1708.03577}}].

\bibitem{Ahlers:2018fkn}
M.~Ahlers and F.~Halzen, \emph{{Opening a new window onto the Universe with
  IceCube}}, \href{http://dx.doi.org/10.1016/j.ppnp.2018.05.001}{\emph{Prog.
  Part. Nucl. Phys.} {\bf 102} (2018) 73--88},
  [\href{http://arxiv.org/abs/1805.11112}{{\tt 1805.11112}}].

\bibitem{Chen:2014gxa}
C.-Y. Chen, P.~S. Bhupal~Dev and A.~Soni, \emph{{Two-component flux explanation
  for the high energy neutrino events at IceCube}},
  \href{http://dx.doi.org/10.1103/PhysRevD.92.073001}{\emph{Phys. Rev.} {\bf
  D92} (2015) 073001}, [\href{http://arxiv.org/abs/1411.5658}{{\tt
  1411.5658}}].

\bibitem{Aartsen:2015knd}
{\scshape IceCube} collaboration, M.~G. Aartsen et~al., \emph{{A combined
  maximum-likelihood analysis of the high-energy astrophysical neutrino flux
  measured with IceCube}},
  \href{http://dx.doi.org/10.1088/0004-637X/809/1/98}{\emph{Astrophys. J.} {\bf
  809} (2015) 98}, [\href{http://arxiv.org/abs/1507.03991}{{\tt 1507.03991}}].

\bibitem{Palladino:2016zoe}
A.~Palladino and F.~Vissani, \emph{{Extragalactic plus galactic model for
  IceCube neutrino events}},
  \href{http://dx.doi.org/10.3847/0004-637X/826/2/185}{\emph{Astrophys. J.}
  {\bf 826} (2016) 185}, [\href{http://arxiv.org/abs/1601.06678}{{\tt
  1601.06678}}].

\bibitem{Vincent:2016nut}
A.~C. Vincent, S.~Palomares-Ruiz and O.~Mena, \emph{{Analysis of the 4-year
  IceCube high-energy starting events}},
  \href{http://dx.doi.org/10.1103/PhysRevD.94.023009}{\emph{Phys. Rev.} {\bf
  D94} (2016) 023009}, [\href{http://arxiv.org/abs/1605.01556}{{\tt
  1605.01556}}].

\bibitem{Palladino:2016xsy}
A.~Palladino, M.~Spurio and F.~Vissani, \emph{{On the IceCube spectral
  anomaly}}, \href{http://dx.doi.org/10.1088/1475-7516/2016/12/045}{\emph{JCAP}
  {\bf 1612} (2016) 045}, [\href{http://arxiv.org/abs/1610.07015}{{\tt
  1610.07015}}].

\bibitem{Palladino:2017qda}
A.~Palladino, C.~Mascaretti and F.~Vissani, \emph{{On the compatibility of the
  IceCube results with a universal neutrino spectrum}},
  \href{http://dx.doi.org/10.1140/epjc/s10052-017-5273-z}{\emph{Eur. Phys. J.}
  {\bf C77} (2017) 684}, [\href{http://arxiv.org/abs/1708.02094}{{\tt
  1708.02094}}].

\bibitem{Winter:2014pya}
W.~Winter, \emph{{Describing the observed cosmic neutrinos by interactions of
  nuclei with matter}},
  \href{http://dx.doi.org/10.1103/PhysRevD.90.103003}{\emph{Phys. Rev.} {\bf
  D90} (2014) 103003}, [\href{http://arxiv.org/abs/1407.7536}{{\tt
  1407.7536}}].

\bibitem{Anchordoqui:2014hua}
L.~A. Anchordoqui et~al., \emph{{End of the cosmic neutrino energy spectrum}},
  \href{http://dx.doi.org/10.1016/j.physletb.2014.10.037}{\emph{Phys. Lett.}
  {\bf B739} (2014) 99--101}, [\href{http://arxiv.org/abs/1404.0622}{{\tt
  1404.0622}}].

\bibitem{Palomares-Ruiz:2015mka}
S.~Palomares-Ruiz, A.~C. Vincent and O.~Mena, \emph{{Spectral analysis of the
  high-energy IceCube neutrinos}},
  \href{http://dx.doi.org/10.1103/PhysRevD.91.103008}{\emph{Phys. Rev.} {\bf
  D91} (2015) 103008}, [\href{http://arxiv.org/abs/1502.02649}{{\tt
  1502.02649}}].

\bibitem{Aartsen:2015zva}
{\scshape IceCube} collaboration, M.~G. Aartsen et~al., \emph{{The IceCube
  Neutrino Observatory - Contributions to ICRC 2015 Part II: Atmospheric and
  astrophysical diffuse neutrino searches of all flavors}},  in
  \emph{{Proceedings of the 34th International Cosmic Ray Conference (ICRC
  2015)}}, 2015.
\newblock \href{http://arxiv.org/abs/1510.05223}{{\tt 1510.05223}}.

\bibitem{Anchordoqui:2016ewn}
L.~A. Anchordoqui, M.~M. Block, L.~Durand, P.~Ha, J.~F. Soriano and T.~J.
  Weiler, \emph{{Evidence for a break in the spectrum of astrophysical
  neutrinos}}, \href{http://dx.doi.org/10.1103/PhysRevD.95.083009}{\emph{Phys.
  Rev.} {\bf D95} (2017) 083009}, [\href{http://arxiv.org/abs/1611.07905}{{\tt
  1611.07905}}].

\bibitem{Ahlers:2013xia}
M.~Ahlers and K.~Murase, \emph{{Probing the galactic origin of the IceCube
  excess with gamma-rays}},
  \href{http://dx.doi.org/10.1103/PhysRevD.90.023010}{\emph{Phys. Rev.} {\bf
  D90} (2014) 023010}, [\href{http://arxiv.org/abs/1309.4077}{{\tt
  1309.4077}}].

\bibitem{Anchordoqui:2014rca}
L.~A. Anchordoqui, H.~Goldberg, T.~C. Paul, L.~H.~M. da~Silva and B.~J. Vlcek,
  \emph{{Estimating the contribution of galactic sources to the diffuse
  neutrino flux}},
  \href{http://dx.doi.org/10.1103/PhysRevD.90.123010}{\emph{Phys. Rev.} {\bf
  D90} (2014) 123010}, [\href{http://arxiv.org/abs/1410.0348}{{\tt
  1410.0348}}].

\bibitem{Troitsky:2015cnk}
S.~Troitsky, \emph{{Search for galactic disk and halo components in the arrival
  directions of high-energy astrophysical neutrinos}},
  \href{http://dx.doi.org/10.1134/S0021364015240133}{\emph{JETP Lett.} {\bf
  102} (2015) 785--788}, [\href{http://arxiv.org/abs/1511.01708}{{\tt
  1511.01708}}].

\bibitem{Kistler:2015oae}
M.~D. Kistler, \emph{{On TeV gamma rays and the search for galactic
  neutrinos}},  \href{http://arxiv.org/abs/1511.05199}{{\tt 1511.05199}}.

\bibitem{Denton:2017csz}
P.~B. Denton, D.~Marfatia and T.~J. Weiler, \emph{{The galactic contribution to
  IceCube's astrophysical neutrino flux}},
  \href{http://dx.doi.org/10.1088/1475-7516/2017/08/033}{\emph{JCAP} {\bf 1708}
  (2017) 033}, [\href{http://arxiv.org/abs/1703.09721}{{\tt 1703.09721}}].

\bibitem{Albert:2017oba}
{\scshape ANTARES} collaboration, A.~Albert et~al., \emph{{New constraints on
  all flavor galactic diffuse neutrino emission with the ANTARES telescope}},
  \href{http://dx.doi.org/10.1103/PhysRevD.96.062001}{\emph{Phys. Rev.} {\bf
  D96} (2017) 062001}, [\href{http://arxiv.org/abs/1705.00497}{{\tt
  1705.00497}}].

\bibitem{Aartsen:2017ujz}
{\scshape IceCube} collaboration, M.~G. Aartsen et~al., \emph{{Constraints on
  galactic neutrino emission with seven years of IceCube data}},
  \href{http://dx.doi.org/10.3847/1538-4357/aa8dfb}{\emph{Astrophys. J.} {\bf
  849} (2017) 67}, [\href{http://arxiv.org/abs/1707.03416}{{\tt 1707.03416}}].

\bibitem{Krings:2017pif}
{\scshape IceCube} collaboration, K.~Krings, \emph{{Using all-flavor and
  all-sky event selections by IceCube to search for neutrino emission from the
  galactic plane}}, \href{http://dx.doi.org/10.22323/1.301.0995}{\emph{PoS}
  {\bf ICRC2017} (2018) 995}.

\bibitem{Pagliaroli:2017fse}
G.~Pagliaroli and F.~L. Villante, \emph{{A multi-messenger study of the total
  galactic high-energy neutrino emission}},
  \href{http://dx.doi.org/10.1088/1475-7516/2018/08/035}{\emph{JCAP} {\bf 1808}
  (2018) 035}, [\href{http://arxiv.org/abs/1710.01040}{{\tt 1710.01040}}].

\bibitem{Neronov:2018ibl}
A.~Neronov, M.~Kachelrieß and D.~V. Semikoz, \emph{{Multimessenger gamma-ray
  counterpart of the IceCube neutrino signal}},
  \href{http://dx.doi.org/10.1103/PhysRevD.98.023004}{\emph{Phys. Rev.} {\bf
  D98} (2018) 023004}, [\href{http://arxiv.org/abs/1802.09983}{{\tt
  1802.09983}}].

\bibitem{Neronov:2016bnp}
A.~Neronov and D.~Semikoz, \emph{{Galactic and extragalactic contributions to
  the astrophysical muon neutrino signal}},
  \href{http://dx.doi.org/10.1103/PhysRevD.93.123002}{\emph{Phys. Rev.} {\bf
  D93} (2016) 123002}, [\href{http://arxiv.org/abs/1603.06733}{{\tt
  1603.06733}}].

\bibitem{Bai:2013nga}
Y.~Bai, R.~Lu and J.~Salvad\'o, \emph{{Geometric compatibility of IceCube
  TeV-PeV neutrino excess and its galactic dark matter origin}},
  \href{http://dx.doi.org/10.1007/JHEP01(2016)161}{\emph{JHEP} {\bf 01} (2016)
  161}, [\href{http://arxiv.org/abs/1311.5864}{{\tt 1311.5864}}].

\bibitem{Esmaili:2014rma}
A.~Esmaili, S.~K. Kang and P.~D. Serpico, \emph{{IceCube events and decaying
  dark matter: hints and constraints}},
  \href{http://dx.doi.org/10.1088/1475-7516/2014/12/054}{\emph{JCAP} {\bf 1412}
  (2014) 054}, [\href{http://arxiv.org/abs/1410.5979}{{\tt 1410.5979}}].

\bibitem{Chianese:2016opp}
M.~Chianese, G.~Miele, S.~Morisi and E.~Vitagliano, \emph{{Low energy IceCube
  data and a possible dark matter related excess}},
  \href{http://dx.doi.org/10.1016/j.physletb.2016.03.084}{\emph{Phys. Lett.}
  {\bf B757} (2016) 251--256}, [\href{http://arxiv.org/abs/1601.02934}{{\tt
  1601.02934}}].

\bibitem{Feldstein:2013kka}
B.~Feldstein, A.~Kusenko, S.~Matsumoto and T.~T. Yanagida, \emph{{Neutrinos at
  IceCube from heavy decaying dark matter}},
  \href{http://dx.doi.org/10.1103/PhysRevD.88.015004}{\emph{Phys. Rev.} {\bf
  D88} (2013) 015004}, [\href{http://arxiv.org/abs/1303.7320}{{\tt
  1303.7320}}].

\bibitem{Esmaili:2013gha}
A.~Esmaili and P.~D. Serpico, \emph{{Are IceCube neutrinos unveiling PeV-scale
  decaying dark matter?}},
  \href{http://dx.doi.org/10.1088/1475-7516/2013/11/054}{\emph{JCAP} {\bf 1311}
  (2013) 054}, [\href{http://arxiv.org/abs/1308.1105}{{\tt 1308.1105}}].

\bibitem{Ema:2013nda}
Y.~Ema, R.~Jinno and T.~Moroi, \emph{{Cosmic-ray neutrinos from the decay of
  long-lived particle and the recent IceCube result}},
  \href{http://dx.doi.org/10.1016/j.physletb.2014.04.021}{\emph{Phys. Lett.}
  {\bf B733} (2014) 120--125}, [\href{http://arxiv.org/abs/1312.3501}{{\tt
  1312.3501}}].

\bibitem{Bhattacharya:2014vwa}
A.~Bhattacharya, M.~H. Reno and I.~Sarcevic, \emph{{Reconciling neutrino flux
  from heavy dark matter decay and recent events at IceCube}},
  \href{http://dx.doi.org/10.1007/JHEP06(2014)110}{\emph{JHEP} {\bf 06} (2014)
  110}, [\href{http://arxiv.org/abs/1403.1862}{{\tt 1403.1862}}].

\bibitem{Zavala:2014dla}
J.~Zavala, \emph{{Galactic PeV neutrinos from dark matter annihilation}},
  \href{http://dx.doi.org/10.1103/PhysRevD.89.123516}{\emph{Phys. Rev.} {\bf
  D89} (2014) 123516}, [\href{http://arxiv.org/abs/1404.2932}{{\tt
  1404.2932}}].

\bibitem{Higaki:2014dwa}
T.~Higaki, R.~Kitano and R.~Sato, \emph{{Neutrinoful Universe}},
  \href{http://dx.doi.org/10.1007/JHEP07(2014)044}{\emph{JHEP} {\bf 07} (2014)
  044}, [\href{http://arxiv.org/abs/1405.0013}{{\tt 1405.0013}}].

\bibitem{Ema:2014ufa}
Y.~Ema, R.~Jinno and T.~Moroi, \emph{{Cosmological implications of high-energy
  neutrino emission from the decay of long-lived particle}},
  \href{http://dx.doi.org/10.1007/JHEP10(2014)150}{\emph{JHEP} {\bf 1410}
  (2014) 150}, [\href{http://arxiv.org/abs/1408.1745}{{\tt 1408.1745}}].

\bibitem{Rott:2014kfa}
C.~Rott, K.~Kohri and S.~C. Park, \emph{{Superheavy dark matter and IceCube
  neutrino signals: Bounds on decaying dark matter}},
  \href{http://dx.doi.org/10.1103/PhysRevD.92.023529}{\emph{Phys. Rev.} {\bf
  D92} (2015) 023529}, [\href{http://arxiv.org/abs/1408.4575}{{\tt
  1408.4575}}].

\bibitem{Fong:2014bsa}
C.~S. Fong, H.~Minakata, B.~Panes and R.~Z. Funchal, \emph{{Possible
  interpretations of IceCube high-energy neutrino events}},
  \href{http://dx.doi.org/10.1007/JHEP02(2015)189}{\emph{JHEP} {\bf 1502}
  (2015) 189}, [\href{http://arxiv.org/abs/1411.5318}{{\tt 1411.5318}}].

\bibitem{Daikoku:2015vsa}
Y.~Daikoku and H.~Okada, \emph{{PeV scale right handed neutrino dark matter in
  $S_4$ flavor symmetric extra U(1) model}},
  \href{http://dx.doi.org/10.1103/PhysRevD.91.075009}{\emph{Phys. Rev.} {\bf
  D91} (2015) 075009}, [\href{http://arxiv.org/abs/1502.07032}{{\tt
  1502.07032}}].

\bibitem{Murase:2015gea}
K.~Murase, R.~Laha, S.~Ando and M.~Ahlers, \emph{{Testing the dark matter
  scenario for PeV neutrinos observed in IceCube}},
  \href{http://dx.doi.org/10.1103/PhysRevLett.115.071301}{\emph{Phys. Rev.
  Lett.} {\bf 115} (2015) 071301}, [\href{http://arxiv.org/abs/1503.04663}{{\tt
  1503.04663}}].

\bibitem{Esmaili:2015xpa}
A.~Esmaili and P.~D. Serpico, \emph{{Gamma-ray bounds from EAS detectors and
  heavy decaying dark matter constraints}},
  \href{http://dx.doi.org/10.1088/1475-7516/2015/10/014}{\emph{JCAP} {\bf 1510}
  (2015) 014}, [\href{http://arxiv.org/abs/1505.06486}{{\tt 1505.06486}}].

\bibitem{Aisati:2015vma}
C.~El~Aisati, M.~Gustafsson and T.~Hambye, \emph{{New search for monochromatic
  neutrinos from dark matter decay}},
  \href{http://dx.doi.org/10.1103/PhysRevD.92.123515}{\emph{Phys. Rev.} {\bf
  D92} (2015) 123515}, [\href{http://arxiv.org/abs/1506.02657}{{\tt
  1506.02657}}].

\bibitem{Roland:2015yoa}
S.~B. Roland, B.~Shakya and J.~D. Wells, \emph{{PeV neutrinos and a 3.5 keV
  x-ray line from a PeV-scale supersymmetric neutrino sector}},
  \href{http://dx.doi.org/10.1103/PhysRevD.92.095018}{\emph{Phys. Rev.} {\bf
  D92} (2015) 095018}, [\href{http://arxiv.org/abs/1506.08195}{{\tt
  1506.08195}}].

\bibitem{Anchordoqui:2015lqa}
L.~A. Anchordoqui et~al., \emph{{IceCube neutrinos, decaying dark matter, and
  the Hubble constant}},
  \href{http://dx.doi.org/10.1103/PhysRevD.92.061301}{\emph{Phys. Rev.} {\bf
  D92} (2015) 061301}, [\href{http://arxiv.org/abs/1506.08788}{{\tt
  1506.08788}}].

\bibitem{Boucenna:2015tra}
S.~M. Boucenna et~al., \emph{{Decaying leptophilic dark matter at IceCube}},
  \href{http://dx.doi.org/10.1088/1475-7516/2015/12/055}{\emph{JCAP} {\bf 1512}
  (2015) 055}, [\href{http://arxiv.org/abs/1507.01000}{{\tt 1507.01000}}].

\bibitem{Ko:2015nma}
P.~Ko and Y.~Tang, \emph{{IceCube events from heavy DM decays through the
  right-handed neutrino portal}},
  \href{http://dx.doi.org/10.1016/j.physletb.2015.10.021}{\emph{Phys. Lett.}
  {\bf B751} (2015) 81--88}, [\href{http://arxiv.org/abs/1508.02500}{{\tt
  1508.02500}}].

\bibitem{EsmailiTaklimi:2016bbx}
A.~Esmaili and P.~Serpico, \emph{{Interpreting the IceCube events by decaying
  dark matter}}, {\emph{PoS} {\bf DSU2015} (2016) 047}.

\bibitem{Esmaili:2016swq}
A.~Esmaili, A.~Palladino and F.~Vissani, \emph{{A discussion of IceCube
  neutrino events, circa 2015}},
  \href{http://dx.doi.org/10.1051/epjconf/201611611002}{\emph{EPJ Web Conf.}
  {\bf 116} (2016) 11002}.

\bibitem{Dev:2016uxj}
P.~S.~B. Dev, D.~K. Ghosh and W.~Rodejohann, \emph{{R-parity violating
  supersymmetry at IceCube}},
  \href{http://dx.doi.org/10.1016/j.physletb.2016.08.066}{\emph{Phys. Lett.}
  {\bf B762} (2016) 116--123}, [\href{http://arxiv.org/abs/1605.09743}{{\tt
  1605.09743}}].

\bibitem{Fiorentin:2016avj}
M.~Re~Fiorentin, V.~Niro and N.~Fornengo, \emph{{A consistent model for
  leptogenesis, dark matter and the IceCube signal}},
  \href{http://dx.doi.org/10.1007/JHEP11(2016)022}{\emph{JHEP} {\bf 11} (2016)
  022}, [\href{http://arxiv.org/abs/1606.04445}{{\tt 1606.04445}}].

\bibitem{Dev:2016qbd}
P.~S.~B. Dev, D.~Kazanas, R.~N. Mohapatra, V.~L. Teplitz and Y.~Zhang,
  \emph{{Heavy right-handed neutrino dark matter and PeV neutrinos at
  IceCube}}, \href{http://dx.doi.org/10.1088/1475-7516/2016/08/034}{\emph{JCAP}
  {\bf 1608} (2016) 034}, [\href{http://arxiv.org/abs/1606.04517}{{\tt
  1606.04517}}].

\bibitem{DiBari:2016guw}
P.~Di~Bari, P.~O. Ludl and S.~Palomares-Ruiz, \emph{{Unifying leptogenesis,
  dark matter and high-energy neutrinos with right-handed neutrino mixing via
  Higgs portal}},
  \href{http://dx.doi.org/10.1088/1475-7516/2016/11/044}{\emph{JCAP} {\bf 1611}
  (2016) 044}, [\href{http://arxiv.org/abs/1606.06238}{{\tt 1606.06238}}].

\bibitem{Chianese:2016smc}
M.~Chianese and A.~Merle, \emph{{A consistent theory of decaying dark matter
  connecting IceCube to the Sesame Street}},
  \href{http://dx.doi.org/10.1088/1475-7516/2017/04/017}{\emph{JCAP} {\bf 1704}
  (2017) 017}, [\href{http://arxiv.org/abs/1607.05283}{{\tt 1607.05283}}].

\bibitem{Chianese:2016kpu}
M.~Chianese, G.~Miele and S.~Morisi, \emph{{Dark matter interpretation of low
  energy IceCube MESE excess}},
  \href{http://dx.doi.org/10.1088/1475-7516/2017/01/007}{\emph{JCAP} {\bf 1701}
  (2017) 007}, [\href{http://arxiv.org/abs/1610.04612}{{\tt 1610.04612}}].

\bibitem{Kuznetsov:2016fjt}
M.~{\relax Yu}. Kuznetsov, \emph{{Hadronically decaying heavy dark matter and
  high-energy neutrino limits}},
  \href{http://dx.doi.org/10.1134/S0021364017090028}{\emph{JETP Lett.} (2017)
  1--7}, [\href{http://arxiv.org/abs/1611.08684}{{\tt 1611.08684}}].

\bibitem{Cohen:2016uyg}
T.~Cohen, K.~Murase, N.~L. Rodd, B.~R. Safdi and Y.~Soreq, \emph{{$\gamma$-ray
  constraints on decaying dark matter and implications for IceCube}},
  \href{http://dx.doi.org/10.1103/PhysRevLett.119.021102}{\emph{Phys. Rev.
  Lett.} {\bf 119} (2017) 021102}, [\href{http://arxiv.org/abs/1612.05638}{{\tt
  1612.05638}}].

\bibitem{Borah:2017xgm}
D.~Borah, A.~Dasgupta, U.~K. Dey, S.~Patra and G.~Tomar, \emph{{Multi-component
  fermionic dark matter and IceCube PeV scale neutrinos in left-right model
  with gauge unification}},
  \href{http://dx.doi.org/10.1007/JHEP09(2017)005}{\emph{JHEP} {\bf 09} (2017)
  005}, [\href{http://arxiv.org/abs/1704.04138}{{\tt 1704.04138}}].

\bibitem{Hiroshima:2017hmy}
N.~Hiroshima, R.~Kitano, K.~Kohri and K.~Murase, \emph{{High-energy neutrinos
  from multibody decaying dark matter}},
  \href{http://dx.doi.org/10.1103/PhysRevD.97.023006}{\emph{Phys. Rev.} {\bf
  D97} (2018) 023006}, [\href{http://arxiv.org/abs/1705.04419}{{\tt
  1705.04419}}].

\bibitem{Bhattacharya:2017jaw}
A.~Bhattacharya, A.~Esmaili, S.~Palomares-Ruiz and I.~Sarcevic, \emph{{Probing
  decaying heavy dark matter with the 4-year IceCube HESE data}},
  \href{http://dx.doi.org/10.1088/1475-7516/2017/07/027}{\emph{JCAP} {\bf 1707}
  (2017) 027}, [\href{http://arxiv.org/abs/1706.05746}{{\tt 1706.05746}}].

\bibitem{Chakravarty:2017hcy}
G.~K. Chakravarty, N.~Khan and S.~Mohanty, \emph{{Dark matter and Inflation in
  PeV scale SUSY}},  \href{http://arxiv.org/abs/1707.03853}{{\tt 1707.03853}}.

\bibitem{Chianese:2017nwe}
M.~Chianese, G.~Miele and S.~Morisi, \emph{{Interpreting IceCube 6-year HESE
  data as an evidence for hundred TeV decaying dark matter}},
  \href{http://dx.doi.org/10.1016/j.physletb.2017.09.016}{\emph{Phys. Lett.}
  {\bf B773} (2017) 591--595}, [\href{http://arxiv.org/abs/1707.05241}{{\tt
  1707.05241}}].

\bibitem{Dhuria:2017ihq}
M.~Dhuria and V.~Rentala, \emph{{PeV scale supersymmetry breaking and the
  IceCube neutrino flux}},
  \href{http://dx.doi.org/10.1007/JHEP09(2018)004}{\emph{JHEP} {\bf 09} (2018)
  004}, [\href{http://arxiv.org/abs/1712.07138}{{\tt 1712.07138}}].

\bibitem{Aartsen:2018mxl}
{\scshape IceCube} collaboration, M.~G. Aartsen et~al., \emph{{Search for
  neutrinos from decaying dark matter with IceCube}},
  \href{http://dx.doi.org/10.1140/epjc/s10052-018-6273-3}{\emph{Eur. Phys. J.}
  {\bf C78} (2018) 831}, [\href{http://arxiv.org/abs/1804.03848}{{\tt
  1804.03848}}].

\bibitem{Sui:2018bbh}
Y.~Sui and P.~S. Bhupal~Dev, \emph{{A combined astrophysical and dark matter
  interpretation of the IceCube HESE and throughgoing muon events}},
  \href{http://dx.doi.org/10.1088/1475-7516/2018/07/020}{\emph{JCAP} {\bf 1807}
  (2018) 020}, [\href{http://arxiv.org/abs/1804.04919}{{\tt 1804.04919}}].

\bibitem{Chianese:2018ijk}
M.~Chianese, G.~Miele, S.~Morisi and E.~Peinado, \emph{{Neutrinophilic dark
  matter in the epoch of IceCube and Fermi-LAT}},
  \href{http://dx.doi.org/10.1088/1475-7516/2018/12/016}{\emph{JCAP} {\bf 1812}
  (2018) 016}, [\href{http://arxiv.org/abs/1808.02486}{{\tt 1808.02486}}].

\bibitem{Blanco:2018esa}
C.~Blanco and D.~Hooper, \emph{{Constraints on decaying dark matter from the
  isotropic gamma-ray background}},
  \href{http://dx.doi.org/10.1088/1475-7516/2019/03/019}{\emph{JCAP} {\bf 2019}
  (2020) 019}, [\href{http://arxiv.org/abs/1811.05988}{{\tt 1811.05988}}].

\bibitem{Aartsen:2014muf}
{\scshape IceCube} collaboration, M.~G. Aartsen et~al., \emph{{Atmospheric and
  astrophysical neutrinos above 1 TeV interacting in IceCube}},
  \href{http://dx.doi.org/10.1103/PhysRevD.91.022001}{\emph{Phys. Rev.} {\bf
  D91} (2015) 022001}, [\href{http://arxiv.org/abs/1410.1749}{{\tt
  1410.1749}}].

\bibitem{Yuan:2018}
{\scshape IceCube} collaboration, T.~Yuan, ``{\it New measurements with
  high-energy neutrinos in IceCube}.'' Poster at Neutrino 2018, Heidelberg,
  Germany, June 4-9, 2018.

\bibitem{Griest:1989wd}
K.~Griest and M.~Kamionkowski, \emph{{Unitarity limits on the mass and radius
  of dark matter particles}},
  \href{http://dx.doi.org/10.1103/PhysRevLett.64.615}{\emph{Phys. Rev. Lett.}
  {\bf 64} (1990) 615}.

\bibitem{Hui:2001wy}
L.~Hui, \emph{{Unitarity bounds and the cuspy halo problem}},
  \href{http://dx.doi.org/10.1103/PhysRevLett.86.3467}{\emph{Phys. Rev. Lett.}
  {\bf 86} (2001) 3467--3470},
  [\href{http://arxiv.org/abs/astro-ph/0102349}{{\tt astro-ph/0102349}}].

\bibitem{Feldman:1997qc}
G.~J. Feldman and R.~D. Cousins, \emph{{A unified approach to the classical
  statistical analysis of small signals}},
  \href{http://dx.doi.org/10.1103/PhysRevD.57.3873}{\emph{Phys. Rev.} {\bf D57}
  (1998) 3873--3889}, [\href{http://arxiv.org/abs/physics/9711021}{{\tt
  physics/9711021}}].

\bibitem{Ackermann:2014usa}
{\scshape Fermi-LAT} collaboration, M.~Ackermann et~al., \emph{{The spectrum of
  isotropic diffuse gamma-ray emission between 100 MeV and 820 GeV}},
  \href{http://dx.doi.org/10.1088/0004-637X/799/1/86}{\emph{Astrophys. J.} {\bf
  799} (2015) 86}, [\href{http://arxiv.org/abs/1410.3696}{{\tt 1410.3696}}].

\bibitem{Fornasa:2015qua}
M.~Fornasa and M.~A. S\'{a}nchez-Conde, \emph{{The nature of the diffuse
  gamma-ray background}},
  \href{http://dx.doi.org/10.1016/j.physrep.2015.09.002}{\emph{Phys. Rept.}
  {\bf 598} (2015) 1--58}, [\href{http://arxiv.org/abs/1502.02866}{{\tt
  1502.02866}}].

\bibitem{Murase:2013rfa}
K.~Murase, M.~Ahlers and B.~C. Lacki, \emph{{Testing the hadronuclear origin of
  PeV neutrinos observed with IceCube}},
  \href{http://dx.doi.org/10.1103/PhysRevD.88.121301}{\emph{Phys. Rev.} {\bf
  D88} (2013) 121301}, [\href{http://arxiv.org/abs/1306.3417}{{\tt
  1306.3417}}].

\bibitem{Hisano:2004ds}
J.~Hisano, S.~Matsumoto, M.~M. Nojiri and O.~Saito, \emph{{Non-perturbative
  effect on dark matter annihilation and gamma ray signature from galactic
  center}}, \href{http://dx.doi.org/10.1103/PhysRevD.71.063528}{\emph{Phys.
  Rev.} {\bf D71} (2005) 063528},
  [\href{http://arxiv.org/abs/hep-ph/0412403}{{\tt hep-ph/0412403}}].

\bibitem{Profumo:2005xd}
S.~Profumo, \emph{{TeV gamma-rays and the largest masses and annihilation cross
  sections of neutralino dark matter}},
  \href{http://dx.doi.org/10.1103/PhysRevD.72.103521}{\emph{Phys. Rev.} {\bf
  D72} (2005) 103521}, [\href{http://arxiv.org/abs/astro-ph/0508628}{{\tt
  astro-ph/0508628}}].

\bibitem{Lattanzi:2008qa}
M.~Lattanzi and J.~I. Silk, \emph{{Can the WIMP annihilation boost factor be
  boosted by the Sommerfeld enhancement?}},
  \href{http://dx.doi.org/10.1103/PhysRevD.79.083523}{\emph{Phys. Rev.} {\bf
  D79} (2009) 083523}, [\href{http://arxiv.org/abs/0812.0360}{{\tt
  0812.0360}}].

\bibitem{Feng:2010zp}
J.~L. Feng, M.~Kaplinghat and H.-B. Yu, \emph{{Sommerfeld enhancements for
  thermal relic dark matter}},
  \href{http://dx.doi.org/10.1103/PhysRevD.82.083525}{\emph{Phys. Rev.} {\bf
  D82} (2010) 083525}, [\href{http://arxiv.org/abs/1005.4678}{{\tt
  1005.4678}}].

\bibitem{Navarro:1995iw}
J.~F. Navarro, C.~S. Frenk and S.~D.~M. White, \emph{{The structure of cold
  dark matter halos}}, \href{http://dx.doi.org/10.1086/177173}{\emph{Astrophys.
  J.} {\bf 462} (1996) 563--575},
  [\href{http://arxiv.org/abs/astro-ph/9508025}{{\tt astro-ph/9508025}}].

\bibitem{Navarro:1996gj}
J.~F. Navarro, C.~S. Frenk and S.~D.~M. White, \emph{{A universal density
  profile from hierarchical clustering}},
  \href{http://dx.doi.org/10.1086/304888}{\emph{Astrophys. J.} {\bf 490} (1997)
  493--508}, [\href{http://arxiv.org/abs/astro-ph/9611107}{{\tt
  astro-ph/9611107}}].

\bibitem{Springel:2008cc}
V.~Springel et~al., \emph{{The Aquarius project: The subhalos of galactic
  halos}}, \href{http://dx.doi.org/10.1111/j.1365-2966.2008.14066.x}{\emph{Mon.
  Not. Roy. Astron. Soc.} {\bf 391} (2008) 1685--1711},
  [\href{http://arxiv.org/abs/0809.0898}{{\tt 0809.0898}}].

\bibitem{Hellwing:2015upa}
W.~A. Hellwing et~al., \emph{{The Copernicus Complexio: A high-resolution view
  of the small-scale Universe}},
  \href{http://dx.doi.org/10.1093/mnras/stw214}{\emph{Mon. Not. Roy. Astron.
  Soc.} {\bf 457} (2016) 3492--3509},
  [\href{http://arxiv.org/abs/1505.06436}{{\tt 1505.06436}}].

\bibitem{Rodriguez-Puebla:2016ofw}
A.~Rodr{\'{\i}}guez-Puebla, P.~Behroozi, J.~Primack, A.~Klypin, C.~Lee and
  D.~Hellinger, \emph{{Halo and subhalo demographics with Planck cosmological
  parameters: Bolshoi–Planck and MultiDark–Planck simulations}},
  \href{http://dx.doi.org/10.1093/mnras/stw1705}{\emph{Mon. Not. Roy. Astron.
  Soc.} {\bf 462} (2016) 893--916},
  [\href{http://arxiv.org/abs/1602.04813}{{\tt 1602.04813}}].

\bibitem{Diemand:2008in}
J.~Diemand et~al., \emph{{Clumps and streams in the local dark matter
  distribution}}, \href{http://dx.doi.org/10.1038/nature07153}{\emph{Nature}
  {\bf 454} (2008) 735--738}, [\href{http://arxiv.org/abs/0805.1244}{{\tt
  0805.1244}}].

\bibitem{Pieri:2009je}
L.~Pieri, J.~Lavalle, G.~Bertone and E.~Branchini, \emph{{Implications of
  high-resolution simulations on indirect dark matter searches}},
  \href{http://dx.doi.org/10.1103/PhysRevD.83.023518}{\emph{Phys. Rev.} {\bf
  D83} (2011) 023518}, [\href{http://arxiv.org/abs/0908.0195}{{\tt
  0908.0195}}].

\bibitem{Moline:2016pbm}
{\'A}.~Molin{\'e}, M.~A. Sánchez-Conde, S.~Palomares-Ruiz and F.~Prada,
  \emph{{Characterization of subhalo structural properties and implications for
  dark matter annihilation signals}},
  \href{http://dx.doi.org/10.1093/mnras/stx026}{\emph{Mon. Not. Roy. Astron.
  Soc.} {\bf 466} (2017) 4974--4990},
  [\href{http://arxiv.org/abs/1603.04057}{{\tt 1603.04057}}].

\bibitem{Serpico:2011in}
P.~D. Serpico, E.~Sefusatti, M.~Gustafsson and G.~Zaharijas,
  \emph{{Extragalactic gamma-ray signal from dark matter annihilation: A power
  spectrum based computation}},
  \href{http://dx.doi.org/10.1111/j.1745-3933.2011.01212.x}{\emph{Mon. Not.
  Roy. Astron. Soc.} {\bf 421} (2012) L87--L91},
  [\href{http://arxiv.org/abs/1109.0095}{{\tt 1109.0095}}].

\bibitem{Sefusatti:2014vha}
E.~Sefusatti, G.~Zaharijas, P.~D. Serpico, D.~Theurel and M.~Gustafsson,
  \emph{{Extragalactic gamma-ray signal from dark matter annihilation: an
  appraisal}}, \href{http://dx.doi.org/10.1093/mnras/stu686}{\emph{Mon. Not.
  Roy. Astron. Soc.} {\bf 441} (2014) 1861--1878},
  [\href{http://arxiv.org/abs/1401.2117}{{\tt 1401.2117}}].

\bibitem{Cooray:2002dia}
A.~Cooray and R.~K. Sheth, \emph{{Halo models of large scale structure}},
  \href{http://dx.doi.org/10.1016/S0370-1573(02)00276-4}{\emph{Phys. Rept.}
  {\bf 372} (2002) 1--129}, [\href{http://arxiv.org/abs/astro-ph/0206508}{{\tt
  astro-ph/0206508}}].

\bibitem{Bergstrom:2001jj}
L.~Bergstrom, J.~Edsjo and P.~Ullio, \emph{{Spectral gamma-ray signatures of
  cosmological dark matter annihilation}},
  \href{http://dx.doi.org/10.1103/PhysRevLett.87.251301}{\emph{Phys. Rev.
  Lett.} {\bf 87} (2001) 251301},
  [\href{http://arxiv.org/abs/astro-ph/0105048}{{\tt astro-ph/0105048}}].

\bibitem{Ullio:2002pj}
P.~Ullio, L.~Bergstrom, J.~Edsjo and C.~G. Lacey, \emph{{Cosmological dark
  matter annihilations into gamma-rays - a closer look}},
  \href{http://dx.doi.org/10.1103/PhysRevD.66.123502}{\emph{Phys. Rev.} {\bf
  D66} (2002) 123502}, [\href{http://arxiv.org/abs/astro-ph/0207125}{{\tt
  astro-ph/0207125}}].

\bibitem{Taylor:2002zd}
J.~E. Taylor and J.~Silk, \emph{{The clumpiness of cold dark matter:
  Implications for the annihilation signal}},
  \href{http://dx.doi.org/10.1046/j.1365-8711.2003.06201.x}{\emph{Mon. Not.
  Roy. Astron. Soc.} {\bf 339} (2003) 505},
  [\href{http://arxiv.org/abs/astro-ph/0207299}{{\tt astro-ph/0207299}}].

\bibitem{Moline:2014xua}
{\'A}.~Moliné, A.~Ibarra and S.~Palomares-Ruiz, \emph{{Future sensitivity of
  neutrino telescopes to dark matter annihilations from the cosmic diffuse
  neutrino signal}},
  \href{http://dx.doi.org/10.1088/1475-7516/2015/06/005}{\emph{JCAP} {\bf 1506}
  (2015) 005}, [\href{http://arxiv.org/abs/1412.4308}{{\tt 1412.4308}}].

\end{thebibliography}\endgroup

\end{document}